\documentclass[journal,onecolumn]{IEEEtran}

\usepackage[T1]{fontenc}
\usepackage[latin1]{inputenc}
\usepackage{color}
\usepackage{psfrag}
\usepackage[dvips]{graphicx}
\usepackage{tikz}
\usepackage{pgfplots}
\pgfplotsset{compat=1.12}
\usepackage{enumerate}
\usepackage{rotating}
\usepackage{srcltx}
\usepackage{booktabs}
\usepackage{multicol}
\usepackage{enumitem}
\usepackage{textcomp}
\usetikzlibrary{shapes,arrows}
\usepackage{array,multirow,graphicx}
\usetikzlibrary{arrows,decorations.pathmorphing,backgrounds,positioning,fit,matrix}

\usepackage[noadjust]{cite}
\usepackage[cmex10]{amsmath}
\usepackage{amsfonts}
\usepackage{amssymb}
\usepackage{ntheorem}
\usepackage{dsfont}    
\usepackage{mathtools} 
\usepackage{stackengine}
\usepackage{enumitem}
\usepackage{xcolor}

\usepackage{algorithmic}
\usepackage{array}
\usepackage{graphicx}
\usepackage{caption}
\usepackage{subcaption}

\theoremseparator{.}
\newtheorem{corollary}{Corollary}
\newtheorem{prop}{Proposition}
\newtheorem{lem}{Lemma}
\newtheorem{theorem}{Theorem}
\theorembodyfont{\upshape}
\theoremseparator{.}

\newtheorem{remark}{Remark}
\newtheorem*{example*}{Example}

\newcommand{\eu}{\mathrm{e}}

\newcommand{\Tr}{\mathrm{Tr}}

\newcommand{\mmse}{\mathrm{mmse}}

\newcommand{\E}{\mathbb{E}}

\newcommand{\X}{\mathbf{X}}
\newcommand{\N}{\mathbf{N}}
\newcommand{\Y}{\mathbf{Y}}
\newcommand{\I}{\mathbf{\mathsf{I}}}
\newcommand{\U}{\mathbf{U}}

\allowdisplaybreaks

\long\def\symbolfootnote[#1]#2{\begingroup%
	\def\thefootnote{\fnsymbol{footnote}}\footnote[#1]{#2}\endgroup} 
\IEEEoverridecommandlockouts    

\begin{document}
\title{A General Derivative Identity for the Conditional Mean Estimator in Gaussian Noise and Some Applications} 
\author{\IEEEauthorblockN{Alex~Dytso\IEEEauthorrefmark{1},  H. Vincent~Poor\IEEEauthorrefmark{2}, and Shlomo Shamai (Shitz)\IEEEauthorrefmark{3}\vspace{5pt}}\\
\IEEEauthorblockA{\IEEEauthorrefmark{1}Department of Electrical and Computer Engineering, New Jersey Institute of Technology\\
\IEEEauthorrefmark{2}Department of Electrical Engineering, Princeton University\\
				  \IEEEauthorrefmark{3}Department of Electrical Engineering, Technion -- Israel Institute of Technology}  \\
				   \IEEEauthorblockA{Email: alex.dytso@njit.edu\IEEEauthorrefmark{1},  poor@princeton.edu\IEEEauthorrefmark{2} and sshlomo@ee.technion.ac.il\IEEEauthorrefmark{3}}

				\thanks{The results in this paper were presented in part at the IEEE International Symposium on Information Theory, Los Angeles, USA, July 2020 \cite{dytso2020general}, and at the IEEE Information Theory Workshop, Riva del Garda, Italy, April 2021 \cite{ITW2020Dytso}.
				}
   \thanks{This work was supported in part by the U. S. National Science Foundation under Grant CCF-1908308 and by the United States-Israel Binational Science Foundation, under Grant BSF-2018710.}

				   }

\maketitle
\begin{abstract}
Consider a channel  $\Y=\X+ \N$ where $\X$ is an $n$-dimensional random vector, and $\N$ is a multivariate Gaussian vector with a full-rank covariance matrix $\boldsymbol{\mathsf{K}}_{\N}$. The object under consideration in this paper is the  conditional mean  of $\X$ given  $\Y={\bf y}$, that is  ${\bf y} \mapsto \E[\X|\Y={\bf y}]$.  
Several identities in the literature connect $ \E[\X|\Y={\bf y}]$ to other quantities such as the conditional variance, score functions, and higher-order conditional moments. The objective of this paper is to provide a unifying view of these identities. 

In the first part of the paper, a general derivative identity for the conditional mean estimator is derived. Specifically, for the Markov chain $\U \leftrightarrow \X  \leftrightarrow \Y$, it is shown that the Jacobian matrix of  $\E[\U|\Y={\bf y}]$ is given by   $\boldsymbol{\mathsf{K}}_{\N}^{-1}   \boldsymbol{\mathsf{Cov}} ( \X, \U | \Y={\bf y})$ where $\boldsymbol{\mathsf{Cov}} ( \X, \U | \Y={\bf y})$ is the conditional covariance. 

In the second part of the paper, via various choices of the random vector $\U$, the new identity is used to recover and generalize many of the known identities and derive some new identities. First, a simple proof of the Hatsel and Nolte identity for the conditional variance is shown. Second, a simple proof of the recursive identity due to Jaffer is provided. The Jaffer identity is then further explored, and several equivalent stamens are derived, such as an identity for the higher-order conditional expectation (i.e., $\E[\X^k|\Y]$) in terms of the derivatives of the conditional expectation. Third, a new fundamental connection between the conditional cumulants and the conditional expectation is demonstrated. In particular, in the univariate case, it is shown that the $k$-th derivative of the conditional expectation is proportional to the $(k+1)$-th conditional cumulant. A similar expression is derived in the multivariate case.

The third part of the paper considers various applications of the derived identities (mostly in the scalar case).  In a first application, using the new identity for higher-order derivatives of the conditional expectation, a power series representation of the conditional expectation is derived. The power series representation, together with the Lagrange inversion theorem, is then used to find an expression for the compositional inverse of $y \mapsto \E[X|Y=y]$. In a second application, the conditional expectation is viewed as a random variable and the probability distribution of $\E[X|Y]$ and probability distribution of the estimator error $(X-\E[X|Y])$ are derived. In the third application, the new identities are used to show that the higher-order conditional expectations and the conditional cumulants depended on the joint distribution only through the marginal of $Y$. This observation is then used to construct consistent estimators (known as the empirical Bayes estimators) of the higher-order conditional expectations and the conditional cumulants from an independent and identically distributed sequence $Y_1,\ldots, Y_n$. 
\end{abstract}  
\begin{IEEEkeywords}
Vector Gaussian noise, conditional mean estimator,  conditional cumulant, minimum mean squared error, empirical Bayes.
\end{IEEEkeywords}

\section{Introduction}
Consider a model given by the following input-output relationship: 
\begin{equation}
\Y=\X+ \N,  \label{eq:Gaussian_Model}
\end{equation}
where $\N \in \mathbb{R}^n$ is a zero mean, normally distributed with the covariance matrix $\boldsymbol{\mathsf{K}}_{\N}$, and independent of $\X \in  \mathbb{R}^n$.  Throughout the paper  $\boldsymbol{\mathsf{K}}_{\N}$ is assumed to be a  positive definite matrix, and   we make no assumptions about the probability distribution of $\X$. 
In the case of $n=1$, we denote $\boldsymbol{\mathsf{K}}_{\N}=\sigma^2$.  Throughout the paper  deterministic scalar quantities are denoted by lowercase letters, scalar random variables are denoted by uppercase letters,  vectors are denoted by bold lowercase letters, random vectors by bold uppercase letters, and matrices by bold uppercase sans serif letters (e.g., $x$, $X$, $\mathbf{x}$, $\mathbf{X}$, $\boldsymbol{\mathsf{X}}$).

In this work, we are interested in studying properties of the conditional mean estimator  of the input $\X$ given the output $\Y$ according to \eqref{eq:Gaussian_Model}, that is 
 \begin{equation}
 \E[ \X | \Y={\bf y} ]=\int {\bf x} \, {\rm d} P_{\X | \Y={\bf y}}({\bf x}), \, {\bf y} \in \mathbb{R}^n. 
 \end{equation}
 The conditional expectation is of  interest in view of the wide range of applications of the conditional expectation in fields such estimation theory and information theory. For example, the conditional expectation is known to be the unique optimal estimator under a very large family of loss functions, namely Bregman divergences \cite{banerjee2005optimality}.   In this work, we will view the conditional mean estimator as a function of channel realizations, that is ${\bf y} \mapsto \E[\X| \Y={\bf y} ]$, and will be interested in characterizing analytical properties the conditional expectation. Specifically, we focus on characterizing various derivative identities involving conditional expectations and will show a few applications of these identities. 
  
There are several derivative identities in the literature that relate  the conditional mean estimator to other quantities such as the score function and the conditional variance. Such identities are often used in information theory to give way to estimation theoretic arguments (e.g., the I-MMSE relationship \cite{guo2013interplay,palomar2006gradient}). In estimation theory such identities are often used  to design new estimation procedures  (e.g., empirical Bayes  \cite{efron2005local,efron2016computer}) or establish connections to detection theory \cite{jaffer1972relations}.      Perhaps   the most  well-known such identity is
 \begin{equation}
\E[\X | \Y={\bf y}]=  {\bf y}+ \boldsymbol{\mathsf{K}}_{\N}  \nabla_{ {\bf y} } \log f_\Y({\bf y}), \,   {\bf y} \in \mathbb{R}^n \label{eq:TweediesFormulaGaussian}
\end{equation}  
where $f_\Y({\bf y})$ is the probability density function (pdf) of $\Y$.  We note that  the quantity $ \nabla_{ {\bf y} } \log f_\Y({\bf y})= \frac{   \nabla_{ {\bf y}}  f_\Y({\bf y})  }{ f_\Y({\bf y})}$ is commonly known as the score function. The scalar version of the identity in \eqref{eq:TweediesFormulaGaussian} has been derived by Robbins in \cite{robbins1956empirical} where he credits  Maurice Tweedie for the derivation.     The vector version  of the identity in \eqref{eq:TweediesFormulaGaussian} was derived by Esposito in \cite{esposito1968relation}.  Therefore, throughout this paper, we refer to the identity in \eqref{eq:TweediesFormulaGaussian} as Tweedie-Robbins-Esposito identity or TRE for short.

 The observation  that, via the TRE identity,   the conditional expectation can be represented  in terms of only the marginal distribution of the output $\Y$ has led to the development of the  empirical Bayes procedure \cite{robbins1956empirical};  the interested reader is referred  to \cite{efron2016computer}  for an overview of the empirical Bayes procedure.  In addition to developing statistical procedures the TRE identity in \eqref{eq:TweediesFormulaGaussian} can considerably simplify the computation of $\E[\X | \Y={\bf y}]$ itself as we do not
need to derive the conditional distribution $P_{\X|\Y}$ and only need to compute $f_\Y({\bf y})$ and the gradient of $f_\Y({\bf y})$. For an example of such an application, the interested reader is referred to  \cite{dytso2019capacity}  where the TRE identity was used to compute $\E[\X | \Y={\bf y}]$ for the case where $\X$ is uniform on a sphere in $\mathbb{R}^n$.  In information theory,  the TRE identity has also been used in the proofs of the scalar and vector versions of the I-MMSE relationship in \cite{I-MMSE} and \cite{palomar2006gradient}, respectively.  
 
Several other such identities, which we will survey through the paper,  find wide applicability in variety of different fields. In this work, it is shown that many of the known identities in the literature can be derived systematically from a single unifying derivative identity. Moreover, we use this new identity to derive several generalizations of the previously known identities and discover some new identities.  Furthermore, we show several application of these new  identities. 

\paragraph*{Contribution}  The contribution and the outline of the paper are as follows: 
\begin{itemize}
\item  Section~\ref{sec:Preliminaries} comments on the notation and presents  several quantities  needed in  this paper.
\item  In Section~\ref{sec:NewIdentity}, Theorem~\ref{thm:MainIdentity} presents a new identity for the Jacobian of the conditional mean. Throughout the paper, this identity will be used for systematic proofs of old and new identities.
\item In Section~\ref{subsec:HatsellNolte}, Proposition~\ref{prop:genHatselNolte} presents a simple proof of  a vector version of the Hatsel-Nolte identity, which relates the Jacobian of the conditional expectation to the conditional variance.
\item In Section~\ref{subsec:jaffersIdentity}, we  study recursive derivative identities for the conditional expectation and show:
\begin{itemize}
\item In Section~\ref{subsect:Proof_jaffer}, Proposition~\ref{prop:Jaffers_identity} shows that the main identity in  Theorem~\ref{thm:MainIdentity} can lead to a simple proof of a recursive identity due to Jaffer. To the best of our knowledge, Jaffer's identity is not well-known and until now has had  no applications;
\item In Section~\ref{subsec:HigherOrderDerivatives}, Proposition~\ref{prop:Jaffer_Integral_Version} provides an alternative integral version of Jaffer's identity.  This new integral identity is shown to be very powerful and leads to simple proofs of old results and several new results. In particular, we have the following three new results. First, in Proposition~\ref{prop:Jaffer_Integral_Version}, the integral version of Jaffer's identity leads to an expression for all higher-order derivatives of the conditional expectation in terms of Bell-polynomials. Second, in Proposition~\ref{prop:Jaffer_Integral_Version}, the integral version of Jaffer's identity leads to a representation of higher-order conditional expectations (i.e., $\E[X^k|Y]$)  in terms of the derivatives of the conditional expectation. 
Third,  in Proposition~\ref{prop:gen_TRE}, the   integral version of Jaffer's identity is used to  generalize the TRE identity to  higher-order conditional expectation. This generalized TRE identity maintains  the property that  $\E[X^k|Y]$ depends on the joint distribution only through the marginal of $Y$.  Finally, in Theorem~\ref{thm:UniqunessOfConditionalExpectation}, the integral version of Jaffer's identity is used to show a simple proof of a known result that the distribution of the input $X$ uniquely determines the conditional expectation; and

\item  In Section~\ref{sec:vectro_gen_vect_jaffer}, Proposition~\ref{prop:JafferVersionTwo} and Proposition~\ref{prop:JafferVector} provide two vector generalizations of Jaffer's identity. 

\end{itemize}   
\item   In Section~\ref{sec:Conditional_Cumulants}, we establish  several new fundamental connections between the conditional expectations and the conditional cumulants and show:
\begin{itemize}
\item In Section~\ref{sec:Cumulatns_Univariet_Case},  Proposition~\ref{prop:Cumulant_derivative_of_CE}, for the univariate case, shows that the $k$-th derivative of the conditional expectation is proportional to the $(k+1)$-th conditional cumulant.  Interestingly,  Proposition~\ref{prop:Cumulant_derivative_of_CE}, in combination with the TRE identity, shows that the conditional cumulants depend on the joint distribution only through the marginal of $Y$. Moreover, the combination of the TRE identity and Proposition~\ref{prop:Cumulant_derivative_of_CE} is used to study the properties of $P_{X|Y=y}$. In particular, it is shown that while $P_{X|Y=y}$ is sub-Gaussian, it  can be  strictly sub-Gaussian only on a set of measure zero. Furthermore,  Proposition~\ref{prop:propositon_bounds_cumulants} establishes an upper bound on the conditional cumulant that would be used in subsequent sections. Finally, Theorem~\ref{thm:KumulantFunction_and_CE} establishes a new derivative identity that connects the higher-order derivative of the conditional expectation and the partial derivatives of the cumulant generating function; and
\item In Section~\ref{sec:Cumulants_Multivariate_Case},  the univariate results of Section~\ref{sec:Cumulatns_Univariet_Case} are generalized to the multivariate case. Specifically, Theorem~\ref{thm:multivariate_CGF_CE} and Theorem~\ref{prop:PartialDerivattives_CE} establish connections between the partial derivatives of the conditional cumulant generating function, the partial derivatives of the  conditional cumulants, and the partial derivatives of the conditional expectation. 
\end{itemize}
\item In Section~\ref{sec:Identities_for_distributions}, we study identities relating the conditional expectation to quantities such as the conditional distribution  $P_{\X|\Y}$, the pdf of $\Y$, and the information density of the pair $(\X,\Y)$ and show:
\begin{itemize}
\item In Section~\ref{sec:info_density_connection},  for the multivariate case, Proposition~\ref{prop:Gradient_log_pmf_pdf} and Proposition~\ref{prop:Hessian_of_log_dist} establish the gradient of the information density and the  Hessian of the information density. In addition, for the univariate case, in Proposition~\ref{prop:kth_derivative_info_density},  every $k$-th order derivative of the information density is determined. Furthermore,  in Proposition~\ref{prop:Hessian_of_log_dist}, the gradient and the Hessian are determined for the quantity $ \log( \mathbb{P}[ \X \in \mathcal{A} | \Y={\bf y}]) $, which is related to the information density and where $\mathcal{A}$ is some measurable set. 
\item In Section~\ref{sec:inverse_TRE_Identity},  the inverse version of the TRE identity is discussed. The inverse identity is then used to show that the conditional expectation uniquely determines the input distribution provided the noise's covariance matrix is full rank. 
\item In Section VII-C, the identities for the information density are used to find two new alternative representations of the minimum mean squared error (MMSE). These representations are then used to find new lower bounds on the MMSE. 

\end{itemize}
\item In Section~\ref{sec:PowerSeries_and_Inverse}, we study analytical properties of the conditional expectation in the univariate case and show:
\begin{itemize}
\item In Section~\ref{sec:Inverse_OF_CE}, Theorem~\ref{thm:Power_Series_ConditionalExpectation}, using the new identity for the conditional expectation derivatives, develops a power series representation of the conditional expectation. 
\item In Section~\ref{sec:Inverse_OF_CE}, Theorem~\ref{thm:Expression_for_inverse_CE}, using the Lagrange inversion theorem and the power series representation in Theorem~\ref{thm:Power_Series_ConditionalExpectation}, finds the compositional inverse of the conditional expectation. 
\end{itemize}
\item In Section~\ref{sec:Distribution_of_CE_and_Error},  we view $\E[X|Y]$ as a random variable and  study the distribution of $\E[X|Y]$ and the distribution of the   estimation error $(X-\E[X|Y])$ and show:
\begin{itemize}
\item In Section~\ref{sec:Distribution_of_CE_and_Error}, Theorem~\ref{thm:pdf_of_CE}, using the expression for the inverse in  Theorem~\ref{thm:Expression_for_inverse_CE},  finds the pdf of $\E[X|Y]$. 
\item In Section~\ref{sec:Distribution_of_X-EXY}, Theorem~\ref{thm:distribution_of_errors} finds the pdf of the estimation error $(X-\E[X|Y])$. Moreover, the pdf is also found in the case when there is a distributional mismatch.  That is,  we consider the distribution of $(X-\E_{Q}[X|Y])$, where $ \E_{Q}[X|Y]$ is the conditional expectation that assumes that $X$ is distributed according to $Q$ instead of $P_X$. 
\end{itemize} 
\item In Section~\ref{sec:Applications}, we show another application of the obtained identities by extending the notion of the empirical Bayes to the higher-order conditional moments and  conditional cumulants and show:
\begin{itemize}
\item In Section~\ref{subsec:EB}, Theorem~\ref{thm:ConsistentEB}  considers an independent and identically distributed (i.i.d.) sequence $Y_1,\ldots, Y_n$ according to $P_Y$ and constructs a consists of estimator $\E[X^k|Y=y]$ for every integer $k$.
\item In Section~\ref{sec:EB_for_CC}, Theorem~\ref{thm:EB_for_CC}  considers an i.i.d. sequence $Y_1,\ldots, Y_n$ according to $P_Y$ and constructs a consists of estimator  of the conditional cumulants for every order $k$. The key technical tool that is used in the construction of the estimators  is the notion of Lanczos' derivatives. 
\end{itemize}
\item Section~\ref{sec:ConclusionANDoutlook}  concludes the paper. All  of the identities are summarized in Table~\ref{table:Identities}. 
  \end{itemize}

\section{Preliminaries: Notation and Lanczos' generalized derivatives}
\label{sec:Preliminaries}

\subsection{Notation} 
 The set of all positive integers is denoted by $\mathbb{N}$,  $[n]$ is the set of integers $\{1,\dots,n\}$, and $\mathbb{R}^n$ is the set of all $n$-dimensional real-valued vectors. All logarithms in the paper are base $\eu$.  All vectors in the paper are column vectors.

For a triplet $(\U,\X,\Y) \in \mathbb{R}^{m \times n \times k}$ we define the \emph{conditional variance matrix} and the \emph{conditional cross-covariance matrix} as follows:
\begin{align}
 \boldsymbol{\mathsf{Var}}(\X | \Y) &= \E \left[  \X   \X^\mathsf{T}   |\Y\right ]    -  \E \left[  \X   |\Y \right ]  \E \left[  \X^\mathsf{T} | \Y\right], \\
\boldsymbol{\mathsf{Cov}}(\X, \U | \Y)  &= \E \left[  \X   \U^\mathsf{T}   |\Y\right ]    -  \E \left[  \X   |\Y \right ]  \E \left[  \U^\mathsf{T} | \Y\right] .
\end{align} 
The \emph{MMSE matrix} is denoted by 
\begin{equation}
\boldsymbol{\mathsf{MMSE}}(\X|\Y)= \E \left[ (\X -\E[\X|\Y) (\X -\E[\X|\Y)^\mathsf{T}  \right]. 
\end{equation}
As usual, we refer to the trace of the MMSE matrix as the MMSE  and denote it by
\begin{equation}
\mmse(\X|\Y)= \E \left[ \| \X -\E[\X|\Y] \|^2 \right] . 
\end{equation}

The  \emph{standard basis vectors} for  $\mathbb{R}^n$  are denoted by  $\mathbf{e}_i, \, i \in [n]$.  For a matrix $\boldsymbol{\mathsf{A}}$,  we use $[\boldsymbol{\mathsf{A}}]_{ij}$ to denote the entry of  row $i$ and  column $j$. 
The \emph{Euclidian norm}  of a vector $\mathbf{x} \in \mathbb{R}^n$ in this paper is denoted by $\| \mathbf{x} \|$.   The \emph{inner product} between vectors ${\bf u}$ and ${\bf v}$ is denoted by ${\bf u}  \boldsymbol{\cdot} {\bf v}$. 

 The \emph{gradient} of a  function $f: \mathbb{R}^n \to \mathbb{R}$  is denoted by 
 \begin{equation}
 \nabla_{\bf x} f(\mathbf{x})  = \left  [     \frac{ \partial  f({\bf x}) }{ \partial  x_1}  , \ldots,  \frac{ \partial   f({\bf x})}{ \partial  x_n}   \right]^\mathsf{T} \in \mathbb{R}^n .
 \end{equation}
 The \emph{Jacobian matrix}  of a function $ \mathbf{f}: \mathbb{R}^n \to \mathbb{R}^m$ is denoted by
 \begin{equation}
   \boldsymbol{\mathsf{J}}_\mathbf{x} \mathbf{f}(\mathbf{x}) =  \begin{pmatrix}
  \frac{ \partial  f_1({\bf x}) }{ \partial  x_1}  &   \frac{ \partial  f_2({\bf x}) }{ \partial  x_1}  & \cdots &   \frac{ \partial  f_m({\bf x}) }{ \partial  x_1}  \\
 \frac{ \partial  f_1({\bf x}) }{ \partial  x_2}  &   \frac{ \partial  f_2({\bf x}) }{ \partial  x_2}  & \cdots &   \frac{ \partial  f_m({\bf x}) }{ \partial  x_2} \\
\vdots  & \vdots  & \ddots & \vdots  \\
 \frac{ \partial  f_1({\bf x}) }{ \partial  x_n}  &   \frac{ \partial  f_2({\bf x}) }{ \partial  x_n}  & \cdots &   \frac{ \partial  f_m({\bf x}) }{ \partial  x_n}
 \end{pmatrix} \in \mathbb{R}^{n \times m} . \label{eq:Def_Jacobian}
   \end{equation} 
    The \emph{Hessian matrix} of a function $f: \mathbb{R}^n \to \mathbb{R}$  is denoted by
  \begin{equation}
  \boldsymbol{\mathsf{D}}^2_\mathbf{x} \mathbf{f}(\mathbf{x}) =  \begin{pmatrix}
  \frac{ \partial^2  f({\bf x}) }{ \partial  x_1^2}  &   \frac{ \partial^2  f({\bf x}) }{ \partial  x_1 \partial  x_2}  & \cdots &   \frac{ \partial  f({\bf x}) }{ \partial  x_1 \partial  x_n }  \\
 \frac{ \partial^2 f_1({\bf x}) }{ \partial  x_2 \partial x_1}  &   \frac{ \partial^2  f({\bf x}) }{ \partial  x_2^2}  & \cdots &   \frac{ \partial^2  f({\bf x}) }{ \partial  x_2 \partial  x_n} \\
\vdots  & \vdots  & \ddots & \vdots  \\
 \frac{ \partial^2  f_1({\bf x}) }{ \partial  x_n \partial  x_1}  &   \frac{ \partial^2  f_2({\bf x}) }{ \partial   x_n \partial  x_2}  & \cdots &   \frac{ \partial^2  f_m({\bf x}) }{ \partial^2  x_n}
 \end{pmatrix} \in \mathbb{R}^{n \times n} .
     \end{equation}

%

The $(n,k)$-th \emph{partial Bell polynomial} is denoted by $\mathsf{B}_{n,k}(x_1,\ldots, x_{n-k+1})$ and the  $n$-th \emph{complete Bell polynomial} is denoted by 
\begin{equation}
\mathsf{B}_{n}(x_1,\ldots, x_n)=\sum_{k=1}^n \mathsf{B}_{n,k}(x_1,\ldots, x_{n-k+1}).  \label{eq:Def_Bell_poly}
\end{equation} 

Finally, the pdf of a zero mean Gaussian random vector with a covariance matrix $\boldsymbol{\mathsf{K}}$ is denoted by $\phi_{\boldsymbol{\mathsf{K}}}(\cdot)$.

\subsection{Lanczos' Generalized Derivatives}
In this paper, we will need to simulate higher-order derivatives of pdfs or logarithms of pdfs.  In order to do this  numerically, we will rely on \emph{Lanczos' generalized derivatives} \cite{lanczos1956applied,groetsch1998lanczos,rangarajan2005lanczos,shen1999generalized}. 
Let $f: \mathbb{R} \to \mathbb{R}$ and define the following operator:  for $h>0$
\begin{equation}
D_h^{(n)} f(x)= \frac{c_n}{h^n }  \int_{-1}^1 f(x+h t)  P_n(t) {\rm d} t,     \label{eq:Lanczos_derivative}
\end{equation} 
where $c_n= \frac{1}{2} \sqrt{ \frac{2^{2n+2}}{\pi}}  \Gamma \left(  n+\frac{3}{2}\right) $, and $P_n$ is the \emph{Legendre polynomial} of order $n$.  The operator in \eqref{eq:Lanczos_derivative} satisfies the following property:
\begin{equation}
D_h^{(n)} f(x)= f^{(n)}(x)+ O(h^2),  \, n \in \mathbb{N}.  \label{eq:Error_Lanczos_derivative}
\end{equation} 

The operator in \eqref{eq:Lanczos_derivative},  via the property in \eqref{eq:Error_Lanczos_derivative},  allows a convenient way of computing derivatives through integration. In addition to being a useful  tool for creating numerical examples, in Section~\ref{sec:EB_for_CC}, the operator in \eqref{eq:Error_Lanczos_derivative} will be used to construct a consistent estimator of the conditional cumulants.   
Next, it is illustrated how the TRE identity and Lanczos' derivative can be used to compute the conditional expectation numerically. 

\begin{figure}[h!]
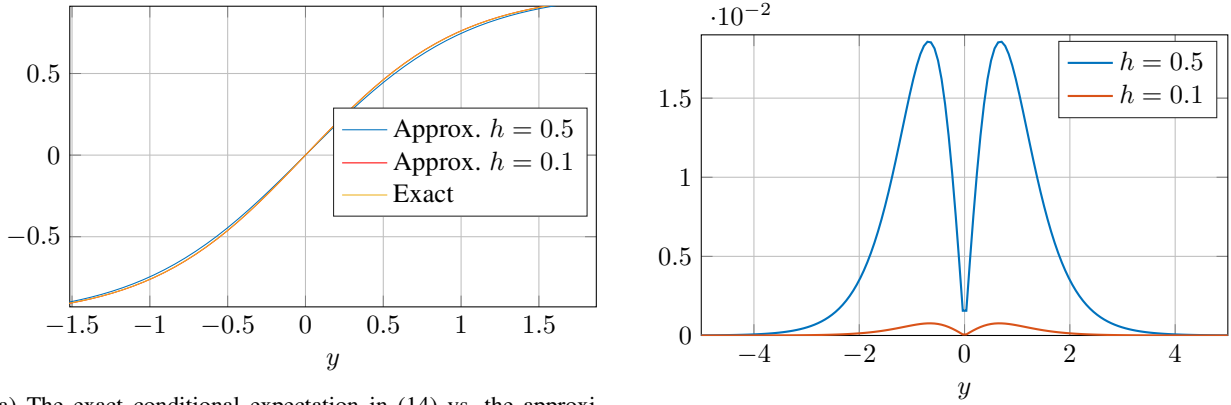
 
	\centering
	  \begin{subfigure}[c]{0.45\textwidth}
	  \center
	\input{tanh_app.tex}
	\caption{The exact conditional expectation in \eqref{eq:Tanh}  vs. the approximation in \eqref{eq:approximation_of_CE_tanh}. } \label{eq:approximation_of_CE_tanh_compare}
	\end{subfigure}
	~
	  \begin{subfigure}[c]{0.45\textwidth}
	  \center
	\input{tanh_Err.tex}
	\caption{The approximation error between  \eqref{eq:Tanh} and \eqref{eq:approximation_of_CE_tanh}.}   \label{eq:approximation_of_CE_tanh_error}
	  \end{subfigure}
		
		\caption{Approximation of the conditional expectation in \eqref{eq:Tanh}  with  \eqref{eq:approximation_of_CE_tanh}  where $\sigma^2=1$.  }
		\label{fig:Conditional_Expectation_tanh}
\end{figure}

\begin{example*}
 Consider $X \in \{-1,1\}$  distributed as $P_X(-1)=P_X(1)=\frac{1}{2}$, then the exact expression for the conditional expectation is given by 
\begin{equation}
\E[X|Y=y]=\tanh \left( \frac{y}{\sigma^2} \right).  \label{eq:Tanh}
\end{equation} 
From the TRE identity in \eqref{eq:TweediesFormulaGaussian} and Lanczos' derivative  in \eqref{eq:Error_Lanczos_derivative} we have that the conditional expectation can be approximated as 
\begin{equation}
\E[X|Y=y]= y + \sigma^2 D_h^{(1)} \log \left( f_Y(y) \right) + O\left(h^2 \right).  \label{eq:approximation_of_CE_tanh}
\end{equation} 
where $f_Y(y)= \frac{1}{2} \phi_{\sigma^2}(y-1)+\frac{1}{2} \phi_{\sigma^2}(y+1)$.
Fig.~\ref{eq:approximation_of_CE_tanh_compare} compares the exact conditional expectation in \eqref{eq:Tanh} to the approximation in \eqref{eq:approximation_of_CE_tanh} for values of $h=0.5$ and $h=0.1$.   The errors for the approximations are shown in Fig.~\ref{eq:approximation_of_CE_tanh_error}.  All of the  forthcoming examples will be of similar flavor where we will use the knowledge of $f_Y(y)$ together with Lanczos' derivatives to compute the desired quantities such as the conditional expectations and the conditional cumulants.  
\end{example*} 

\begin{remark} Some caution must be taken when computing \eqref{eq:Error_Lanczos_derivative}. For example, the integral in \eqref{eq:Error_Lanczos_derivative} must be computed  to within an error that is well below the value of $h^n$.   Moreover, if the function  $f$ is known only within an error $\epsilon$, then taking $h \to 0$ will lead to a wrong result.  In this case, the optimal choice of $h$ is not $h \to 0$ but $h=O \left( \epsilon^{ \frac{1}{n+2}} \right)$. The error bounds with the explicit constants that explain this choice of $h$ are shown in Lemma~\ref{lem:Lancos_approximations} of Section~\ref{sec:EB_for_CC}. 
\end{remark}

\section{A New Identity for the Conditional Expectation} 
\label{sec:NewIdentity}
The first main result of this paper is the following general identity.

\begin{theorem}\label{thm:MainIdentity} Suppose that  random vectors $\U \in \mathbb{R}^m ,  \X \in \mathbb{R}^n,$ and $\Y\in \mathbb{R}^n$  satisfy the following conditions: 
\begin{align}
&\text{$\U \leftrightarrow \X  \leftrightarrow \Y$ form a Markov chain, in that order}; \\
&\E \left[ \left \|  \U  \|  \|    \X  \right \|  | \Y= {\bf y}\right]<\infty, {\bf y}\in \mathbb{R}^n; \text{and}   \\
&\E \left[ \left \|  \U   \right \| | \Y={\bf y}  \right ]<\infty,  {\bf y}\in \mathbb{R}^n. 
\end{align} 
Then, 
\begin{equation}
 \boldsymbol{\mathsf{J}}_\mathbf{y}  \E \left[  \U | \Y= {\bf y} \right ]=   \boldsymbol{\mathsf{K}}_{\N}^{-1}   \boldsymbol{\mathsf{Cov}} ( \X, \U | \Y={\bf y}), \,  {\bf y} \in \mathbb{R}^n .   \label{eq:MainIdentity}
\end{equation} 
\end{theorem} 

\begin{IEEEproof}  See  Appendix~\ref{app:sec:MainThm}.
\end{IEEEproof}

Alternatively, Theorem~\ref{thm:MainIdentity}  can be re-written as 
\begin{equation}
\E[ U_i | \Y={\bf y}] 
= \E[ U_i | \Y={\bf 0}] +    \oint_{\bf 0}^{\bf y}  \boldsymbol{\mathsf{K}}_{\N}^{-1} \boldsymbol{\mathsf{Cov}}(\X, U_i | \Y={\bf t})       \boldsymbol{\cdot} {\rm \mathbf{d} } \mathbf{t},   \label{eq:IntegralVersionOfTheMainIdentity}
\end{equation} 
where $\,  i\in [m], $ and  $ \oint$ is a line integral over an arbitrary path between  ${\bf 0}$ and ${\bf y}$.

It is not difficult to see that conditions in Theorem~\ref{thm:MainIdentity} are rather  mild.    For example, by using Bayes' formula, we have that 
\begin{equation}
\E \left[ \left \|  \U   \right \| | \Y={\bf y}  \right ]= \E \left[ \left \|  \U   \right \|   \frac{\phi_{\boldsymbol{\mathsf{K}}_{\N}}({\bf y}-\X )}{f_{\Y}({\bf y}) }  \right ] =\frac{  \E \left[ \left \|  \U   \right \|   \phi_{\boldsymbol{\mathsf{K}}_{\N}}({\bf y}-\X ) \right ] }{f_{\Y}({\bf y}) } .  \label{eq:agument_for_finitness} 
\end{equation} 
Consequently, since $f_{\Y}({\bf y}) >0$, $\E \left[ \left \|  \U   \right \| | \Y={\bf y}  \right ]<\infty$  if and only if  $\E \left[ \left \|  \U   \right \|   \phi_{\boldsymbol{\mathsf{K}}_{\N}}({\bf y}-\X ) \right ]< \infty$. Therefore, in order to violate the conditions in Theorem~\ref{thm:MainIdentity} one needs to find $\left \|  \U   \right \|$ that goes to infinity faster than a Gaussian density.   In particular, by setting $\U=\X$, the above discussion shows that  $ \E \left[  \X | \Y= {\bf y} \right ]$  always exists.

In the rest of the paper, it is shown that many of the known identities in the literature can be derived systematically  from the identity in Theorem~\ref{thm:MainIdentity}.    Moreover, we use this new identity to derive several generalizations of the previously known identities and discover some new identities.  Specifically, this will be done by evaluating  Theorem~\ref{thm:MainIdentity} with different choices of $\U$  such as $\U=\X$, $U  = \mathsf{1}_{ \mathcal{A} }(\X)$, $\U= (\X \X^\mathsf{T} )^{k-1} \X, \, k \in \mathbb{N}$ and $\U=\eu^{{\bf t}^\mathsf{T} \X}, \, {\bf t} \in \mathbb{R}^n$.  For all these choices the conditional expectations will be finite.

 \section{The Variance Identity of Hatsell and Nolte} 
\label{subsec:HatsellNolte}

Our first application is to use  \eqref{eq:MainIdentity} to recover a variance identity shown by Hatsell and Nolte in  \cite{hatsell1971some}. 
By setting  $\U=\X$  in \eqref{eq:MainIdentity}, we arrive at the following  identity. 
\begin{prop}\label{prop:genHatselNolte} For ${\bf y} \in \mathbb{R}^n$ 
\begin{equation}
 \boldsymbol{\mathsf{J}}_\mathbf{y}  \E \left[  \X | \Y= {\bf y} \right ]  =  \boldsymbol{\mathsf{K}}_{\N}^{-1}  \boldsymbol{\mathsf{Var}}(\X | \Y={\bf y})   .  \label{eq:Hatsell-Nolte-general}
\end{equation} 
\end{prop}


The identity in \eqref{eq:Hatsell-Nolte-general} has been first derived by Hatsell and Nolte in  \cite{hatsell1971some}  for the case of $\boldsymbol{\mathsf{K}}_{\N}=\I$.   The general version  in \eqref{eq:Hatsell-Nolte-general} was  first derived in \cite{palomar2006gradient}.  In terms of applications, in  \cite{palomar2006gradient},  the identity in  \eqref{eq:Hatsell-Nolte-general}
 was  used, together with the TRE identity in \eqref{eq:TweediesFormulaGaussian}, to give a proof of the vector version of the I-MMSE relationship; in \cite{FunctionalPropMMSE}, the scalar version of the identity in  \eqref{eq:Hatsell-Nolte-general}, was used to show that the minimum mean squared error is Lipschitz continuous with respect to the Wasserstein distance; and   in \cite{cai2014optimal}, the identity in  \eqref{eq:Hatsell-Nolte-general} was used to show log-convexity of the function akin to the  log-likelihood ratio.

The identity of Hatsell and Nolte in \eqref{eq:Hatsell-Nolte-general} can  be used to make various statements about the minimum  and maximum `slope' of $ \E \left[  \X | \Y = {\bf y} \right ]$.
 \begin{corollary} For every $ {\bf y} \in \mathbb{R}^n$
\begin{equation}
0  \le   \mathsf{Tr} \left( \boldsymbol{\mathsf{J}}_\mathbf{y}  \E \left[  \X | \Y= {\bf y} \right ]  \right). 
  \end{equation}  
In addition, if  $\| \X\| \le R$, then 
  \begin{equation}
  \mathsf{Tr} \left( \boldsymbol{\mathsf{J}}_\mathbf{y}  \E \left[  \X | \Y= {\bf y} \right ]  \right)    \le  R^2  \mathsf{Tr} \left(     \boldsymbol{\mathsf{K}}_{\N}^{-1} \right). \label{eq:UpperBoundOnSlope}
  \end{equation}  
\end{corollary} 
\begin{IEEEproof}
The proof of the lower bound follows by using \eqref{eq:Hatsell-Nolte-general} together with the properties that both variance is a positive definite matrices and that the trace of the product of two positive definite matrices is non-negative. 
To show the upper bound in \eqref{eq:UpperBoundOnSlope}, we use \eqref{eq:Hatsell-Nolte-general} together with the Cauchy-Schwarz inequality
\begin{align}
  \mathsf{Tr} \left( \boldsymbol{\mathsf{J}}_\mathbf{y}  \E \left[  \X | \Y= {\bf y} \right ]  \right)  & =   \mathsf{Tr} \left(  \boldsymbol{\mathsf{Var}}(\X | \Y={\bf y})   \boldsymbol{\mathsf{K}}_{\N}^{-1} \right)\\
    & \le  R^2  \mathsf{Tr} \left(     \boldsymbol{\mathsf{K}}_{\N}^{-1} \right). 
\end{align} 
\end{IEEEproof} 

Several new applications of  \eqref{eq:Hatsell-Nolte-general}  will be given in subsequent sections. For example, in Section~\ref{sec:Identities_for_distributions}  the identity in \eqref{eq:Hatsell-Nolte-general}  will be used to show concavity of the information density, and in Section~\ref{sec:PowerSeries_and_Inverse} will be used to argue that the conditional expectation has an inverse.

\section{Recursive Identities for Higher-Order Conditional Moments} 
\label{subsec:jaffersIdentity}

In this section, we study  recursive identities for higher-order conditional moments.  The treatment of the scalar and vector case will be done separately. We begin by first showing a simple proof of the recursive identity due to Jaffer.

\subsection{Jaffer's Identity} 
\label{subsect:Proof_jaffer}

In \cite{jaffer1972note},  Jaffer has shown the following  identity, which now easily follows from our main result in  Theorem~\ref{thm:MainIdentity}. 

\begin{prop}\label{prop:Jaffers_identity}
 For  $y \in \mathbb{R}$ and  $k~\in~\mathbb{N}~\cup~\{ 0\}$
\begin{align}
\E  \hspace{-0.05cm} \left[ X^{k+1} |Y =y  \right] 
&= \sigma^2 \frac{{\rm d}}{ {\rm d} y }  \E  \hspace{-0.05cm}\left[ X^{k} |Y =y \right]  \hspace{-0.05cm}+   \E  \hspace{-0.05cm}\left[ X^{k} |Y =y \right]  \E  \hspace{-0.05cm}\left [ X |Y =y \right ] . \label{eq:RecursiveScalarForm}
\end{align} 
\end{prop} 
\begin{IEEEproof}
Letting $U=X^k$ in Theorem~\ref{thm:MainIdentity}, we arrive at
\begin{align}
\frac{ {\rm d}}{ {\rm d} y} \E[X^k|Y=y]  
&= 
 \frac{1}{\sigma^2} {\rm Cov}(X, X^k| Y=y)\\
&= \frac{1}{\sigma^2} \left( \E[X^{k+1}|Y=y] -  \E[X^{k}|Y=y]   \E[X|Y=y]  \right). 
\end{align}
This concludes the proof.  
\end{IEEEproof} 

 To the best of our knowledge, Jaffer's identity in \eqref{eq:RecursiveScalarForm} has had no applications  and is not well-known.  In what follows, we develop several alternative representations and  generalizations of Jaffer's identity and also show the utility of Jaffar's identity.   Specifically, we will first derive an alternative but equivalent integral version of Jaffer's identity and show how this new identity can be used to prove  the uniqueness of the conditional mean estimator.  In Section~\ref{subsec:EB},  we also use this integral identity to extend the empirical Bayes procedure to higher-order conditional moments. Furthermore, in Section~\ref{sec:Conditional_Cumulants},  this identity will be used to show a new fundamental connection between conditional cumulants and conditional expectations.

\subsection{A New Perspective on Jaffer's Identity}
\label{subsec:HigherOrderDerivatives}

Next, we show that Jaffer's identity has an alternative integral version.  This new integral representation leads to several interesting consequences.    The following lemma will be useful in deriving this new representation. 

\begin{lem}\label{lem:recurenceRelationship}  Let  $f_k: \mathbb{R} \to \mathbb{R}$ be a sequence of functions with $k\in \mathbb{N} \cap \{0\}$  such that \begin{equation}
f_k(x) = \frac{{\rm d}}{ {\rm d} x}  f_{k-1}(x)+ f_{k-1}(x) f_{1}(x),   k=1,2,\ldots,  \label{eq:RecurcenceRelationship}
\end{equation} 
with $f_0 	\equiv 1$.

Then, the following statements hold: 
\begin{itemize}
\item The solution to \eqref{eq:RecurcenceRelationship} is given by 
 \begin{align}
f_k(x)&=  \eu^{- \int_0^x f_1(t) {\rm d} t}  \frac{  {\rm d}^k} { {\rm d} x^k } \eu^{ \int_0^x f_1(t) {\rm d} t} ,  \label{eq:SolutionToRecursion}\\
&=\mathsf{B}_k \left(f_1^{(0)}(x), \ldots, f_1^{(k-1)}(x)  \right ). \label{eq:Bell-polynomial_representation} 
\end{align}  
\item The derivatives of $f_1$ are given by 
\begin{equation}
f_1^{(k)}(x) = \sum_{m=1}^{k+1} c_m\mathsf{B}_{k+1,m} \left(f_1(x),\ldots, f_{k-m+2}(x) \right),  \label{eq:Derivative_Solution_Bell}
\end{equation} 
where $c_m= (-1)^{m-1}(m-1)! $. 
\end{itemize} 

\end{lem}
\begin{IEEEproof}
See Appendix~\ref{app:lem:recurenceRelationship}. 
\end{IEEEproof}

Using Lemma~\ref{lem:recurenceRelationship}, we can now present an alternative integral version of Jaffer's identity.  In addition, we also provide an expression for all higher-order derivatives of the conditional expectation.  
\begin{prop}\label{prop:Jaffer_Integral_Version}
For $y \in \mathbb{R}$  and $k \in \mathbb{N}$
\begin{align}
 \E  \hspace{-0.05cm} \left[  X^{k}     |Y= y\right ]  
& =   \hspace{-0.05cm} \sigma^{2k} \eu^{-   \frac{1}{\sigma^2}  \hspace{-0.05cm} \int_0^{y}   \hspace{-0.05cm} \E \left[      X    |Y= t \right ] {\rm d} t}   \hspace{-0.05cm} \frac{  {\rm d}^k} { {\rm d} y^k }   \eu^{ \frac{1}{\sigma^2}  \hspace{-0.05cm} \int_0^{y}   \hspace{-0.05cm} \E \left[     X      |Y= t\right ] {\rm d} t} \label{eq:ScalarVersionOfTheIdenity}\\
&= \sigma^{2k} \mathsf{B}_{k}\left (   \E^{(0)}  \hspace{-0.05cm} \left[  \frac{X}{\sigma^2}    |Y= y\right ],\ldots,   \E^{(k-1)}  \hspace{-0.05cm} \left[  \frac{X}{\sigma^2}    |Y= y\right ]    \right), \label{eq:Higher_Moments_Bell_Polynomial}
\end{align} 
where $ \E^{(k)}  \hspace{-0.05cm} \left[  X     |Y= y\right ] =\frac{ {\rm d}^k}{ {\rm d} y^k}\E \hspace{-0.05cm} \left[  X     |Y= y\right ] $. 
Moreover, 
\begin{equation}
\frac{ {\rm d}^k}{ {\rm d} y^k}\E \hspace{-0.05cm} \left[  X     |Y= y\right ] = \hspace{-0.05cm}  \sigma^2 \sum_{m=1}^{k+1}  c_m \mathsf{B}_{k+1,m} \left( \E  \hspace{-0.05cm} \left[  \frac{X}{\sigma^2}    |Y= y\right ] ,\ldots,  \E  \hspace{-0.05cm} \left[ \left (\frac{X}{\sigma^2} \right)^{k-m+2}     |Y= y\right ]  \right), \label{eq:Higher_Order_Derivatives_of_CE}
 \end{equation}
 where $c_m= (-1)^{m-1}(m-1)! $.
\end{prop}
\begin{IEEEproof}
First, observe that Jaffer's identity in  \eqref{eq:RecursiveScalarForm} can be re-written as
\begin{equation}
\E  \hspace{-0.05cm} \left[ \left(\frac{X}{\sigma^2} \right)^{k} |Y =y  \right] 
 =  \frac{{\rm d}}{ {\rm d} y }  \E  \hspace{-0.05cm}\left[  \left(\frac{X}{\sigma^2} \right)^{k-1} |Y =y \right]  +   \E  \hspace{-0.05cm}\left[  \left(\frac{X}{\sigma^2} \right)^{k-1} |Y =y \right]  \E  \hspace{-0.05cm}\left [  \left(\frac{X}{\sigma^2} \right) |Y =y \right ] . 
\end{equation} 
Hence, if we take $f_k(y)= \E  \hspace{-0.05cm} \left[ \left(\frac{X}{\sigma^2} \right)^{k} |Y =y  \right] $, then the   Jaffer's identity is of the same the form as the recurrence relationship in \eqref{eq:RecurcenceRelationship}. In view of this observation, the proof now follows by   applying the results in Lemma~\ref{lem:recurenceRelationship}.  
\end{IEEEproof}

\begin{example*}
The second and third Bell polynomials age given by 
\begin{align}
\mathsf{B}_{2}(x_1,x_2)&=x_1^2+x_2,\\
\mathsf{B}_{3}(x_1,x_2,x_3)&=x_1^2+3x_1x_2+x_3.
\end{align} 
Therefore,  using \eqref{eq:Higher_Moments_Bell_Polynomial} we have that 
\begin{equation}
  \E  \hspace{-0.05cm} \left[  X^{2}     |Y= y\right ] = \E^2[X|Y=y]+ \sigma^2 \E^{(1)}  \hspace{-0.05cm} \left[  X     |Y= y\right ]   ,
  \end{equation}
  and 
  \begin{equation}
   \E  \hspace{-0.05cm} \left[  X^{3}     |Y= y\right ] = \sigma^2 \E^2[X|Y=y] +3 \sigma^2 \E[X|Y=y] \E^{(1)}  \hspace{-0.05cm} \left[  X     |Y= y\right ] +  \sigma^4\E^{(2)}  \hspace{-0.05cm} \left[  X     |Y= y\right ] . 
\end{equation} 
\end{example*}

\begin{example*}  We now show how  \eqref{eq:Bell-polynomial_representation} and Lanczo's derivative can be used to compute the $k$-th order conditional expectation.  Note that \eqref{eq:Bell-polynomial_representation} requires the computation of the $k$-th derivative of the conditional expectation. This can be achieved by using Lanczo's derivative and the TRE identity as follows: 
\begin{equation}
\frac{ {\rm d}^k}{ {\rm d} y^k}\E \hspace{-0.05cm} \left[  X     |Y= y\right ] \approx \frac{ {\rm d}^k}{ {\rm d} y^k} y+ \sigma^2   D_h^{(k+1)}   \log(    f_Y(y) ) . 
\end{equation} 
As an illustration of this procedure the $k$-th order conditional expectation for $\sigma^2=1$ of a random variable  $X$ uniformly distributed on $\{-2,0,2\}$ is plotted in Fig.~\ref{fig:Conditional_Moments}. 

\begin{figure}[h!]  
	\centering   
%
%
\definecolor{mycolor1}{rgb}{0.00000,0.44700,0.74100}%
\definecolor{mycolor2}{rgb}{0.85000,0.32500,0.09800}%
\definecolor{mycolor3}{rgb}{0.92900,0.69400,0.12500}%
\definecolor{mycolor4}{rgb}{0.49400,0.18400,0.55600}%
\begin{tikzpicture}

\begin{axis}[%
width=7cm,
height=4cm,
at={(1.011in,0.642in)},
scale only axis,
xmin=-5,
xmax=5,
ymin=-10,
ymax=20,
xlabel={$y$},
ylabel={$\E[X^k|Y=y]$},
axis background/.style={fill=white},
xmajorgrids,
ymajorgrids,
legend style={legend cell align=left, align=left, draw=white!15!black}
]
\addplot [color=mycolor1,thick]
  table[row sep=crcr]{%
-5	-1.99932926469584\\
-4.79591836734694	-1.99899134302556\\
-4.59183673469388	-1.9984832982092\\
-4.38775510204082	-1.99771964034891\\
-4.18367346938776	-1.99657211684\\
-3.97959183673469	-1.99484857047515\\
-3.77551020408163	-1.99226165681508\\
-3.57142857142857	-1.98838294843795\\
-3.36734693877551	-1.98257647742252\\
-3.16326530612245	-1.97390446247275\\
-2.95918367346939	-1.96099792272231\\
-2.75510204081633	-1.94188866462141\\
-2.55102040816327	-1.91381201397919\\
-2.3469387755102	-1.87302082298056\\
-2.14285714285714	-1.81471243334049\\
-1.93877551020408	-1.73326458554005\\
-1.73469387755102	-1.62306522575842\\
-1.53061224489796	-1.4801707611971\\
-1.3265306122449	-1.30461065244419\\
-1.12244897959184	-1.10230737743686\\
-0.918367346938776	-0.884909513340393\\
-0.714285714285714	-0.666545491930081\\
-0.510204081632653	-0.458749213889993\\
-0.306122448979592	-0.266465511968024\\
-0.102040816326531	-0.087162509227732\\
0.102040816326531	0.0871625092277549\\
0.306122448979592	0.266465511968082\\
0.510204081632653	0.458749213890018\\
0.714285714285714	0.666545491930104\\
0.918367346938776	0.884909513340414\\
1.12244897959184	1.1023073774369\\
1.3265306122449	1.3046106524442\\
1.53061224489796	1.4801707611971\\
1.73469387755102	1.62306522575843\\
1.93877551020408	1.73326458554002\\
2.14285714285714	1.81471243334049\\
2.3469387755102	1.87302082298061\\
2.55102040816327	1.91381201397921\\
2.75510204081633	1.94188866462147\\
2.95918367346939	1.96099792272233\\
3.16326530612245	1.97390446247278\\
3.36734693877551	1.98257647742252\\
3.57142857142857	1.988382948438\\
3.77551020408163	1.99226165681515\\
3.97959183673469	1.99484857047516\\
4.18367346938776	1.99657211684007\\
4.38775510204082	1.99771964034899\\
4.59183673469388	1.99848329820933\\
4.79591836734694	1.99899134302563\\
5	1.99932926469583\\
};
\addlegendentry{$k=1$}

\addplot [color=mycolor2,thick]
  table[row sep=crcr]{%
-5	3.9986585305104\\
-4.79591836734694	3.99798270021129\\
-4.59183673469388	3.99696664609742\\
-4.38775510204082	3.99543941924269\\
-4.18367346938776	3.99314458644361\\
-3.97959183673469	3.9896979976076\\
-3.77551020408163	3.98452533819553\\
-3.57142857142857	3.97677060172622\\
-3.36734693877551	3.96516376766287\\
-3.16326530612245	3.94783357900615\\
-2.95918367346939	3.9220517019796\\
-2.75510204081633	3.88390312567747\\
-2.55102040816327	3.82790548588086\\
-2.3469387755102	3.74666627587768\\
-2.14285714285714	3.63079618700621\\
-1.93877551020408	3.46949590984737\\
-1.73469387755102	3.2524228328459\\
-1.53061224489796	2.97334477539663\\
-1.3265306122449	2.63522814542362\\
-1.12244897959184	2.25464849338835\\
-0.918367346938776	1.86201331152538\\
-0.714285714285714	1.4955371566548\\
-0.510204081632653	1.19149839434428\\
-0.306122448979592	0.976586020428876\\
-0.102040816326531	0.866014945988007\\
0.102040816326531	0.866014945992174\\
0.306122448979592	0.976586020432029\\
0.510204081632653	1.19149839433701\\
0.714285714285714	1.49553715665587\\
0.918367346938776	1.86201331151813\\
1.12244897959184	2.25464849337804\\
1.3265306122449	2.63522814542053\\
1.53061224489796	2.97334477539663\\
1.73469387755102	3.25242283283761\\
1.93877551020408	3.46949590984521\\
2.14285714285714	3.63079618699998\\
2.3469387755102	3.74666627587679\\
2.55102040816327	3.82790548587366\\
2.75510204081633	3.88390312566939\\
2.95918367346939	3.92205170198383\\
3.16326530612245	3.94783357900625\\
3.36734693877551	3.9651637676587\\
3.57142857142857	3.97677060172225\\
3.77551020408163	3.98452533819582\\
3.97959183673469	3.98969799760763\\
4.18367346938776	3.99314458642308\\
4.38775510204082	3.99543941925135\\
4.59183673469388	3.99696664606881\\
4.79591836734694	3.99798270028653\\
5	3.99865853050204\\
};
\addlegendentry{$k=2$}

\addplot [color=mycolor3,thick]
  table[row sep=crcr]{%
-5	-7.99731701700929\\
-4.79591836734694	-7.99596530149855\\
-4.59183673469388	-7.99393310140647\\
-4.38775510204082	-7.99087842373617\\
-4.18367346938776	-7.98628824151078\\
-3.97959183673469	-7.97939394185148\\
-3.77551020408163	-7.96904611982593\\
-3.57142857142857	-7.95353103020414\\
-3.36734693877551	-7.93030476255436\\
-3.16326530612245	-7.89561613346155\\
-2.95918367346939	-7.84398914508655\\
-2.75510204081633	-7.76755087663791\\
-2.55102040816327	-7.65524249305526\\
-2.3469387755102	-7.49207518383595\\
-2.14285714285714	-7.25883809963642\\
-1.93877551020408	-6.93304198752676\\
-1.73469387755102	-6.49223862737946\\
-1.53061224489796	-5.92065399672707\\
-1.3265306122449	-5.21840700735194\\
-1.12244897959184	-4.40918933551166\\
-0.918367346938776	-3.53959721581403\\
-0.714285714285714	-2.66614528436893\\
-0.510204081632653	-1.83496845206483\\
-0.306122448979592	-1.06584437417343\\
-0.102040816326531	-0.34864407649483\\
0.102040816326531	0.348644080867481\\
0.306122448979592	1.06584437927618\\
0.510204081632653	1.83496844768339\\
0.714285714285714	2.66614528437111\\
0.918367346938776	3.53959721360905\\
1.12244897959184	4.40918933547755\\
1.3265306122449	5.21840700733983\\
1.53061224489796	5.92065399745566\\
1.73469387755102	6.49223862515327\\
1.93877551020408	6.93304198897286\\
2.14285714285714	7.25883809960246\\
2.3469387755102	7.49207518601632\\
2.55102040816327	7.65524249374228\\
2.75510204081633	7.7675508736758\\
2.95918367346939	7.84398914948274\\
3.16326530612245	7.89561613783335\\
3.36734693877551	7.93030475815809\\
3.57142857142857	7.9535310301799\\
3.77551020408163	7.96904611545527\\
3.97959183673469	7.97939394185158\\
4.18367346938776	7.98628822972961\\
4.38775510204082	7.99087840338672\\
4.59183673469388	7.99393310123334\\
4.79591836734694	7.99596530777753\\
5	7.99731701404491\\
};
\addlegendentry{$k=3$}

\addplot [color=mycolor4,thick]
  table[row sep=crcr]{%
-5	15.9946495439742\\
-4.79591836734694	15.9919570930317\\
-4.59183673469388	15.987881509866\\
-4.38775510204082	15.9817675975859\\
-4.18367346938776	15.9725833132006\\
-3.97959183673469	15.95879925266\\
-3.77551020408163	15.9381084165241\\
-3.57142857142857	15.9070926773167\\
-3.36734693877551	15.8606610193726\\
-3.16326530612245	15.7913462506741\\
-2.95918367346939	15.6882093077806\\
-2.75510204081633	15.5356132293578\\
-2.55102040816327	15.31162098886\\
-2.3469387755102	14.9866566058408\\
-2.14285714285714	14.523181059212\\
-1.93877551020408	13.8779722319058\\
-1.73469387755102	13.0096763664342\\
-1.53061224489796	11.8933571479189\\
-1.3265306122449	10.5408819341656\\
-1.12244897959184	9.01855763470839\\
-0.918367346938776	7.44800698212766\\
-0.714285714285714	5.98210209902224\\
-0.510204081632653	4.76595159911168\\
-0.306122448979592	3.90630192618325\\
-0.102040816326531	3.4640225280418\\
0.102040816326531	3.46402384103854\\
0.306122448979592	3.90630258736326\\
0.510204081632653	4.76594568950881\\
0.714285714285714	5.98210406620824\\
0.918367346938776	7.44801222011357\\
1.12244897959184	9.0185602574703\\
1.3265306122449	10.5408799669401\\
1.53061224489796	11.8933584636836\\
1.73469387755102	13.0096776635325\\
1.93877551020408	13.8779755205918\\
2.14285714285714	14.5231777804487\\
2.3469387755102	14.9866631794492\\
2.55102040816327	15.311618371369\\
2.75510204081633	15.5356145179794\\
2.95918367346939	15.6882158994239\\
3.16326530612245	15.7913462851906\\
3.36734693877551	15.86066623041\\
3.57142857142857	15.907090054317\\
3.77551020408163	15.9380992015331\\
3.97959183673469	15.958804498464\\
4.18367346938776	15.9725937112128\\
4.38775510204082	15.9817635004186\\
4.59183673469388	15.9878801977204\\
4.79591836734694	15.991947961281\\
5	15.9946547662711\\
};
\addlegendentry{$k=4$}

\end{axis}

\end{tikzpicture}%
	\caption{Plot of  $\E[X^k|Y=y]$  vs. $y$ for  $k=1,2,3$ and $4$.  }
		\label{fig:Conditional_Moments}
\end{figure}
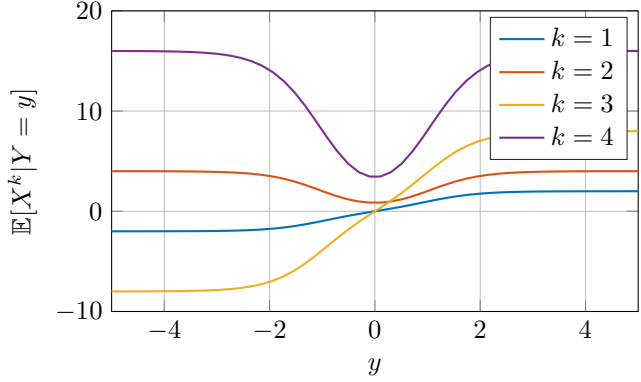%
\end{example*}

An important feature of the integral version of Jaffer's identity in \eqref{eq:ScalarVersionOfTheIdenity} is that every higher-order conditional moment is determined by the first-oder conditional moment. This observation can be used to  show that the conditional expectation  is uniquely determined by the distribution of the input $X$.

\begin{theorem}\label{thm:UniqunessOfConditionalExpectation}  The conditional  expectation $\E[X|Y]$  is a bijective operator of the distribution of $X$. In other words,  for $Y_1=X_1+N_1$ and $Y_2=X_2+N_2$ where $N_1$ and $N_2$ are Gaussian with the same variance 
\begin{equation}
\E[X_1|Y_1  \hspace{-0.05cm} =y] \hspace{-0.05cm}= \hspace{-0.05cm} \E[X_2|Y_2  \hspace{-0.05cm} =y],  \forall y \in \mathbb{R}   \Longleftrightarrow    P_{X_1}  \hspace{-0.05cm}= \hspace{-0.05cm} P_{X_2}. 
\end{equation} 
\end{theorem} 
\begin{IEEEproof}  
Let $P_{X_1}=P_{X_2} $; then it is immediate that
\begin{align}
 \E[X_1|Y_1=y]= \E[X_2|Y_2=y],  \forall y \in \mathbb{R} . 
\end{align} 
Now suppose that $ \E[X_1|Y_1=y]= \E[X_2|Y_2=y],  \forall y \in \mathbb{R}$, and the goal is to show that $P_{X_1}=P_{X_2} $.  First, observe that by the identity in \eqref{eq:ScalarVersionOfTheIdenity} we have that for all $k\in \mathbb{N}$ and all $y \in \mathbb{R} $
\begin{equation}
\E[X_1^k|Y_1=y]= \E[X_2^k|Y_2=y] .  \label{eq: higherConditionalMoments}
\end{equation}
We now use \eqref{eq: higherConditionalMoments} to establish that $P_{X_1|Y_1=y}=P_{X_2|Y_2=y}$ for all $y\in \mathbb{R}$. First, without loss of generality assume that $\sigma^2=1$.  Second, fix some $y\in \mathbb{R}$ in \eqref{eq: higherConditionalMoments} and let
\begin{equation}
m_{1,k}=\E[X_1^k|Y_1=y],  \, m_{2,k}=\E[X_2^k|Y_1=y].
\end{equation} 
The identity \eqref{eq: higherConditionalMoments} implies that  $m_{1,k}=m_{2,k}, \forall k\in \mathbb{N}$. The question of whether a distribution of a real-valued random variable is determined by its moment is known as the Hamburger moment problem \cite{shohat1943problem}. We now use Carleman's sufficient condition to check whether the moments uniquely determine   the distribution, which requires verifying that the following sum of moments is divergent: 
\begin{equation}
\sum_{k=1}^\infty m_{i,2k}^{-\frac{1}{2k}}=\infty.  \label{eq:CarlemanCondition}
\end{equation} 
To show that the left side of \eqref{eq:CarlemanCondition} diverges, we  will need the following upper bound on the conditional moments shown in \cite{GuoMMSEprop}: for $i \in \{1,2\}$
\begin{equation}
m_{i,2k} \le c_i k 2^{k+1} \sqrt{ (2k-1)!},  k \in [n] \label{eq:BoundOnMoments}
\end{equation}
with $c_i=  \frac{  \eu^{\frac{y^2}{2}}}{f_{Y_i}(y)} $. Applying  \eqref{eq:BoundOnMoments} to \eqref{eq:CarlemanCondition} we have that 
\begin{align}
\sum_{k=1}^\infty m_{i,2k}^{-\frac{1}{2k}}  &\ge  \sum_{k=1}^\infty   \frac{1}{c_i^{ \frac{1}{2k}  }   k^{  \frac{1}{2k} } 2^{  \frac{k+1}{2k} }  ( (2k-1)! )^{   \frac{1}{4k}  }   }\\
& \ge \frac{1}{ 2 c_i^{ \frac{1}{2}  }}  \sum_{k=1}^\infty   \frac{1}{  k^{  \frac{1}{2k} }  ( (2k)! )^{   \frac{1}{4k}  }   },
\end{align}
which diverges by the comparison test.  Therefore, the Carleman condition in \eqref{eq:CarlemanCondition} is satisfied, and the moments determine the distribution. In other words, we have demonstrated that \eqref{eq: higherConditionalMoments} implies that  for all $y \in \mathbb{R}$
 \begin{equation}
 P_{X_1|Y_1=y}= P_{X_2|Y_2=y}.  \label{eq:EqualityOfConditionalDistributions}
 \end{equation}  
To see that the equality in \eqref{eq:EqualityOfConditionalDistributions} implies that  $P_{X_1}= P_{X_2}$,  choose some measurable set $\mathcal{A} \subseteq \mathbb{R}$ and observe that
\begin{align}
P_{X_1}(\mathcal{A})&= \E[  1_{\mathcal{A}}(X_1) ]\\
&= \E \left[  \E \left[ 1_{\mathcal{A}}(X_1) | Y_1 \right]  \right ]\\
&= \E \left[  \E \left[ 1_{\mathcal{A}}(X_2) | Y_2 \right]  \right ]=P_{X_2}(\mathcal{A}). 
\end{align} 
This concludes the proof. 
\end{IEEEproof} 

Theorem~\ref{thm:UniqunessOfConditionalExpectation} has been previously shown in \cite{FunctionalPropMMSE}. However, our proof here is different than that in \cite{FunctionalPropMMSE}, and our method is more akin to the proof of the uniqueness of the conditional expectation for the Poisson noise used in \cite{poissonDytso2020}.  A simpler proof of Theorem~\ref{thm:UniqunessOfConditionalExpectation} that also holds for vectors and discussion  surrounding statistical  applications of Theorem~\ref{thm:UniqunessOfConditionalExpectation} will be provided in Section~\ref{sec:inverse_TRE_Identity}.

Another identity, equivalent to that in \eqref{eq:ScalarVersionOfTheIdenity} and which allows expressing higher-order conditional moments in terms of the pdf of $Y$, is shown next. 
\begin{prop} \label{prop:gen_TRE}
For any $k \in \mathbb{N}$ and $y \in\mathbb{R}$
\begin{equation}
\E[   X^k |Y=y ]  = \sigma^{2k}     \frac{  \frac{{\rm d}^k}{ {\rm d} y^k }  \Big(  f_Y(y)  \phi_{\sigma^2}^{-1}(y)   \Big) }{ f_Y(y)   \phi_{\sigma^2}^{-1}(y)  }.  \label{eq: GeneralizedHigherOrderTweedy}
\end{equation} 
Alternatively,  let $t \mapsto H_{e_m}(t), \, m   \in  \mathbb{N} \cup \{0\} $  be a \emph{probabilistic Hermite polynomial};  then
\begin{equation}
\E[   X^k |Y=y ]  = \sigma^{2k}   \frac{ \sum_{m=0}^k  {{k} \choose {m}}   
 f_Y^{(k-m)}(y)   \frac{(-i)^m}{\sigma^m}   H_{e_m} \left( i \frac{y}{\sigma} \right) }{  f_Y(y)  },  \label{eq:generalTREhermite}
\end{equation} 
where $i=\sqrt{-1}$. 
\end{prop}

\begin{IEEEproof}
First, observe that using the scalar TRE identity in \eqref{eq:TweediesFormulaGaussian} we have that 
\begin{align}
   \int_0^{y}   \hspace{-0.05cm}  \frac{ \E \left[    X   |Y \hspace{-0.05cm} = \hspace{-0.05cm} t\right ] }{\sigma^2}   {\rm d} t  \hspace{-0.05cm} &= \hspace{-0.05cm}   \int_0^{y} \hspace{-0.05cm}    \left( \frac{t}{\sigma^2}  +\frac{\rm d}{ {\rm d} t } \log (f_Y(t))  \right)   {\rm d} t    \\
&= \hspace{-0.05cm}       \frac{y^2}{2 \sigma^2}  +  \log (f_Y(y)) -   \log (f_Y(0)). \label{eq:IntegrationOfOFScoreFunction}
\end{align} 
Inserting \eqref{eq:IntegrationOfOFScoreFunction} into \eqref{eq:ScalarVersionOfTheIdenity} leads to \eqref{eq: GeneralizedHigherOrderTweedy}.  The proof of  \eqref{eq:generalTREhermite} follows by applying  the generalized product rule to \eqref{eq: GeneralizedHigherOrderTweedy} and the following derivative  \cite[eq.~19.13.3]{abramowitz1965handbook}:
\begin{equation}
 \frac{ {\rm d}^m }{ {\rm d} y^m }  \eu^{ \frac{x^2}{2 \sigma}} =   \frac{(-i)^m}{\sigma^m}   H_{e_m} \left( i \frac{y}{\sigma} \right)   \eu^{\frac{x^2}{2 \sigma^2}} . 
\end{equation}    
\end{IEEEproof} 

The identity in \eqref{eq: GeneralizedHigherOrderTweedy} can be thought of as a generalization of the TRE identity in  \eqref{eq:TweediesFormulaGaussian} to the higher-order moments.  Indeed for $k=1$, we recover the TRE identity.      Similarly to the TRE identity, the important feature of the identity in \eqref{eq: GeneralizedHigherOrderTweedy} is that  $\E[   X^k |Y]$ depends on the joint distribution  $P_{X,Y}$ only through the marginal pdf  of $Y$.  In Section~\ref{subsec:EB}, the identity in \eqref{eq: GeneralizedHigherOrderTweedy} will be used to extend the empirical Bayes procedure to higher-order conditional moments.

\subsection{Vector Generalizations of Jaffer's Identity} 
\label{sec:vectro_gen_vect_jaffer}
Given the fact that there is no unique generalization of higher-order moments to the vector case,  several vector generalizations of the identity in \eqref{eq:RecursiveScalarForm} are possible.  Next, we present two such generalizations.

The first generalization of \eqref{eq:RecursiveScalarForm} is in terms of powers of a matrix.
\begin{prop}\label{prop:JafferVersionTwo} For~$k~\in~\mathbb{N}$ and $ {\bf y} \in \mathbb{R}^n$
\begin{align}
&\E \left[ (\X \X^\mathsf{T} )^{k}  |\Y ={\bf y} \right]  = \boldsymbol{\mathsf{K}}_{\N} \boldsymbol{\mathsf{J}}_\mathbf{y}  \E \left[ (\X \X^\mathsf{T} )^{k-1} \X |\Y ={\bf y} \right]  +  \E \left[  \X  |\Y ={\bf y} \right]  \E \left[   \X^\mathsf{T}  (\X \X^\mathsf{T} )^{k-1}  |\Y ={\bf y} \right] .   \label{eq:Vector_Jaffer_Ver1}
\end{align}
\end{prop} 
\begin{IEEEproof}
The proof follows by evaluating  \eqref{eq:MainIdentity} with  $\U= (\X \X^\mathsf{T} )^{k-1} \X $ and noting that $\U^\mathsf{T} =\X^\mathsf{T}  (\X \X^\mathsf{T} )^{k-1} $. 
\end{IEEEproof}

The second generalization of \eqref{eq:RecursiveScalarForm} allows for different exponents across elements of $\X$. 

\begin{prop}\label{prop:JafferVector}
For every $m \in [n]$,   $v_i \in \mathbb{N} \cup \{0\}, \, i \in [n]$  and $ {\bf y} \in \mathbb{R}^n$
\begin{align}
\frac{{\rm d}}{ {\rm d} y_m} \E \left[ \prod_{i=1}^n \left(    {\bf e}_i ^\mathsf{T}   \boldsymbol{\mathsf{K}}_{\N}^{-1} \X \right)^{v_i}     |\Y= {\bf y} \right ] 
&= \E \left[ \prod_{i=1 : i \neq m}^n \left(    {\bf e}_i ^\mathsf{T}    \boldsymbol{\mathsf{K}}_{\N}^{-1} \X \right)^{v_i}  \left( {\bf e}_m ^\mathsf{T}   \boldsymbol{\mathsf{K}}_{\N}^{-1} \X \right)^{v_m+1} |\Y= {\bf y} \right ] \notag\\
&-  \E \left[ \prod_{i=1}^n \left(    {\bf e}_i ^\mathsf{T}   \boldsymbol{\mathsf{K}}_{\N}^{-1} \X \right)^{v_i}     |\Y= {\bf y} \right ]  \hspace{-0.05cm}   \E \left[  {\bf e}_m^\mathsf{T} \boldsymbol{\mathsf{K}}_{\N}^{-1} \X | \Y={\bf y} \right].  \label{eq:IntermediateIdentity}
\end{align} 
\end{prop} 
\begin{IEEEproof}
The proof follows by evaluating  \eqref{eq:MainIdentity} with $U=  \prod_{i=1}^n \left(    {\bf e}_i ^\mathsf{T}   \boldsymbol{\mathsf{K}}_{\N}^{-1} \X \right)^{v_i}   $.
\end{IEEEproof} 

In the case when $  \boldsymbol{\mathsf{K}}_{\N}$  is a diagonal matrix  with $[ \boldsymbol{\mathsf{K}}_{\N}]_{ii}= \sigma_{ii}^2$  the identity \eqref{eq:IntermediateIdentity} reduces to 
\begin{align}
&\E  \hspace{-0.05cm} \left[ X_m^{v_m+1} \hspace{-0.05cm} \prod_{i=1 : i \neq m}^n X_i^{v_i}   | \Y = {\bf y} \right] \hspace{-0.05cm} = \sigma_{mm}^2 \frac{{\rm d}}{ {\rm d} y_m }  \E  \hspace{-0.05cm}\left[  \prod_{i=1}^n  X_i^{v_i}    |\Y ={\bf y}  \hspace{-0.05cm} \right]  \hspace{-0.05cm}  +   \E  \hspace{-0.05cm}\left[  \prod_{i=1}^n  X_i^{v_i}  |\Y ={\bf y} \right]  \E  \hspace{-0.05cm}\left [ X_m |\Y ={\bf y} \right ] .
\end{align}

\section{Identities for the Conditional  Cumulants}
\label{sec:Conditional_Cumulants}

This section establishes a new connection between the conditional cumulants and the conditional expectation.  
For ease of exposition, we first focus on the univariate case and then generalize the results to the multivariate case. 

\subsection{The Univariate Case} 
\label{sec:Cumulatns_Univariet_Case} 

Consider the  \emph{conditional cumulant generating function}
\begin{equation}
K_X(t|Y=y)=\log\left( \E[\eu^{tX}| Y=y] \right),   \, y \in \mathbb{R},t \in \mathbb{R}. 
\end{equation}
The $k$-th conditional cumulant is given by 
\begin{equation}
\kappa_{X|Y=y}(k) =\frac{{\rm d}^{k}}{ {\rm d} t^{k} } K_X(t|Y=y) \Big |_{t=0}, \,  k\in \mathbb{N},  \,  t \in \mathbb{R}.
\end{equation}

\begin{remark} 
The conditional moment generating (i.e., $\E[\eu^{tX}| Y=y]$)  is well-defined in view of \eqref{eq:agument_for_finitness}. An alternative way to argue this is to use the fact all $t\in \mathbb{R}$,  the conditional distribution $P_{X|Y=y}$ is sub-Gaussian \cite{GuoMMSEprop}. 
\end{remark}

It is well-known that the cumulants and the moments  of a random variable $U$ have a one-to-one correspondence with the inverse relationship given by
\begin{equation}
\kappa_U(k) =   \sum_{m=1}^{k} c_m\mathsf{B}_{k,m} \left(\mu_1,\ldots, \mu_{k-m+1} \right),  \label{eq:Cumulnats+Moments}
\end{equation} 
where $\mu_{m}=\E[U^m]$ \cite[Example 11.4]{charalambides2018enumerative}.    This expression together with  the integral version of Jaffer's identity in \eqref{eq:Higher_Order_Derivatives_of_CE}  lead to the following simple relationship between the conditional expectation and the conditional cumulants. 
\begin{prop}\label{prop:Cumulant_derivative_of_CE} For $y \in \mathbb{R}$ and $k \in \mathbb{N} \cup\{0\}$
\begin{equation}
 \sigma^{2k} \frac{{\rm d}^{k}}{ {\rm d} y^{k} } \E[ X | Y=y]=\kappa_{X|Y=y}(k+1).  \label{eq:Cumulants_and_Conditional}
\end{equation}
\end{prop} 
\begin{IEEEproof}
First, let    $X_y \sim P_{X|Y=y}$ and let $U =\frac{X_y}{ \sigma^2} $.  Second, by using the scaling property of cumulants, we have that
\begin{equation}
\kappa_U(k) =   \frac{1}{\sigma^{2k}} \kappa_{X|Y=y}(k).  
\end{equation}
Next, by using \eqref{eq:Cumulnats+Moments}, for $k \ge 1$ we have that
\begin{align}
 \frac{1}{\sigma^{2k}} \kappa_{X|Y=y}(k)  
 &=    \sum_{m=1}^{k} c_m\mathsf{B}_{k,m} \left(\E \left[ \frac{X_y}{ \sigma^2} \right] ,\ldots, \E \left[  \left(\frac{X_y}{ \sigma^2} \right)^{k-m+1}  \right]\right)\\
 &= \frac{1}{\sigma^2}  \frac{ {\rm d}^{k-1}}{ {\rm d} y^{k-1}} \E \hspace{-0.05cm} \left[  X_y \right ],
\end{align} 
where the last step follows by using \eqref{eq:Higher_Order_Derivatives_of_CE}.  This concludes the proof. 
\end{IEEEproof}

From Proposition~\ref{prop:Cumulant_derivative_of_CE}, we make the following two observations:
\begin{itemize}
\item  For $k \in \mathbb{N} $
\begin{equation}
\sigma^2 \frac{{\rm d}}{ {\rm d} y } \kappa_{X|Y=y}(k)=  \kappa_{X|Y=y}(k+1); \text{ and}  \label{eq:Cumulants_Recursive_Identity}
\end{equation} 
\item  By using the TRE identity, we arrive at a new the representation of  cumulants in terms of  only $f_Y$
\begin{align}
\kappa_{X|Y=y}(1)&=  y+ \sigma^2 \frac{{\rm d}}{ {\rm d} y }  \log f_Y(y),\\
\kappa_{X|Y=y}(2)&=   \sigma^2+ \sigma^4 \frac{{\rm d}^2}{ {\rm d} y^2 }  \log f_Y(y),  \\
\kappa_{X|Y=y}(k)&=   \sigma^{2k} \frac{{\rm d}^k}{ {\rm d} y^k }  \log f_Y(y) , k \ge 3. \label{eq:TRE_for_Cumulants}
\end{align} 
In other words, the conditional cumulants depend on the joint distribution $P_{XY}$ only through the marginal distribution $P_Y$.  These formulas suggest that conditional cumulants can be estimated based  only on observations $Y$. This fact will be explored in Section~\ref{sec:EB_for_CC} where such estimators will be constructed.  
\end{itemize}

\begin{example*}In the case when  $X$ is standard Gaussian the conditional expectation  $\E[ X | Y=y]$ is a linear function of $y$. Therefore,  by using  \eqref{eq:Cumulants_and_Conditional}, we have that
\begin{align}
\kappa_{X|Y=y}(1)&= \frac{1}{1+\sigma^2} y,  \\
\kappa_{X|Y=y}(2)&= \frac{\sigma^2}{1+\sigma^2},  \\
\kappa_{X|Y=y}(k)&=0, \, k \ge 3. 
\end{align}  
Note that this is as  expected since $P_{X|Y}$ is Gaussian, and for the Gaussian distribution only the first and the second cumulants are non-zero. 
\end{example*}

\begin{example*}
Consider an example of a random variable $X \in \{-3,0,3\}$ with a uniform distribution.  In order to compute the conditional cumulants for this random variable we combine \eqref{eq:Cumulants_and_Conditional}, the TRE formula in \eqref{eq:TweediesFormulaGaussian} and Lanczos' derivative in \eqref{eq:Lanczos_derivative}, which results in 
\begin{equation}
\kappa_{X|Y=y}(k+1)= \frac{{\rm d}^k}{ {\rm d}  y^k} y+ \sigma^2 D_h^{(k+1)} \log f_Y(y) + O(h^2),
\end{equation} 
where we set $h=0.1$.  Fig.~\ref{fig:ConditionalCumulants} shows plots of $\kappa_{X|Y=y}$  vs. $y$ for several values of $k$. 

\begin{figure}[h!] 
	\centering   
	\input{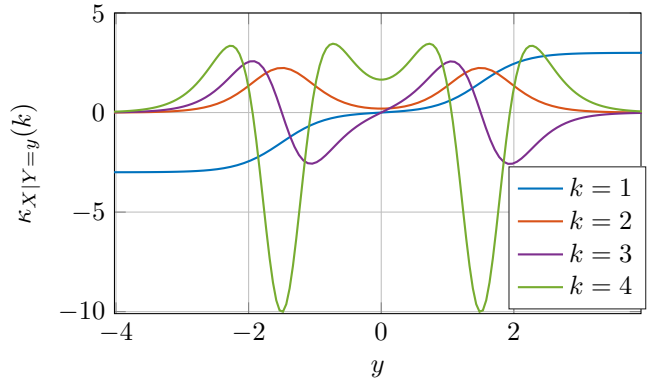}
	\caption{Plot of  $\kappa_{X|Y=y}(k)$  vs. $y$ for  $k=1,2,3$ and $4$.  }
		\label{fig:ConditionalCumulants}
\end{figure}%
One more example of the expression for the conditional cumulants  will be given in Section~\ref{sec:Cumulants_Multivariate_Case} for  the case of $X$ distributed uniformly on $\{-R,R\}$. 
\end{example*} 

We next show as small application of the new identity in \eqref{eq:Cumulants_and_Conditional}. 
\begin{example*}
A random variable $U$ with mean $\mu=\E[U]$ is said to have a \emph{sub-Gaussian} distribution if there exists a $\gamma^2$ such that
\begin{equation}
\E \left[ \eu^{\lambda(X-\mu)}\right] \le \eu^{\frac{\lambda^2 \gamma^2}{2}},   \, \lambda \in \mathbb{R}. 
\end{equation} 
The quantity $\gamma^2$ is known as  \emph{a proxy variance}. 
The distribution is said to be \emph{strictly} sub-Gaussian if ${\rm Var}(U) =\gamma$.   In \cite{GuoMMSEprop}, it was shown that for an arbitrary distribution of $X$, the conditional distribution $P_{X|Y=y}$ is sub-Gaussian. We now use the identity in \eqref{eq:TRE_for_Cumulants} to answer if $P_{X|Y=y}$ is strictly sub-Gaussian. As shown in \cite[Prop.~4.3]{arbel2019strict} a necessary condition for strict sub-Gaussianity requires that the third cumulant is zero. Therefore, a necessary condition for $P_{X|Y=y}$ to be strictly sub-Gaussian is 
\begin{equation}
0=\kappa_{X|Y=y}(3).
\end{equation}
This  certainly holds if $X$ is Gaussian. Moreover, as shown in Fig.~\ref{fig:ConditionalCumulants}, for a non-Gaussian example, the above can be zero for some  values of $y$.  However, this set of $y$'s must necessarily be of  Lebesgue measure zero  for a non-Gaussian  $X$. To see this, suppose that $0=\kappa_{X|Y=y}(3), y \in \mathcal{S}$ where $\mathcal{S}$ is a set of positive  Lebesgue measure. Then,  since $\kappa_{X|Y=y}(k)$ is real-analytic (see  Lemma~\ref{lem:AnalyticLemma}  below), by the identity theorem for real-analytic functions \cite{krantz2002primer}, it follows that  $0=\kappa_{X|Y=y}(3)$ for all $y\in \mathbb{R}$. Next, using this and the identity in \eqref{eq:TRE_for_Cumulants}, we arrive at a differential equation 
\begin{equation}
0=\kappa_{X|Y=y}(3)= \sigma^6 \frac{{\rm d}^3}{ {\rm d} y^3 }  \log f_Y(y), y\in \mathbb{R}. \label{eq:Differential_Equation_stric}
 \end{equation} 
The solution to \eqref{eq:Differential_Equation_stric} states that $ \log f_Y(y)$  must be a quadratic function and implies that $Y$ must be Gaussian. This further implies that  $X$ is Gaussian and contradicts our assumption that $X$ is non-Gaussian.  In conclusion, \emph{for a non-Gaussian $X$, the distribution $P_{X|Y=y}$ can only be strictly sub-Gaussian on a set of Lebesque measure zero.}  
 \end{example*}

We now show two bounds on the absolute value of the conditional cumulants.  
 Together with the identity in \eqref{eq:Cumulants_and_Conditional}, these bounds produce bounds on the rate of growth of the conditional expectation.  In addition, in Section~\ref{sec:Power_Series_Expansion},  these bounds will be used   to characterize the Taylor series expansion of the conditional expectation, and in Section~\ref{sec:EB_for_CC}, these bounds will be used to construct  consistent estimators of the conditional cumulants. 
\begin{prop} \label{prop:propositon_bounds_cumulants} For $y\in \mathbb{R}$ and $k \in \mathbb{N} $
\begin{align}
|\kappa_{X|Y=y}(k)| &\le 2^{k-1} k^k  \E[ |X |^k |Y =y]   \label{eq:FirstBound_on_cumulant}\\
&\le     a_k |y|^k  + b_k, 
\end{align} 
where 
\begin{align}
a_k&=k^k 2^{k-1}(2^{ \max (\frac{k}{2}-1,1)}+2), \\
b_k&=k^k(2^{ \max (\frac{k}{2}-1,1)+k} \E^{ \frac{k}{2}}[X^2]  +\E[|X|^k] ). 
\end{align}  \end{prop} 
\begin{IEEEproof}
See Appendix~\ref{app:proof_propositon_bounds_cumulants}.
\end{IEEEproof}

We conclude this section by  showing that  Theorem~\ref{thm:MainIdentity}  can be used to get a more general identity than that in \eqref{eq:Cumulants_and_Conditional}.

\begin{theorem}\label{thm:KumulantFunction_and_CE} For $ y\in \mathbb{R}$ and $k \in \mathbb{N} \cup\{0\}$
\begin{align}
\frac{{\rm d}^{k+1}}{ {\rm d} t^{k+1} }  K_X(t|Y=y) 
&  = \sigma^{2 (k+1)}\frac{{\rm d}^{k+1}}{ {\rm d} y^{k+1} }  K_X(t|Y=y) + \sigma^{2 k}  \frac{{\rm d}^{k}}{ {\rm d} y^{k} } \E[ X | Y=y].   \label{eq:Cum_gen_function_and_Cond_Expect}
\end{align}
\end{theorem} 

\begin{IEEEproof}
First, consider the case of $k=0$.  By setting $U=\eu^{tX}, \, t \in \mathbb{R}$ in Theorem~\ref{thm:MainIdentity}, we arrive at  
\begin{align}
\frac{{\rm d}}{ {\rm d} y } K_X(t|Y=y) 
 &=\frac{\frac{{\rm d}}{ {\rm d} y }  \E[\eu^{tX}| Y=y] }{ \E[ \eu^{tX}| Y=y]}\\
&=  \frac{1}{\sigma^2} \frac{ \E[ X \eu^{tX}| Y=y]-  \E[ \eu^{tX}| Y=y]  \E[ X | Y=y]}{\E[ \eu^{tX}| Y=y]}\\
&= \frac{1}{\sigma^2}  \frac{ \frac{{\rm d}}{ {\rm d} t }  \E[  \eu^{tX}| Y=y]-  \E[ \eu^{tX}| Y=y]  \E[ X | Y=y]}{\E[ \eu^{tX}| Y=y]}\\
&= \frac{1}{\sigma^2} \left( \frac{{\rm d}}{ {\rm d} t }  \log( \E[\eu^{tX}| Y=y]  )   -  \E[ X | Y=y] \right)\\
&= \frac{1}{\sigma^2} \left(\frac{{\rm d}}{ {\rm d} t }  K_X(t|Y=y) -  \E[ X | Y=y] \right).  \label{eq:cumulants_k=0}
\end{align} 
The rest of the proof follows by using  \eqref{eq:cumulants_k=0} together with a simple induction.
\end{IEEEproof} 

Alternatively, we could have used Theorem~\ref{thm:KumulantFunction_and_CE}  to show  the derivative identity in \eqref{eq:Cumulants_and_Conditional}.  This approach has the benefit of avoiding the use of Bell polynomials.
This alternative view will be taken in the next section to derive a multivariate generalization of the identity in \eqref{eq:Cumulants_and_Conditional}.

\subsection{The Multivariate Case} 
\label{sec:Cumulants_Multivariate_Case}
 Consider the  \emph{multivariate conditional  cumulant generating function}
\begin{equation}
K_{\X}( {\bf t} |\Y={\bf y})=\log\left( \E[\eu^{ {\bf t}^\mathsf{T} \X}| \Y={\bf y} ] \right),   \, {\bf y} \in \mathbb{R}^n, {\bf t}  \in \mathbb{R}^n. 
\end{equation}
The conditional cumulants  are now defined as
\begin{equation}
\kappa_{\X|\Y={\bf y}}(s_1,\ldots,s_j) =\frac{\partial^j}{\partial t_{s_1} \dots  \partial t_{s_j}  } K_{\X}( {\bf t} |\Y={\bf y}) \big |_{{\bf t}={\bf 0}}, \label{eq:Cumulant_definition_mulitivariate_case}
\end{equation} 
where $j \in \mathbb{N}$ and   $ s_1,\ldots,s_j \in  [n]$. 
Note that the cumulants are the same for all permutations of the sequence $s_1,\ldots,s_j$.    The above definitions follow the conventions of \cite{kolassa2006series}. 

It is instructive to consider the following example.  

\begin{example*}  Let $\X \sim \mathcal{N}({\bf 0}, \I)$.   Then, $\X | \Y={\bf y}  \sim \mathcal{N}\left( \E[\X|\Y={\bf y}],  \boldsymbol{\mathsf{Var}}(\X | \Y={\bf y}) \right)$  
where 
\begin{align}
 \E[\X|\Y={\bf y}]  &=   (\I+\boldsymbol{\mathsf{K}}_{\N} )^{-1}  {\bf y},  {\bf y} \in \mathbb{R}^n, \\
 \boldsymbol{\mathsf{\Sigma}}= \boldsymbol{\mathsf{Var}}(\X | \Y={\bf y})&=  \boldsymbol{\mathsf{K}}_{\N} (\I+\boldsymbol{\mathsf{K}}_{\N} )^{-1}, \,  {\bf y} \in \mathbb{R}^n ,
 \end{align}
 with the moment generating function given by 
\begin{equation}
\E[\eu^{ {\bf t}^\mathsf{T} \X}| \Y={\bf y} ]=   \eu^{ {\bf t}^\mathsf{T } \E[\X|\Y={\bf y}]  }   \eu^{ \frac{{\bf t}^\mathsf{T}    \boldsymbol{\mathsf{\Sigma}} {\bf t}   }{2}},  {\bf t} \in \mathbb{R}^n,  {\bf y} \in \mathbb{R}^n,
\end{equation} 
and the cumulant generating function given by
\begin{equation}
K_{\X}( {\bf t} |\Y={\bf y})= {\bf t}^\mathsf{T } \E[\X|\Y={\bf y}]  + \frac{{\bf t}^\mathsf{T}   \boldsymbol{\mathsf{\Sigma}}  {\bf t}   }{2}, {\bf t} \in \mathbb{R}^n,  {\bf y} \in \mathbb{R}^n.
\end{equation} 
Using the definition in \eqref{eq:Cumulant_definition_mulitivariate_case}, the cumulants are calculated to be 
\begin{align*}
j&=1: \, \kappa_{\X|\Y={\bf y}}(s_1)=   \E[X_{s_1}|\Y={\bf y}],  \,  s_1 \in [n], \\
j&=2: \,  \kappa_{\X|\Y={\bf y}}(s_1,s_2)= [   \boldsymbol{\mathsf{\Sigma}}  ]_{s_1,s_2}, \,   s_1,s_2 \in [n], \\
j&\ge 3: \,  \kappa_{\X|\Y={\bf y}}(s_1,\ldots, s_j)= 0,  \,   s_1,\ldots, s_j \in [n]. 
\end{align*}
\end{example*} 

We first show a multivariate generalization of Theorem~\ref{thm:KumulantFunction_and_CE}, which  follows by letting $U= \eu^{ {\bf t}^\mathsf{T} \X}$ in Theorem~\ref{thm:MainIdentity}. 

\begin{theorem}\label{thm:multivariate_CGF_CE} Let  ${\bf k}_i^\mathsf{T} =\mathbf{e}_i^\mathsf{T} \boldsymbol{\mathsf{K}}_{\N}^{-1} $  (i.e., the $i$-th row of $\boldsymbol{\mathsf{K}}_{\N}^{-1}$). Then,   for $j \in \mathbb{N}$ and ${\bf y} \in \mathbb{R}^n$
\begin{equation}
\frac{\partial^{j}  K_{\X}( {\bf t} |\Y={\bf y})  }{ \partial y_{s_1} \dots  \partial y_{s_{j}}   }
=  {\bf k}_{s_1}^\mathsf{T}   \nabla_{\bf t} {\bf k}_{s_2}^\mathsf{T}   \nabla_{\bf t}  \dots   {\bf k}_{s_j}^\mathsf{T}   \nabla_{\bf t} K_{\X}( {\bf t} |\Y={\bf y}) -  {\bf k}_{s_j}^\mathsf{T}   \frac{\partial^{j-1}   \E[ \X |\Y={\bf y}]  }{ \partial y_{s_1} \dots  \partial y_{s_{j-1}}    }  . \label{eq:Multivariate_Cumulant_Function_CE}
\end{equation} 
\end{theorem} 
\begin{IEEEproof}
See Appendix~\ref{app:proof:thm:multivariate_CGF_CE}.
\end{IEEEproof}

\begin{remark} The first term on the right side of \eqref{eq:Multivariate_Cumulant_Function_CE} can be equivalently written as
\begin{align}
{\bf k}_{s_1}^\mathsf{T}   \nabla_{\bf t} {\bf k}_{s_2}^\mathsf{T}   \nabla_{\bf t}  \dots   {\bf k}_{s_j}^\mathsf{T}   \nabla_{\bf t} K_{\X}( {\bf t} |\Y={\bf y}) =\sum_{p_1=1}^n \dots  \sum_{p_j=1}^n  \prod_{i=1}^j   k_{s_i,p_i}  \frac{\partial^{j}  K_{\X}( {\bf t} |\Y={\bf y})  }{ \partial t_{p_1} \dots  \partial t_{p_{j}}   },
\end{align}  
where  $k_{s_j,p_j}$ is $p_j$'s entry of  ${\bf k}_{s_j}$.  This representation will be useful in the next proof, which relates the conditional cumulants and the conditional moments. 
\end{remark}

The next result generalizes the derivative identity in \eqref{eq:Cumulants_and_Conditional} to the multivariate case.

\begin{prop}\label{prop:PartialDerivattives_CE}  Let  ${\bf k}_m^\mathsf{T} =\mathbf{e}_m^\mathsf{T} \boldsymbol{\mathsf{K}}_{\N}^{-1} $ and $k_{m,i}= \mathbf{e}_m^\mathsf{T} \boldsymbol{\mathsf{K}}_{\N}^{-1} \mathbf{e}_i $. Then,  for $j \in \mathbb{N}$ and ${\bf y} \in \mathbb{R}^n$
\begin{align}
  {\bf k}_{s_j}^\mathsf{T}   \frac{\partial^{j-1}   \E[ \X |\Y={\bf y}]  }{ \partial y_{s_1} \dots  \partial y_{s_{j-1}}   }   =\sum_{p_1=1}^n \dots  \sum_{p_j=1}^n  \prod_{i=1}^j   k_{s_i,p_i}   \kappa_{\X|\Y={\bf y}}(p_1,\ldots, p_j).  \label{eq:Multivaraite_Cumulant_Expecations_identity}
\end{align} 
\end{prop} 
\begin{IEEEproof}
The proof  follows by showing that for every $j$  we have that  $\frac{\partial^{j}  K_{\X}( {\bf t} |\Y={\bf y})  }{ \partial y_{s_1} \dots  \partial y_{s_{j}}   } |_{{\bf t}={\bf 0}} =0$.  See Appendix~\ref{app:prop:PartialDerivattives_CE} for the details.  
\end{IEEEproof}

In the case when $\boldsymbol{\mathsf{K}}_{\N}$ is a diagonal matrix the above simplifies to 
\begin{align}
   \kappa_{\X|\Y={\bf y}}(s_1,\ldots, s_j)  = \left(  \prod_{i=1}^{j-1}  \sigma^2_{s_i,s_i}   \right)    \frac{\partial^{j-1} \E[ X_{s_j} |\Y={\bf y}]  }{ \partial y_{s_1} \dots  \partial y_{s_{j-1}}   }   ,
\end{align} 
where we let $\prod_{i=1}^{0}  \sigma^2_{s_i,s_i}  =1$.

Closed-form expressions for the conditional expectation are rare in the univariate case and even more so in the multivariate case. In particular, not many examples of $\E[ \X |\Y={\bf y}]$ are known when $\X$ has a non-product distribution.  The next example computes the first two cumulants for one of the rare cases when we do have a closed-form expression for the conditional expectation.

\begin{example*} Consider a case when $\X$ is distributed  uniformly on $\{{\bf x} \in \mathbb{R}^n:  \|{\bf x} \|=R   \}$ (i.e., $(n-1)$-sphere of radius $R$)   and  let $\boldsymbol{\mathsf{K}}_{\N}=\I$.    This distribution has several applications in information theory and estimation theory.  For example,  in information theory,  this distribution  is the capacity-achieving distribution for an amplitude constrained channel  \cite{dytso2019capacity}.  In estimation theory, this distribution has been shown to be the least favorable distribution for  the problem of estimating a bounded normal mean \cite{berry1990minimax}.   The conditional expectation for this distribution is given by  \cite{dytso2019capacity} 
\begin{align*}
\E[\X| \Y={\bf y} ]=  \frac{R  {\bf y}}{ \| {\bf y} \| }   \frac{I_{\frac{n}{2}}( R \| {\bf y}\|  ) }{ I_{\frac{n}{2}-1}(R \| {\bf y}\|) }, \, {\bf y} \in \mathbb{R}^n, 
\end{align*}
where $I_{\nu}(\cdot)$ is the modified   Bessel function of the first kind of order $\nu$.  Next, using Proposition~\ref{prop:PartialDerivattives_CE}, we characterize the first two conditional cumulants of this distribution.  For   $j=1,$   $ \kappa_{\X|\Y={\bf y}}(s_1)=   \E[ X_{s_2} |\Y={\bf y}] $, 
and for $j=2$ the expression for  $\kappa_{\X|\Y={\bf y}}(s_1,s_2)$ is given in \eqref{eq:Kappa_for_X_uniform_sphere}.  The proofs for $j=2$ can be found in Appendix~\ref{app:Proof_Cumulatns_shell}.    

For the case of  $n=1$, the distribution under consideration becomes uniform on $\{-R,R\}$ and as expected  the expression in \eqref{eq:Kappa_for_X_uniform_sphere} reduces to $k_{X|Y=y}(2)= \left( \frac{R}{\cosh(Ry)} \right)^2 $, which is the derivative of $\E[X|Y=y]=R \tanh( R y)$. 
\end{example*} 

\begin{figure*}[h]
\begin{equation}
 \kappa_{\X|\Y={\bf y}}(s_1,s_2) =
 \left \{ \begin{array}{ll}
 \frac{ R  y_{s_2} y_{s_1} }{ \| {\bf y} \| }   \frac{I_{\frac{n}{2}}( R \| {\bf y}\|  ) }{ I_{\frac{n}{2}-1}(R \| {\bf y}\|) }   +    \frac{ R^2  y_{s_2}  y_{s_1} }{ \| {\bf y} \| ^2}   \left(1 -\frac{n-1}{R \| {\bf y}\|}  \frac{I_{\frac{n}{2}}( R \| {\bf y}\|  ) }{ I_{\frac{n}{2}-1}(R \| {\bf y}\|) }  - \left( \frac{I_{\frac{n}{2}}( R \| {\bf y}\|  ) }{ I_{\frac{n}{2}-1}(R \| {\bf y}\|) }  \right)^2  \right) ,   & s_1 \neq s_2 \\
  R \frac{  \| {\bf y} \|  -   \frac{ y_{s_2}^2  }{ \| {\bf y} \|}  }{ \| {\bf y} \|^2 }   \frac{I_{\frac{n}{2}}( R \| {\bf y}\|  ) }{ I_{\frac{n}{2}-1}(R \| {\bf y}\|) } +    \frac{ R^2  y_{s_2}^2 }{ \| {\bf y} \| ^2}   \left(1 -\frac{n-1}{R \| {\bf y}\|}  \frac{I_{\frac{n}{2}}( R \| {\bf y}\|  ) }{ I_{\frac{n}{2}-1}(R \| {\bf y}\|) }  - \left( \frac{I_{\frac{n}{2}}( R \| {\bf y}\|  ) }{ I_{\frac{n}{2}-1}(R \| {\bf y}\|) }  \right)^2  \right),  & s_1 =s_2 
\end{array}  \right.  \label{eq:Kappa_for_X_uniform_sphere}
\end{equation} 
\end{figure*}

\section{Identities for   Distributions, the Information Density and the MMSE}
\label{sec:Identities_for_distributions}

In this section, we study identities between the conditional expectation and quantiles such the conditional  distribution $P_{\X|\Y}$, the pdf of $\Y$ and the information density of the pair $(\X,\Y)$.  As a small application, we show how such identities can be used to find lower bounds on the MMSE.

\subsection{An Alternative View  and Generalization of the TRE Identity, and Higher-Order Derivative of the Information Density}
\label{sec:info_density_connection}

Let  \emph{the information density} be defined as 
\begin{equation}
\iota_{P_{\X \Y}}({\bf x};{\bf y})=\log \frac{ {\rm d} P_{\X \Y} }{ {\rm d} (P_{\X } \otimes P_{\Y }  ) }({\bf x}, {\bf y}),  \,  {\bf x}, {\bf y} \in   \mathbb{R}^n.  \label{eq:Info_Density_def}
\end{equation} 
In this section, we are interested in characterizing derivatives of the information density with respect to the variable ${\bf y}$. 

We start this section by  observing the following alternative version  of the TRE identity, which establishes the gradient of information density. 

\begin{prop}\label{prop:Gradient_log_pmf_pdf}   For ${\bf x}, {\bf y} \in \mathbb{R}^n$
\begin{equation}
 \nabla_{\bf y} \iota_{P_{\X \Y}}({\bf x};{\bf y})= \boldsymbol{\mathsf{K}}_{\N}^{-1}  ( {\bf x} -  \E[  \X   | \Y={\bf y}]     ).   \label{eq:Gradient_Info_Density}
\end{equation}

\end{prop} 
\begin{IEEEproof}
Fix some ${\bf x}$ and ${\bf y}$. Then, 
\begin{align}
 \nabla_{\bf y} \iota({\bf x};{\bf y})  &=  \nabla_{\bf y} \log \frac{f_{\Y|\X}({\bf y}| {\bf x}) }{ f_{\Y}( {\bf y})} \\
 & =\nabla_{\bf y} \log \frac{\phi_{ \boldsymbol{\mathsf{K}}_{\N}} ({\bf y}-{\bf x}) }{ f_{\Y}( {\bf y})} \\
  & =  -    \boldsymbol{\mathsf{K}}_{\N}^{-1} ({\bf y}-{\bf x})  - \nabla_{\bf y} \log  f_{\Y}( {\bf y})   \label{Eq:ApplyingDerivativeOfGaussian}\\\
    & =  -    \boldsymbol{\mathsf{K}}_{\N}^{-1} ({\bf y}-{\bf x})  - \boldsymbol{\mathsf{K}}_{\N}^{-1} (\E[\X|\Y={\bf y}]-{\bf y} ) \label{Eq:ApplyingTRE_proof_info_density}    \\
    &=\boldsymbol{\mathsf{K}}_{\N}^{-1}  ( {\bf x} -  \E[  \X   | \Y={\bf y}]     ),
\end{align} 
where in \eqref{Eq:ApplyingDerivativeOfGaussian}  we have used   $\nabla_{\bf y}   \log  \phi_{ \boldsymbol{\mathsf{K}}_{\N}} ({\bf y}-{\bf x})= -    \boldsymbol{\mathsf{K}}_{\N}^{-1} ({\bf y}-{\bf x})$; and \eqref{Eq:ApplyingTRE_proof_info_density}   follows by using the TRE identity in \eqref{eq:TweediesFormulaGaussian}. This concludes the proof. 
\end{IEEEproof} 

Using Theorem~\ref{thm:MainIdentity} and setting  $U = \mathsf{1}_{ \mathcal{A} }(\X), \, \mathcal{A} \subseteq \mathbb{R}^n$, the  TRE identity can be generalized as follows.

\begin{prop} \label{prop:ConditionaTweedie}  Let $\mathcal{A} \subseteq \mathbb{R}^n$ be a measurable set  such that $\mathbb{P}[ \X  \in \mathcal{A} ]>0$. 
Then,  for ${\bf y} \in \mathbb{R}^n$
\begin{equation}
\nabla_{\bf y} \log \left(   \mathbb{P}[\X \in \mathcal{A}| \Y={\bf y}] \right) = \boldsymbol{\mathsf{K}}_{\N}^{-1}    (  \E[  \X  | \Y={\bf y}, \X \in \mathcal{A}] -  \E[  \X   | \Y={\bf y}]     )   .  \label{eq:Indentity_Conditiona_Probaiblity}
\end{equation} 
\end{prop} 
\begin{IEEEproof}
Let $U  = \mathsf{1}_{ \mathcal{A} }(\X)$ in Theorem~\ref{thm:MainIdentity}. 
This choice of $\U$ implies   the following: 
\begin{align}
\E[ U   | \Y={\bf y}] &= \mathbb{P}[\X \in \mathcal{A}| \Y={\bf y}],  \label{eq:FirstExpressionOfU}\\
\E[  \X  U | \Y={\bf y}] &= \E[  \X   | \Y={\bf y}, \X \in \mathcal{A}]  \mathbb{P}[\X \in \mathcal{A}| \Y={\bf y}]. \label{eq:SecondExpressionOfU} 
\end{align} 
Combining \eqref{eq:FirstExpressionOfU} and \eqref{eq:SecondExpressionOfU}  with the identity in \eqref{eq:MainIdentity} we arrive at 
\begin{align}
& \frac{\nabla_{\bf y} \mathbb{P}[\X \in \mathcal{A}| \Y={\bf y}] }{\mathbb{P}[\X \in \mathcal{A}| \Y={\bf y}]} =  \boldsymbol{\mathsf{K}}_{\N}^{-1}    (  \E[  \X  | \Y={\bf y}, \X \in \mathcal{A}] -  \E[  \X   | \Y={\bf y}]     )  ,\label{eq:TheExpressionFromTheMain}
\end{align} 
where we have used that  $\mathbb{P}[ \X  \in \mathcal{A} ]>0$. 
The proof is concluded by observing that $\nabla_{\bf y} \log (   \mathbb{P}[\X \in \mathcal{A}| \Y={\bf y}] )  =\frac{ \nabla_{{\bf y}}  \mathbb{P}[\X \in \mathcal{A}| \Y={\bf y}]}{ \mathbb{P}[\X \in \mathcal{A}| \Y={\bf y}]}$.
\end{IEEEproof}

To see that \eqref{eq:Indentity_Conditiona_Probaiblity} is a generalization of \eqref{eq:Gradient_Info_Density} suppose that $\X$ is a discrete random vector. Then,  by setting $\mathcal{A}=\{ {\bf x} \}$ where ${\bf x}$  is a point of support of $\X$, the identity in \eqref{eq:Indentity_Conditiona_Probaiblity} reduces to 
\begin{align}
\nabla_{\bf y} \log \left(   \mathbb{P}[\X={\bf x} | \Y={\bf y}] \right)  
&= \nabla_{\bf y}  \iota_{P_{\X \Y}}({\bf x};{\bf y})\\ 
 &=     \boldsymbol{\mathsf{K}}_{\N}^{-1}  ( {\bf x} -  \E[  \X|\Y={\bf y}    ]     ),  \label{eq:error_and_via_conditional_Expectation}
\end{align} 
where we have used that  $\E[  \X   | \Y={\bf y}, \X ={\bf x}]={\bf x}$.

%

As an application of Proposition~\ref{prop:Gradient_log_pmf_pdf} and Proposition~\ref{prop:ConditionaTweedie}, we now  characterizing the Hessian of the information density and  the Hessian of the log of the posterior distribution.  Again, the key ingredient in the proof will be the identity in Theorem~\ref{thm:MainIdentity}. 

\begin{prop} \label{prop:Hessian_of_log_dist}   For $({\bf x},{\bf y} ) \in  \mathbb{R}^n\times  \mathbb{R}^n$:
\begin{itemize}[leftmargin=*]
\item (Hessian of the Information Density) 
\begin{equation}
 \boldsymbol{\mathsf{D}}^2_{\bf y} \iota_{P_{\X \Y}}({\bf x};{\bf y}) =  -   \boldsymbol{\mathsf{K}}_{\N}^{-1}  \boldsymbol{\mathsf{Var}}(\X | \Y={\bf y})  \boldsymbol{\mathsf{K}}_{\N}^{-1}.  \label{eq:Hessian_of_Conditional_distribuiton_continious}
 \end{equation}
\item (General Case)  
Let $\mathcal{A} \subseteq \mathbb{R}^n$ be a measurable set  such that $\mathbb{P}[ \X  \in \mathcal{A} ]>0$. 
Then, 
\begin{equation}
\boldsymbol{\mathsf{D}}^2_{\bf y}  \log \left(   \mathbb{P}[ \X \in \mathcal{A} | \Y={\bf y}] \right) =\boldsymbol{\mathsf{K}}_{\N}^{-1} \left( \boldsymbol{\mathsf{Var}}(\X | \Y={\bf y}, \X \in \mathcal{A}) - \boldsymbol{\mathsf{Var}}(\X | \Y={\bf y})     \right) \boldsymbol{\mathsf{K}}_{\N}^{- 1 }.   \label{eq:Hessian_of_Conditional_General_Case}
\end{equation}
\end{itemize} 

\end{prop} 
\begin{IEEEproof}
See Appendix~\ref{app:prop:Hessian_of_log_dist}. 
\end{IEEEproof}

By choosing a specific family of sets $\mathcal{A}$, we next illustrate an example of how the identity in \eqref{eq:Hessian_of_Conditional_distribuiton_continious} can be used to study other quantities such the conditional cumulative distribution function (cdf).

\begin{figure}[h!]
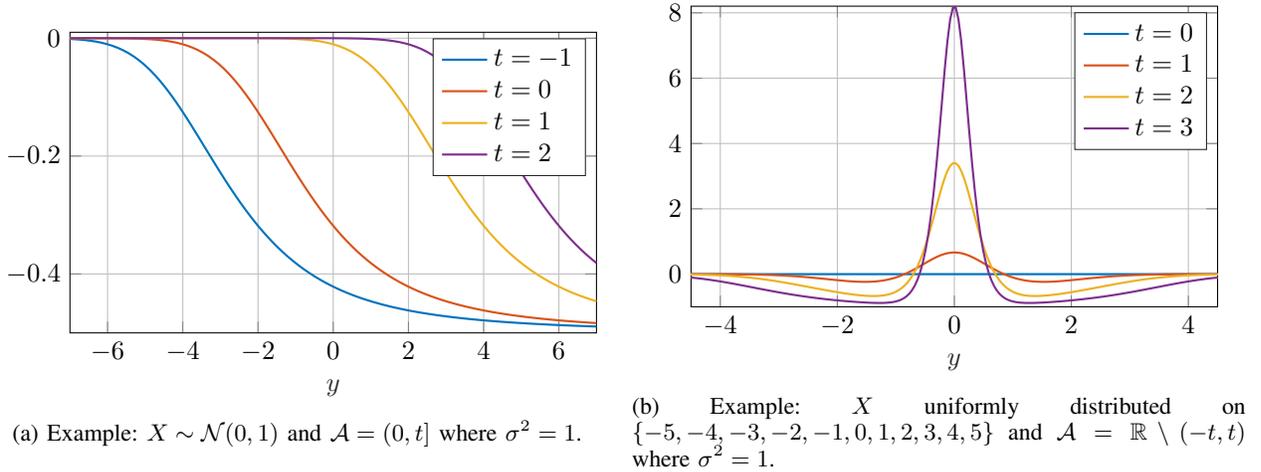
 
	\centering
	  \begin{subfigure}[c]{0.45\textwidth}
	  \center
	\input{Gaussian_Log_Second_Derivative.tex}
	\caption{Example: $X\sim \mathcal{N}(0,1)$ and $\mathcal{A}=(0,t]$ where $\sigma^2=1$.}
	\label{fig:Gaussian_Log_P_Second_Der}
	\end{subfigure}
	~
	  \begin{subfigure}[c]{0.45\textwidth}
	  \center
	\input{Discrete_Log_P_Second_Der.tex}
	\caption{Example: $X$ uniformly distributed on  $\{- 5, -4, -3, -2, -1, 0, 1, 2, 3,4,5  \}$  and $\mathcal{A}=\mathbb{R}\setminus(-t,t)$ where $\sigma^2=1$. }  
	\label{fig:Discrete_Log_P_Second_Der}
	  \end{subfigure}
		
		\caption{Examples of $\frac{\rm d^2}{ {\rm d} y^2}   \log \left(   \mathbb{P}[ X \in \mathcal{A} | Y=y] \right)$. }
		
\end{figure}

\begin{example*}   Consider a family of sets given by $\mathcal{A}=(-\infty,t] , \, t>0$.  With this choice, we have that  $\mathbb{P}[X \in \mathcal{A}| Y=y]$ is equal to the conditional cumulative distribution function  $F(X \le t|Y=y)$ and by using \eqref{eq:Indentity_Conditiona_Probaiblity},  we arrive at 
\begin{equation}
\frac{\rm d^2}{ {\rm d} y^2}   \log \left(  F(X \le t|Y=y) \right) =\frac{\mathsf{Var} \left(X | Y=y, X \le t \right) - \mathsf{V ar} \left(X | Y=y\right)  }{\sigma^4} . 
\end{equation}

Now, consider the case of $X \sim \mathcal{N}(0,1)$. Then, $X|Y=y \sim \mathcal{N}( \frac{y}{1+\sigma^2} ,  \frac{\sigma^2}{1+\sigma^2})$. Moreover, since  $X|\{X \le t  \}$ is a truncated Gaussian \cite{johnson1995continuous}, it can be shown that 
\begin{equation}
\mathsf{Var} \left(X | Y=y, X\le t  \right) =  \frac{\sigma^2}{1+\sigma^2} \left(1-  \frac{\beta(y) \phi( \beta(y))}{\Phi(\beta(y)) } -  \left(\frac{  \phi( \beta(y))}{\Phi(\beta(y)) } \right)^2 \right),
\end{equation}
where
\begin{equation} 
 \beta(y)=\frac{t  -\frac{y}{1+\sigma^2}}{  \sqrt{  \frac{\sigma^2}{1+\sigma^2}} } . 
\end{equation} 
Consequently, from \eqref{eq:Hessian_of_Conditional_General_Case} we have that 
\begin{equation}
\frac{\rm d^2}{ {\rm d} y^2}   \log \left(  F(X \le t|Y=y) \right) 
=- \frac{1}{\sigma^2 (1+\sigma^2)} \left( \frac{\beta(y) \phi( \beta(y))}{\Phi(\beta(y)) } +  \left(\frac{  \phi( \beta(y))}{\Phi(\beta(y)) } \right)^2 \right).  
\end{equation} 
Fig.~\ref{fig:Gaussian_Log_P_Second_Der} plots  $\frac{\rm d^2}{ {\rm d} y^2}   \log \left(  F(X \le t|Y=y) \right)$. 
\end{example*}

From Proposition~\ref{prop:Hessian_of_log_dist} we have the following corollary. 
\begin{corollary}  The mapping $ ({\bf x},{\bf y} ) \to  \boldsymbol{\mathsf{D}}^2_{\bf y} \iota_{P_{\X \Y}}({\bf x};{\bf y})$   is only a function of the variable  ${\bf y} $. Moreover,   $   {\bf y} \to  \iota_{P_{\X \Y}}({\bf \cdot} ;{\bf y})$ is a concave function.   
\end{corollary}

 We were not able to locate an explicit statement of the above result in the literature.    The equivalent statement was produced as an intermediate step in a proof  used in \cite[p.~2229]{cai2014optimal} where  convexity of the function akin to the  log-likelihood ratio was shown.   

It is also interesting to ask whether $ {\bf y} \to  \log( \mathbb{P}[ \X \in \mathcal{A} | \Y={\bf y}])$ is also concave?  The following example shows that this is not the case and one can find a non-trivial  family of sets for which $ {\bf y} \to  \log( \mathbb{P}[ \X \in \mathcal{A} | \Y={\bf y}])$ is  not concave. 

\begin{example*} Consider the case of $X$ uniformly distributed on $\{- 5, -4, -3, -2, -1, 0, 1, 2, 3,4,5  \}$.  Moreover, we let  $\mathcal{A}=\mathbb{R}\setminus(-t,t)$. For example,  if $t=3$, then $X|\{X \in  \mathcal{A} \}$ is uniformly distributed on  $\{- 5, -4, -3, 3,4,5  \}$.   Fig.~\ref{fig:Discrete_Log_P_Second_Der}  shows the plot of  $\frac{\rm d^2}{ {\rm d} y^2}   \log \left(   \mathbb{P}[ X \in \mathcal{A} | Y=y] \right)$ for $t=0,1,2$ and $t=3$ where we set $\sigma^2=1$.  From  Fig.~\ref{fig:Discrete_Log_P_Second_Der}  we see that the second derivative can be positive. Hence,  $ {y} \to \mathbb{P}[ X \in \mathcal{A} | Y=y]$ is not log-concave for this family of sets.
\end{example*} 

So far we have characterized the first and the second order derivatives of the information density and discussed generalizations of these identities.  The next results provides an expression for all the higher-order derivatives  in the scalar case. 
\begin{prop}\label{prop:kth_derivative_info_density} For $k\ge 2$ and $(x,y)\in \mathbb{R} \times \mathbb{R}$
\begin{equation}
\frac{{\rm d}^k}{ {\rm d} y^k} \iota_{P_{X Y}}(x;y)= - \frac{1}{\sigma^{2k}} \kappa_{X|Y=y}(k). \label{eq:higher_order_derivatives_of_Information_density}
\end{equation} 
\end{prop}  
\begin{IEEEproof}
\begin{align}
\frac{{\rm d}^k}{ {\rm d} y^k} \iota_{P_{X Y}}(x;y)&= \frac{{\rm d}^{k-1}}{ {\rm d} y^{k-1}} 
\frac{1}{\sigma^2}  ( x -  \E[  X   | Y=y]     ) \label{eq:Using_TRE_info_density}\\
&=-  \frac{   1 }{\sigma^2}   \frac{\kappa_{X|Y=y}(k+1)}{ \sigma^{2(k-1)}} \label{eq:Using_cumulant_identity}\\
&=- \frac{1}{\sigma^{2k}} \kappa_{X|Y=y}(k),
\end{align} 
where in \eqref{eq:Using_TRE_info_density} we have used \eqref{eq:Gradient_Info_Density}; and in \eqref{eq:Using_cumulant_identity} we have used \eqref{eq:Cumulants_and_Conditional}. 
\end{IEEEproof}


\subsection{An Inverse TRE Identity}
\label{sec:inverse_TRE_Identity}

The TRE identity shows that the conditional expectation is completely determined by $f_{\Y}$. It is also possible to have an inverse statement  that shows that   $\E[\X | \Y]$ completely determines  $f_{\Y}$.
\begin{prop}\label{prop:Inverse_TRE} For ${\bf y} \in \mathbb{R}^n$
\begin{equation}
f_\Y({\bf y})= c  \exp \left(   \oint_{\bf 0}^{\bf y}  \boldsymbol{\mathsf{K}}_{\N}^{-1} (\mathbf{t}-   \E[\X | \Y={\bf t}])      \boldsymbol{\cdot} {\rm \mathbf{d} } \mathbf{t} \right) , \label{eq:Invers_TRE}
\end{equation} 
where $0<c<\infty$ is the normalization constant and is given by
\begin{equation}
c^{-1} = \int_{\mathbb{R}^n }   \exp \left(   \oint_{\bf 0}^{\bf y}  \boldsymbol{\mathsf{K}}_{\N}^{-1} (\mathbf{t}-   \E[\X | \Y={\bf t}])      \boldsymbol{\cdot} {\rm \mathbf{d} } \mathbf{t} \right) {\rm \mathbf{d} } {\bf y}. 
\end{equation} 
\end{prop} 
\begin{IEEEproof}
The  TRE expression in \eqref{eq:TweediesFormulaGaussian} can be rewritten as
\begin{equation}
 \nabla_{ {\bf y}}   \log (   f_\Y({\bf y})  ) = \boldsymbol{\mathsf{K}}_{\N}^{-1} (  {\bf y}- \E[\X | \Y={\bf y}]).  
\end{equation} 
Using the fundamental theorem of calculus for the line integral, we have that
\begin{equation}
 \log (   f_\Y({\bf y})  )-  \log (   f_\Y({\bf 0})  )=  \oint_{\bf 0}^{\bf y}  \boldsymbol{\mathsf{K}}_{\N}^{-1} (\mathbf{t}-   \E[\X | \Y={\bf t}])      \boldsymbol{\cdot} {\rm \mathbf{d} } \mathbf{t} ,
\end{equation}  
or equivalently 
\begin{equation}
f_\Y({\bf y})=   f_\Y({\bf 0})  \exp \left(   \oint_{\bf 0}^{\bf y}  \boldsymbol{\mathsf{K}}_{\N}^{-1} (\mathbf{t}-   \E[\X | \Y={\bf t}])      \boldsymbol{\cdot} {\rm \mathbf{d} } \mathbf{t} \right) .\label{eq:intermediateStep}
\end{equation} 
After integrating both sides of \eqref{eq:intermediateStep}, we have arrive at 
\begin{equation}
 f_\Y({\bf 0})=\frac{1}{ \int_{\mathbb{R}^n }   \exp \left(   \oint_{\bf 0}^{\bf y}  \boldsymbol{\mathsf{K}}_{\N}^{-1} (\mathbf{t}-   \E[\X | \Y={\bf t}])      \boldsymbol{\cdot} {\rm \mathbf{d} } \mathbf{t} \right) {\rm \mathbf{d} } {\bf y}} . 
\end{equation} 
Letting $c= f_\Y({\bf 0})$ and noting that 
\begin{equation}
 0<f_\Y({\bf 0})=\frac{1}{ (2\pi)^{\frac{n}{2}} {\rm det}^{\frac{1}{2}}(\boldsymbol{\mathsf{K}}_{\N})} \E\left[ \eu^{-\frac{\X^\mathsf{T}\boldsymbol{\mathsf{K}}_{\N}^{-1}\X  }{ 2}} \right ]<\infty
 \end{equation} concludes the proof. 
\end{IEEEproof} 


We now use  the representation of  $f_\Y$ in  Proposition~\ref{prop:Inverse_TRE} to show the following important result, which generalizes Theorem~\ref{thm:UniqunessOfConditionalExpectation} to the vector case. 
\begin{corollary}\label{cor:UniqunessOfConditionalExpectation}  The conditional  expectation $\E[\X|\Y]$  is a bijective operator of $P_{\X}$ or $f_{\Y}$. In other words,  for $\Y_1=\X_1+\N_1$ and $\Y_2=\X_2+\N_2$ where $\N_1$ and $\N_2$ are Gaussian with the same covariance matrix, 
\begin{align}
&\E[\X_1|\Y_1  \hspace{-0.05cm} ={\bf y}] \hspace{-0.05cm}= \hspace{-0.05cm} \E[\X_2|\Y_2  \hspace{-0.05cm} ={\bf y}],  \forall {\bf y} \in \mathbb{R}^n    \notag\\ &\quad \Longleftrightarrow   \hspace{-0.05cm} P_{\X_1}  \hspace{-0.05cm}= \hspace{-0.05cm} P_{\X_2}  \notag\\
&\quad \Longleftrightarrow    f_{\Y_1}({\bf y})=  f_{\Y_2}({\bf y}),  \, \forall {\bf y} \in \mathbb{R}^n.
\end{align} 
\end{corollary} 
\begin{IEEEproof} First, let $P_{\X_1}=P_{\X_2} $, then it is immediate that
\begin{equation}
 \E[\X_1|\Y_1={\bf y}]= \E[\X_2|\Y_2={\bf y}],   \, \forall {\bf y} \in \mathbb{R}^n  . 
\end{equation} 
Now, suppose that 
\begin{equation}
 \E[\X_1|\Y_1={\bf y}]= \E[\X_2|\Y_2={\bf y}],   \, \forall {\bf y} \in \mathbb{R}^n  . 
 \end{equation} 
 Then, by using \eqref{eq:Invers_TRE} we have that
 \begin{equation}
 f_{\Y_1}({\bf y})=  f_{\Y_2}({\bf y}),  \, \forall {\bf y} \in \mathbb{R}^n.   \label{eq:fy=fy}
 \end{equation} 
 The fact that \eqref{eq:fy=fy} implies that $P_{\X_1}=P_{\X_2} $  follows from the standard argument that uses characteristics functions. 
\end{IEEEproof}

 Corollary~\ref{cor:UniqunessOfConditionalExpectation} has important ramification in estimation theory.  In particular, combining   Corollary~\ref{cor:UniqunessOfConditionalExpectation} and \cite[Thm.~1]{FunctionalPropMMSE} leads to a conclusion that the $ \mmse(\X| \Y)$ is a \emph{strictly} convex function of the input distribution $P_\X$.  This can further be used to argue that optimization problems of the following form have  unique maximizers: 
 \begin{equation}
 \max _{P_\X \in \mathcal{P}} \mmse(\X| \Y), 
 \end{equation}
 where $\mathcal{P}$ is some compact set of probability distributions. The interested reader is referred to \cite{dytso2018structure} for more details.

\subsection{Representations of the MMSE} 
\label{sec:MMSE_representation}

In this section, we use properties developed for the information density   to find  alternative representations of the MMSE. These representations are then used to find lower bounds on the MMSE.

\begin{prop}\label{prop:MMSE_rep}    \text{} 
\begin{itemize}[leftmargin=*]
\item \emph{(Gradient Representation)}
\begin{align}
&\E \left[     \nabla_{\bf Y} \iota_{P_{\X \Y}}(\X;\Y)  \left(  \nabla_{\bf Y}\iota_{P_{\X \Y}}(\X;\Y)   \right)^\mathsf{T}   \right]   =   \boldsymbol{\mathsf{K}}_{\N}^{-1} \boldsymbol{\mathsf{MMSE}}(\X|\Y)  \boldsymbol{\mathsf{K}}_{\N}^{-1} . \label{eq:MMSE_Gradient_Expression}
\end{align} 
Consequently,
\begin{equation}
 \mmse(\X| \Y)=\E \left[   \| \boldsymbol{\mathsf{K}}_{\N} \iota_{P_{\X \Y}}(\X;\Y)   \|^2   \right].  \label{eq:Gradient_Rep_MMSE_discrete}
\end{equation} 
\item \emph{(Hessian Representation)}  
\begin{equation}
\E \left[   \boldsymbol{\mathsf{D}}^2_{\Y} \iota_{P_{\X \Y}}(\X;\Y) \right] =   -\boldsymbol{\mathsf{K}}_{\N}^{- 1 } \boldsymbol{\mathsf{MMSE}}(\X|\Y)  \boldsymbol{\mathsf{K}}_{\N}^{-1}.  \label{eq:MMSE_Hessian_connection}
 \end{equation} 
 Consequently,
 \begin{equation}
  \mmse(\X|\Y) =- \Tr \left( \boldsymbol{\mathsf{K}}_{\N}^{2} \E \left[ \boldsymbol{\mathsf{D}}^2_{\Y}   \iota_{P_{\X \Y}}(\X;\Y)  \right]  \right). 
 \end{equation} 

\end{itemize}  
\end{prop} 
\begin{IEEEproof}
First, observe that the MMSE matrix can be represented as
\begin{equation}
\boldsymbol{\mathsf{MMSE}}(\X|\Y)  =\E \left[   \boldsymbol{\mathsf{Var}}(\X | \Y)  \right]. 
\end{equation}
The proof of \eqref{eq:Gradient_Rep_MMSE_discrete} now follows by using  the expression for the conditional variance in Proposition~\ref{prop:Gradient_log_pmf_pdf}, and the proof of  \eqref{eq:MMSE_Hessian_connection} follows from the expression in Proposition~\ref{prop:Hessian_of_log_dist}.
This concludes the proof. 
\end{IEEEproof}

Since the MMSE seldom has a closed-form expression, finding lower bounds on the MMSE is of great importance.  We next show how the identities in Proposition~\ref{prop:MMSE_rep} can  be used to find a lower bound on the MMSE.  The key step of the bound is the Gaussian Poincar\'e inequality which is given by the following statement \cite{ledoux1999concentration}: for $\mathbf{Z} \sim \mathcal{N}({\bf 0}, \sigma^2 \boldsymbol{\mathsf{I}})$ and a smooth function $f: \mathbb{R}^n \to \mathbb{R}$
\begin{equation}
 \mathsf{Var}\left(  f(\mathbf{Z} ) \right)  \le \sigma^2  \E \left[  \| \nabla_{\mathbf{Z} } f(\mathbf{Z} )  \|^2 \right]. \label{eq:Gaussian_Poincare_inequality}
\end{equation}

\begin{theorem} Let $\boldsymbol{\mathsf{K}}_{\N}=\sigma^2 \boldsymbol{\mathsf{I}}$.  Then, 
\begin{equation}
\mmse(\X| \Y)\ge \sigma^2 \E \left[    \mathsf{Var}\left( \iota_{P_{\X \Y}}(\X;\Y) | \X\right)    \right] .  \label{eq:MMSE_LowerBOund} 
\end{equation} 
\end{theorem} 
\begin{IEEEproof}
\begin{align}
\mmse(\X| \Y) &=   \sigma^4 \E \left[   \|     \nabla_{\bf Y}\iota_{P_{\X \Y}}({\bf X};{\bf Y})  \|^2   \right] \label{eq:Applying_New_rep}\\
& =  \sigma^4 \E \left[  \E[  \|   \nabla_{\bf Y} \iota_{P_{\X \Y}}({\bf X};{\bf Y})  \|^2  |\X]  \right]\\
&\ge \sigma^2 \E \left[    \mathsf{Var}\left(  \iota_{P_{\X \Y}}({\bf X};{\bf Y})  | \X   \right)   \right],   \label{eq:apply_PI}
\end{align} 
where \eqref{eq:Applying_New_rep} follows from \eqref{eq:Gradient_Rep_MMSE_discrete};  and \eqref{eq:apply_PI} follows from the fact that  $\Y$ conditioned on $\X$ is a normal distribution and the Poincar\'e  inequality in \eqref{eq:Gaussian_Poincare_inequality}.
This concludes the proof. 
\end{IEEEproof}

%

The quantity on the right side of \eqref{eq:MMSE_LowerBOund}  is not new and  has appeared in the past; see for example  \cite[Eq. 237]{polyanskiy2010channel} where it is referred to as the \emph{conditional divergence variance}.

While evaluating the tightness of the lower bound in \eqref{eq:MMSE_LowerBOund} is outside of the scope of this paper and is left for  future work, the next example evaluates it for the case when $X$ is standard normal and shows that the bound can be  tight in the high and low noise regimes. 

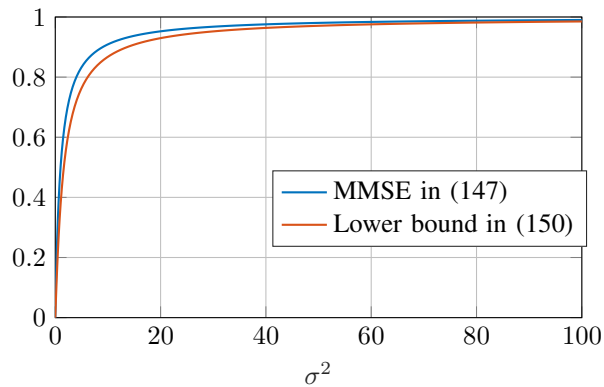
\begin{figure}[h!] 
	\centering   
%
%
\definecolor{mycolor1}{rgb}{0.00000,0.44700,0.74100}%
\definecolor{mycolor2}{rgb}{0.85000,0.32500,0.09800}%
\begin{tikzpicture}

\begin{axis}[%
width=7cm,
height=4cm,
at={(1.011in,0.642in)},
scale only axis,
xmin=0,
xmax=100,
xlabel style={font=\color{white!15!black}},
xlabel={$\sigma^2$},
ymin=0,
ymax=1,
axis background/.style={fill=white},
xmajorgrids,
ymajorgrids,
legend style={legend cell align=left, align=left, draw=white!15!black, at={(1.01,0.49)}}
]
\addplot [color=mycolor1,thick]
  table[row sep=crcr]{%
0	0\\
0.0102030405060708	0.0100999899000101\\
0.0408121620242832	0.0392118419762768\\
0.0918273645546373	0.0841042893187553\\
0.163248648097133	0.140338566792387\\
0.25507601265177	0.203235509308186\\
0.367309458218549	0.268636668905306\\
0.49994898479747	0.333310659138834\\
0.652994592388532	0.395037343373866\\
0.826446280991735	0.452488687782805\\
1.02030405060708	0.505024998737438\\
1.23456790123457	0.552486187845304\\
1.4692378328742	0.595016734845668\\
1.72431384552597	0.632935096063818\\
1.99979593918988	0.666643991700963\\
2.29568411386593	0.696572861521315\\
2.61197836955413	0.723143414027852\\
2.94867870625446	0.746750729955298\\
3.30578512396694	0.767754318618042\\
3.68329762269156	0.786475240190845\\
4.08121620242832	0.80319672295737\\
4.49954086317723	0.818166638837869\\
4.93827160493827	0.831600831600832\\
5.39740842771146	0.843686703561347\\
5.87695133149679	0.854586727199893\\
6.37690031629425	0.864441708966681\\
6.89725538210387	0.873373729021589\\
7.43801652892562	0.881488736532811\\
7.99918375675951	0.888878810897836\\
8.58075706560555	0.895624114759161\\
9.18273645546373	0.901794571196681\\
9.80512192633405	0.907451298854591\\
10.4479134782165	0.912647837363303\\
11.1111111111111	0.917431192660551\\
11.7947148250179	0.921842728526886\\
12.4987246199367	0.925918927294578\\
13.2231404958678	0.929692039511912\\
13.9679624528109	0.933190639463944\\
14.7331904907662	0.936440100907257\\
15.5188246097337	0.939463005169826\\
16.3248648097133	0.942279491875784\\
17.151311090705	0.944907560946819\\
17.9981634527089	0.947363333172217\\
18.8654218957249	0.949661275494219\\
19.7530864197531	0.951814396192742\\
20.6611570247934	0.95383441434567\\
21.5896337108458	0.955731907263291\\
22.5385164779104	0.957516439027139\\
23.5078053259871	0.959196672786541\\
24.497500255076	0.96078046906575\\
25.507601265177	0.962274971997798\\
26.5381083562902	0.963686685117877\\
27.5890215284155	0.965021538110142\\
28.6603407815529	0.966284945700221\\
29.7520661157025	0.96748185971513\\
30.8641975308642	0.968616815187912\\
31.9967350270381	0.969693971261684\\
33.1496786042241	0.970717147543628\\
34.3230282624222	0.971689856470663\\
35.5167840016325	0.972615332172863\\
36.7309458218549	0.973496556255932\\
37.9655137230895	0.974336280868602\\
39.2204877053362	0.975137049373289\\
40.495867768595	0.975901214897431\\
41.791653912866	0.976630957007733\\
43.1078461381492	0.977328296719184\\
44.4444444444444	0.97799511002445\\
45.8014488317519	0.978633140106518\\
47.1788593000714	0.979244008377788\\
48.5766758494031	0.979829224471652\\
49.994898479747	0.980390195297729\\
51.433527191103	0.980928233258935\\
52.8925619834711	0.981444563717221\\
54.3720028568513	0.981940331784905\\
55.8718498112438	0.982416608509852\\
57.3921028466483	0.982874396515121\\
58.932761963065	0.983314635147029\\
60.4938271604938	0.983738205179683\\
62.0752984389348	0.98414593311884\\
63.6771757983879	0.984538595143406\\
65.2994592388532	0.984916920718805\\
66.9421487603306	0.985281595912906\\
68.6052443628201	0.985633266442002\\
70.2887460463218	0.985972540471532\\
71.9926538108356	0.98629999119375\\
73.7169676563616	0.986616159202295\\
75.4616875828997	0.986921554681672\\
77.2268135904499	0.98721665942786\\
79.0123456790124	0.987501928714705\\
80.8182838485869	0.987777793019338\\
82.6446280991736	0.988044659618615\\
84.4913784307724	0.988302914067414\\
86.3585348433833	0.98855292156865\\
88.2460973370064	0.988795028243937\\
90.1540659116417	0.989029562313004\\
92.082440567289	0.989256835189263\\
94.0312213039486	0.989477142498236\\
96.0004081216202	0.989690765024966\\
97.9900010203041	0.989897969595991\\
100	0.99009900990099\\
};
\addlegendentry{MMSE in \eqref{eq:ExactMMSE}}

\addplot [color=mycolor2, thick]
  table[row sep=crcr]{%
0	0\\
0.0102030405060708	0.0051009998479952\\
0.0408121620242832	0.0203747052637247\\
0.0918273645546373	0.0455889104002841\\
0.163248648097133	0.0800167400608639\\
0.25507601265177	0.122270090775972\\
0.367309458218549	0.170401164392922\\
0.49994898479747	0.222203327317199\\
0.652994592388532	0.275545923016874\\
0.826446280991735	0.328617350177105\\
1.02030405060708	0.380037624043593\\
1.23456790123457	0.42886358780257\\
1.4692378328742	0.474530824796033\\
1.72431384552597	0.516770965946566\\
1.99979593918988	0.555529101685978\\
2.29568411386593	0.590893306464654\\
2.61197836955413	0.623039905639855\\
2.94867870625446	0.652193691322035\\
3.30578512396694	0.678600506187348\\
3.68329762269156	0.702509271812047\\
4.08121620242832	0.724160849363415\\
4.49954086317723	0.743781643872563\\
4.93827160493827	0.761580387360013\\
5.39740842771146	0.777746978663779\\
5.87695133149679	0.792452600753059\\
6.37690031629425	0.805850588583958\\
6.89725538210387	0.818077699783332\\
7.43801652892562	0.829255564583511\\
7.99918375675951	0.839492175680493\\
8.58075706560555	0.848883334848646\\
9.18273645546373	0.857514009918244\\
9.80512192633405	0.865459579323737\\
10.4479134782165	0.872786956203609\\
11.1111111111111	0.879555592963555\\
11.7947148250179	0.88581837233239\\
12.4987246199367	0.89162239360846\\
13.2231404958678	0.897009663921865\\
13.9679624528109	0.902017704523534\\
14.7331904907662	0.906680081747226\\
15.5188246097337	0.911026871626273\\
16.3248648097133	0.915085066342735\\
17.151311090705	0.918878929840642\\
17.9981634527089	0.922430309105695\\
18.8654218957249	0.925758906833763\\
19.7530864197531	0.928882520496249\\
20.6611570247934	0.931817252167908\\
21.5896337108458	0.93457769291221\\
22.5385164779104	0.937177085017176\\
23.5078053259871	0.939627464935656\\
24.497500255076	0.941939789401976\\
25.507601265177	0.944124046865581\\
26.5381083562902	0.946189356095679\\
27.5890215284155	0.948144053563304\\
28.6603407815529	0.94999577099355\\
29.7520661157025	0.951751504296489\\
30.8641975308642	0.953417674926342\\
31.9967350270381	0.95500018458147\\
33.1496786042241	0.956504464039434\\
34.3230282624222	0.95793551681932\\
35.5167840016325	0.959297958275297\\
36.7309458218549	0.960596050649046\\
37.9655137230895	0.961833734542731\\
39.2204877053362	0.963014657216867\\
40.495867768595	0.964142198067856\\
41.791653912866	0.965219491596786\\
43.1078461381492	0.966249448143603\\
44.4444444444444	0.967234772628093\\
45.8014488317519	0.968177981510631\\
47.1788593000714	0.969081418160793\\
48.5766758494031	0.969947266800185\\
49.994898479747	0.970777565166823\\
51.433527191103	0.971574216031715\\
52.8925619834711	0.972338997683654\\
54.3720028568513	0.973073573485378\\
55.8718498112438	0.973779500592926\\
57.3921028466483	0.974458237920041\\
58.932761963065	0.975111153420682\\
60.4938271604938	0.975739530754913\\
62.0752984389348	0.976344575396596\\
63.6771757983879	0.976927420235178\\
65.2994592388532	0.977489130718508\\
66.9421487603306	0.978030709578794\\
68.6052443628201	0.978553101179566\\
70.2887460463218	0.97905719551771\\
71.9926538108356	0.979543831911271\\
73.7169676563616	0.980013802400692\\
75.4616875828997	0.980467854888481\\
77.2268135904499	0.980906696039882\\
79.0123456790124	0.981330993964983\\
80.8182838485869	0.981741380700746\\
82.6446280991736	0.98213845450974\\
84.4913784307724	0.982522782010778\\
86.3585348433833	0.982894900155282\\
88.2460973370064	0.983255318061933\\
90.1540659116417	0.983604518721028\\
92.082440567289	0.98394296057897\\
94.0312213039486	0.984271079012356\\
96.0004081216202	0.984589287700334\\
97.9900010203041	0.984897979903128\\
100	0.985197529653955\\
};
\addlegendentry{Lower bound in \eqref{eq:MMSE_LowerBOund_evalutation_normal}}

\end{axis}

\end{tikzpicture}%
	\caption{Comparison of the exact MMSE in \eqref{eq:ExactMMSE} and the lower bound in \eqref{eq:MMSE_LowerBOund}  }
		\label{fig:mmse_lower_gaussian}
\end{figure}

\begin{example*}  Let $X$ be a standard normal. Then, on the one hand, it is well-known that 
\begin{equation}
\mmse(X|Y) =\frac{\sigma^2}{1+\sigma^2}.  \label{eq:ExactMMSE}
\end{equation} 
On the other hand, the terms on the right side of \eqref{eq:MMSE_LowerBOund} can be computed as follows:  the information density is given by 
\begin{equation}
 \iota_{P_{X Y}}(x;y)=   c-  \frac{ (y-x)^2}{2 \sigma^2} +  \frac{y^2}{2( 1+\sigma^2)}, \label{eq:Info_density_Gaussian}
\end{equation}
where $c$ is a constant, and the conditional variance is given by  
\begin{equation}
\mathsf{Var}\left(   \iota_{P_{X Y}}(X; Y )  | X =x\right)=\frac{1}{2} \frac{1}{(1+\sigma^2)^2}+ \frac{\sigma^2 x^2 }{(1+\sigma^2)^2}. \label{eq:Conditional_Variance_of_info_gaussian}
\end{equation}
The above computations are provided in Appendix~\ref{app:sec:Computation_of_conditional_var_of_info-dens}. Consequently,  the lower bound in \eqref{eq:MMSE_LowerBOund} evaluates to 
\begin{equation}
  \sigma^2 \E \left[    \mathsf{Var}\left( \iota_{P_{X Y}}(X;Y) | X \right)    \right] = \frac{\sigma^2}{1+\sigma^2}- \frac{1}{2}\frac{ \sigma^2}{(1+\sigma^2)^2} . \label{eq:MMSE_LowerBOund_evalutation_normal}
\end{equation} 
The exact MMSE in \eqref{eq:ExactMMSE} and the lower bound in \eqref{eq:MMSE_LowerBOund_evalutation_normal} are compared in Fig.~\ref{fig:mmse_lower_gaussian}. 
\end{example*}

  There exists a large collection of lower  bounds on estimation error  for continuous distributions \cite{ziv1969some,weinstein1988general,dytso2019mmse} of which the Cram\'er-Rao bound  is arguably the most popular. 
However, lower bounds on the MMSE of other distributions are rare.  The interesting feature of the bound in \eqref{eq:MMSE_LowerBOund} is that it holds for all distributions and does not require any regularity conditions.

\section{Conditional Expectation and Its Inverse: Analytic Properties} 
\label{sec:PowerSeries_and_Inverse}
In this section, we use the new derivative identities to study  the power series expansion of the conditional expectation and its inverse.

\subsection{Power Series Expansion of the Conditional Expectation} 
\label{sec:Power_Series_Expansion}
In this section, we find the power series expansion of the conditional expectation in terms of the conditional cumulants. 
Furthermore, in Section~\ref{sec:Inverse_OF_CE}, this power series representation, together with the Lagrange inversion theorem, will lead to a representation of the inverse of the conditional expectation. 
  The fact that  a power series expansion exists follows  from the next result. 
\begin{lem}\label{lem:AnalyticLemma} The functions $y \mapsto \E[X^k|Y=y], k \in \mathbb{N}$ and $ y \mapsto \kappa_{X|Y=y}(k), k \in \mathbb{N}$ are  real-analytic.  
\end{lem} 
\begin{IEEEproof}
Note that by the TRE identity  in \eqref{eq:TweediesFormulaGaussian}
\begin{equation}
\E[X|Y=y]= y + \sigma^2 \frac{ \frac{ {\rm d}}{ {\rm d} y}  f_Y(y)}{f_Y(y)}. 
\end{equation} 
Hence,  since the ratios and sums of analytic functions are analytic,  $\E[X|Y=y]$ is real-analytic provided that  $ f_Y(y)$ is real-analytic.  The analyticity of $f_Y$ is a known consequence of convolution with  Gaussian measures (see e.g.,  \cite{folland2013real}).   Since $\E[X|Y=y]$ is real-analytic, the identities in \eqref{eq:Higher_Moments_Bell_Polynomial} and \eqref{eq:Cumulants_and_Conditional} imply that $\E[X^k|Y=y]$ and  $\kappa_{X|Y=y}(k)$ are also real-analytic. 
\end{IEEEproof} 

Before studying the Taylor series of the conditional expectation, it is instructive to consider the following example.

\begin{example*} For  $X$ uniformly distributed on $\{-1,1\}$, the conditional expectation is given by 
\begin{equation*}
\E[X|Y=y]=\tanh \left( \frac{y}{\sigma^2} \right), y\in \mathbb{R} . \label{eq:Binary_CE}
\end{equation*}
By using the Taylor series of $\tanh(\cdot)$ around zero \cite[Eq.~4.5.64]{abramowitz1965handbook}, we have that
\begin{equation*}
\E[X|Y=y]= \sum_{k=1}^\infty   \frac{2^{2k} (2^{2k}-1) b_{2k}}{(2k)!}  \left( \frac{y}{\sigma^2} \right)^{2k-1}, \,  |y| < \frac{ \sigma^2  \pi}{2}, 
\end{equation*} 
where  $b_{n}$ is  the $n$-th Bernoulli number.  
\end{example*}

The key observation here is that even in this simple case, the Taylor expansion  has a finite radius of convergence. 
Therefore, in general, we cannot expect to get a power series representation of  $\E[X|Y=y]$ that converges for all $\mathbb{R}$ (i.e., the power series with an infinite radius of convergence). 

The next result provides a power series representation for the conditional expectation. In addition, it also finds a lower bound on the radius of converges in the case when $X$ is bounded.  
\begin{theorem}\label{thm:Power_Series_ConditionalExpectation} Fix some $a\in \mathbb{R}$.  Then,  for every $X$ there exists some $r_{\sigma, a}>0$ such that 
\begin{equation}
\E[X|Y=y] \hspace{-0.1cm} = \hspace{-0.1cm} \sum_{k=0}^\infty  \frac{\kappa_{X|Y=a}(k+1)}{k! \sigma^{2k}} (y-a)^k , \,   | y-a| <  r_{\sigma, a}    .  \label{eq:Power_Series_of_CE}
\end{equation} 
In addition, if $|X| \le A$, then $r_{\sigma, a} \ge  \frac{\sigma^2}{2A \eu}$. 

\end{theorem} 
\begin{IEEEproof}
From Lemma~\ref{lem:AnalyticLemma}, we have that  for every $a \in \mathbb{R}$ there exits an $r_{\sigma, a}>0$ such that $\E[X|Y=y]$ has a power series representation on $(a-r_{\sigma, a}, a+r_{\sigma, a})$. Moreover, by using \eqref{eq:Cumulants_and_Conditional}, this power series is given by 
\begin{align}
\E[X|Y=y]&= \sum_{k=0}^\infty  \frac{ \E^{(k)}[X|Y=a]}{k! } (y-a)^k \\
&= \sum_{k=0}^\infty  \frac{\kappa_{X|Y=a}(k+1)}{k! \sigma^{2k}} (y-a)^k.
\end{align} 
Finally, the radius of convergence for $|X| \le A$  can be found as follows by using the root test: 
\begin{align}
r_{\sigma, a}&= \limsup_{n \to \infty}   \left|   \frac{\kappa_{X|Y=a}(k+1)}{ k! \sigma^{2k}} \right|^{-  \frac{1}{k}}\\
&\ge\limsup_{n \to \infty}   \left|   \frac{ 2^{k} (k+1)^{k+1}  \E[ |X |^{k+1} |Y =a] }{k! \sigma^{2k}} \right|^{-  \frac{1}{k}}  \label{eq:First_lower_bound_on_cumulant_apply}\\
&\ge\limsup_{n \to \infty}   \left|   \frac{ 2^{k} (k+1)^{k+1}  A^{k+1} }{k! \sigma^{2k}} \right|^{-  \frac{1}{k}}\label{eq:Using_|X| <A} \\
&=       \frac{\sigma^2}{2 A  \eu}  , \label{eq:factorial_limit}
\end{align} 
where \eqref{eq:First_lower_bound_on_cumulant_apply} follows from the bound in \eqref{eq:FirstBound_on_cumulant};   \eqref{eq:Using_|X| <A} follows by using $  \E[ |X |^{k+1} |Y ]\le A^{k+1}$; and \eqref{eq:factorial_limit} follows from the limit  $ \limsup_{n \to \infty}  \left|   \frac{ (k+1)^{k+1}  }{k!} \right|^{-  \frac{1}{k}} =\frac{1}{\eu}$. 
\end{IEEEproof}

\begin{remark}  It is important to note that since   the conditional cumulants can be expressed in terms of $f_Y$ and derivatives of $f_Y$ (see \eqref{eq:TRE_for_Cumulants}), the power series can also be expressed in terms of $f_Y$ only. 
\end{remark}

The approximation of the conditional expectation with polynomials of degree $k$ has been recently considered in \cite{alghamdi2021polynomial}. Using results from Bernstein's approximation theory, the authors of \cite{alghamdi2021polynomial} were able to quantify the average approximation error as a function of the degree $k$.

\subsection{On the Inverse of the Conditional Expectation}
\label{sec:Inverse_OF_CE}  
In this section, we study the inverse of the conditional expectation.  For  ease of notation, we let $\hat{X}(y)=\E[X|Y=y]$.
The results of this section will be useful in Section~\ref{sec:Distribution_of_CE_and_Error}  where the distribution of $\E[X|Y]$ and $X-\E[X|Y]$ are considered. 

\begin{lem} Suppose that $X$ is non-constant random variable.  Then,  $\hat{X}(y)$ has an inverse $\hat{X}^{-1}$. Moreover, the inverse $\hat{X}^{-1}$ is a real-analytic  function.  
\end{lem} 
\begin{IEEEproof}
By using the Hatsell and Nolte identity in \eqref{eq:Hatsell-Nolte-general} we have that 
\begin{equation}
  \sigma^2 \frac{{\rm d}}{{\rm d} y} \E[X|Y=y]=  \mathsf{Var}(X|Y=y),  \,y \in \mathbb{R}  \label{eq:HN_identity}
\end{equation} 
which is non-vanishing provided that $X$ is not  a constant.  
From \eqref{eq:HN_identity} we have that  $\hat{X}(y)$ is a strictly increasing function for non-constant random variables. Therefore, $\hat{X}(y)$  has a proper inverse.  
The proof is concluded by using  \cite[Thm.~1.5.3]{krantz2002primer}, which states that   the inverse of an analytic function with a non-vanishing derivative is also analytic.  
\end{IEEEproof}

We next give two examples for which the inverse has a closed-form expression. 

\begin{example*} Suppose that $X$ is a standard Gaussian random variable.  Then, the conditional expectation is given by 
\begin{equation}
\hat{X}(y)=   \frac{1}{1+\sigma^2} y, \, y \in \mathbb{R}, 
\end{equation}
and the inverse is given by 
\begin{equation}
 \hat{X}^{-1}(x)= (1+\sigma^2) x,  \,    x\in \mathbb{R}. 
 \end{equation} 
\end{example*}

\begin{example*} Suppose that $X$  is distributed according to $P_X(1)=p=1-P_X(-1)$.   Then, the conditional expectation is given by
\begin{equation}
\E[X|Y=y]=  \tanh \left(  \frac{y}{\sigma^2} + \frac{1}{2} \log \left( \frac{p}{1-p} \right)   \right), y\in \mathbb{R},    \label{eq:tahn_p_classifier}
\end{equation} 
and the inverse is given by 
\begin{equation}
\hat{X}^{-1}(x)=  \frac{\sigma^2}{2} \log \left( \frac{1+x}{1-x} \frac{1-p}{p} \right)  , x  \in  (-1,1).  \label{eq:tahn_p_classifier_inverse} 
\end{equation} 
\end{example*}

To find a general expression for the inverse of the conditional expectation, we use  the power series  expansion  of the conditional expectation in Theorem~\ref{thm:Power_Series_ConditionalExpectation} and the Lagrange inversion theorem \cite{charalambides2018enumerative}; the latter is presented next. 
\begin{theorem}\label{thm:Inverse_CE} \emph{(Lagrange Inversion Theorem)}
The Taylor coefficients of a formal power series  $f^{-1}(t)= \sum_{n=1}^\infty  b_n \frac{t^n}{n!}$, which is the inverse of $f(t) =   \sum_{n=1}^\infty  a_n \frac{t^n}{n!}$, can be expressed as a function of the Taylor coefficients of  $f$ in the following manner: 
\begin{align}
 b_1 & = \hspace{-0.05cm}  \frac{1}{a_1},\\
 b_n& =  \hspace{-0.05cm}  b_1^n  \hspace{-0.05cm} \sum_{k=1}^{n-1} (-1)^k  n^{(k)}  \mathsf{B}_{n-1,k} \left(c_1,  c_{2},\ldots, c_{n-k} \right) ,  n \ge 2, \label{eq:Coefficients_Lagrange_Inversion}
\end{align} 
where $n^{(k)}$ is the rising factorial\footnote{The rising factorial is define as $n^{(k)}=n(n+1)\ldots (n+k-1)$.} and $c_n=\frac{a_{n+1}}{(n+1) a_1} $. 
\end{theorem} 

The  main result of this section is the following theorem that characterizes the power series of  the inverse of the conditional expectation. 
\begin{theorem}\label{thm:Expression_for_inverse_CE}   Fix an $a \in \mathbb{R}$. Then,  for every non-constant $X$ there exists a $\tau_{\sigma,a}>0$ such that 
\begin{equation}
\hat{X}^{-1}(x) =  a+   \hspace{-0.1cm}\sum_{k=1}^\infty  b_k  \frac{ \left(x- \hat{X}(a) \right)^k}{k!}, \, | x- \hat{X}(a)| < \tau_{\sigma,a} \label{eq:Inverse_of_CE}
\end{equation} 
where
\begin{align}
b_1 & = \hspace{-0.05cm} \frac{\sigma^2}{\kappa_{X|Y=a}(2)},\\
 b_n& = \hspace{-0.05cm} b_1^n \sum_{k=1}^{n-1} (-1)^k  n^{(k)} \mathsf{B}_{n-1,k} \left(c_1,  c_{2},\ldots, c_{n-k} \right) ,  n \ge 2, \\
 c_k& = \hspace{-0.05cm} \frac{\kappa_{X|Y=a}(k+2)}{ (k+1) \sigma^{2(k+1)} \kappa_{X|Y=a}(2)}, k \ge 1.  \label{eq:Coeffiencet_inverse_ck}
\end{align} 
\end{theorem} 
\begin{IEEEproof}
First, since $\hat{X}(y)$ is real-analytic, it has a power-series expansion around $\hat{X}(a)$ with some positive radius of convergence  $\tau_{\sigma,a}$. 
Second, by using \eqref{eq:Power_Series_of_CE}, we have that 
\begin{align}
f(y) &= \hat{X}(y+a) - \hat{X}(a) \\
&= \sum_{k=1}^\infty  \frac{\kappa_{X|Y=a}(k+1)}{k! \sigma^{2k}} y^k \\
&= \sum_{k=1}^\infty  a_k  \frac{y^k}{k!},
\end{align} 
where $a_k= \frac{\kappa_{X|Y=a}(k+1)}{ \sigma^{2k}}$.   
Therefore,  by the Lagrange inversion theorem, we have that 
\begin{equation}
f^{-1}(x)=   \sum_{k=1}^\infty b_k \frac{x^k}{k!},
\end{equation} 
where the expression for  the $b_k$'s is given in \eqref{eq:Coefficients_Lagrange_Inversion}.   Next, by noting that $f^{-1}(x)= \hat{X}^{-1}\left(x+ \hat{X}(a)\right) -a$,
we arrive at
\begin{equation}
\hat{X}^{-1}(x)= a+  \sum_{k=1}^\infty  b_k  \frac{ \left(x- \hat{X}(a) \right)^k}{k!}.
\end{equation} 
This concludes the proof. 
\end{IEEEproof}

\begin{remark} We remark that the inverse of the conditional expectation in \eqref{eq:Inverse_of_CE}  depends on the  joint distribution $P_{XY}$ only through the marginal $f_Y$. Indeed, by using \eqref{eq:TRE_for_Cumulants}  we have that  the coefficients in \eqref{eq:Coeffiencet_inverse_ck} can  be expressed   in terms of only $f_Y$ or derivatives of $f_Y$
\begin{equation}
c_k= \frac{ \frac{{\rm d}^{k+2}}{ {\rm d} y^{k+2} }  \log f_Y(y)   }{ (k+1)   (    1+ \sigma^2 \frac{{\rm d}^2}{ {\rm d} y^2 }  \log f_Y(y)   ) }  |_{y=a}, k \ge 1. 
\end{equation} 
\end{remark} 

 While the  formula for the coefficient in Theorem~\ref{thm:Expression_for_inverse_CE} is algebraically involved, in principle, it is not difficult to implement numerically.  A few non-trivial numerical examples will be given in the next section.

\section{On the Distributions of $\E[X|Y]$ and $X-\E[X|Y]$ }
\label{sec:Distribution_of_CE_and_Error}

In this section, we study the distribution of   $\E[X|Y]$ and the distribution of the estimation error $X-\E[X|Y]$. 
The distribution of a given estimator typically contains more information than measures like variance and is more useful.  
The question of finding the distribution of $\E[X|Y]$ or $X-\E[X|Y]$ is akin to \emph{the information spectrum method}   \cite{koga2013information}, where the objective is to find the distribution of  the information density $\iota_{P_{XY}}(X;Y)$.   Indeed, from \eqref{eq:Gradient_Info_Density} 
 the derivative of the information density can be expressed as
\begin{equation}
\frac{{\rm d}}{ {\rm d} y }   \iota_{P_{XY}}(x;y)= \frac{x-\E[X|Y=y]}{\sigma^2}. 
\end{equation}  

\subsection{On the Distribution of $\E[X|Y]$}

As before, for ease of notation, we let $\hat{X}=\hat{X}(Y)=\E[X|Y]$.
The distribution of $\hat{X}(Y)$ is closely related to $P_X$ and $P_{X|Y}$. However, while the distributions  $P_X$ or $P_{X|Y}$ can be arbitrary (discrete, continuous, or singular),  the  random variable  $\hat{X}$ is always continuous.  This follows from the fact that   $Y$ is a continuous random variable,  and, as was shown in Lemma~\ref{lem:AnalyticLemma}, the fact that the function $y \mapsto \hat{X}(y)$ is real-analytic.  

Our starting place for finding the distribution of $\hat{X}(Y)$  is the following well-known change of variable formulas: for the random variable $V$ with the  cdf $F_V$ and the pdf $f_V$,  let  $W=g(V)$; then
\begin{align}
F_W(w) &= F_V( g^{-1} (w) ), \text{ and} \label{eq:Change_of_variable_cdf} \\
f_W(w)&= f_V( g^{-1} (w))   \left |  \frac{{\rm d}}{ {\rm d} w} g^{-1}(w) \right| , \label{eq:Change_of_variable_pdf}
\end{align}
where $g^{-1}$ the  inverse of $g$.   

   The following theorem provides expressions for the cdf and the pdf of the conditional expectation. 
\begin{theorem}\label{thm:pdf_of_CE}   Suppose that  $X$ is non-constant. Then, the cdf of  $\hat{X}$ is given by 
\begin{equation}
 F_{\hat{X}}(x)= F_Y \left(  \hat{X}^{-1}(x)  \right), 
 \end{equation} 
 and pdf of $\hat{X}$ has the following three characterizations:
 \begin{align} 
f _{\hat{X}}(x)&=  f_Y\left( \hat{X}^{-1}(x) \right)  \left |  \frac{{\rm d}}{ {\rm d} x} \hat{X}^{-1}(x)\right| \\
&= \frac{ \sigma^2 f_Y\left( \hat{X}^{-1}(x) \right) }{  \mathsf{Var}\left(X|Y= \hat{X}^{-1}(x) \right)  }\\
  &=   \frac{f^3_Y \left(  \hat{X}^{-1}(x) \right)}{  f^2_Y \left(  \hat{X}^{-1}(x) \right)+ \sigma^2f_Y \left(  \hat{X}^{-1}(x) \right) f^{''}_Y \left(  \hat{X}^{-1}(x) \right)- \sigma^2 \left( f^{'}_Y \left(  \hat{X}^{-1}(x) \right) \right)^2}, \label{eq:pdf_general_formula_Hat_X}
 \end{align}
Furthermore,  $\hat{X}^{-1}(x)$ can be expressed  in terms of only  $f_Y$ and derivatives of $f_Y$. 
\end{theorem} 
\begin{IEEEproof}
Combining Theorem~\ref{thm:Expression_for_inverse_CE} with  \eqref{eq:Change_of_variable_cdf} we  arrive at the following expression for the cdf and the pdf of $\hat{X}$:
\begin{align}
 F_{\hat{X}}(x)&= F_Y(  \hat{X}^{-1}(x)  ), \\ 
 f_{\hat{X}}(x)&=  f_Y\left( \hat{X}^{-1}(x) \right)  \left |  \frac{{\rm d}}{ {\rm d} x} \hat{X}^{-1}(x)\right|,  
\end{align}
To simplify the expression of the pdf further, recall that the derivative of the inverse of $g$ is given by $\frac{{\rm d}}{ {\rm d} t} g^{-1}(t)=\frac{1}{g'( g^{-1}(t) ) }$. Therefore,
\begin{align}
 \left |  \frac{{\rm d}}{ {\rm d} x} \hat{X}^{-1}(x)\right| &= \frac{1}{ \frac{1}{\sigma^2} \mathsf{Var}\left(X|Y= \hat{X}^{-1}(x) \right)  } \label{eq:Using_HN_identity_inverse}\\
 &=   \frac{1}{  1+ \sigma^2 \frac{{\rm d}^2}{ {\rm d} y^2 }  \log f_Y(y) |_{y= \hat{X}^{-1}(x)}}  \label{eq:Using_cumulant_identity_inverse} \\
  &=   \frac{f^2_Y \left(  \hat{X}^{-1}(x) \right)}{  f^2_Y \left(  \hat{X}^{-1}(x) \right)+ \sigma^2f_Y \left(  \hat{X}^{-1}(x) \right) f^{''}_Y \left(  \hat{X}^{-1}(x) \right)- \sigma^2 \left( f^{'}_Y \left(  \hat{X}^{-1}(x) \right) \right)^2} \label{eq:SEcond_Derivative_log}
 \end{align} 
where  \eqref{eq:Using_HN_identity_inverse} follows by using the Hatsell and Nolte identity in  \eqref{eq:Hatsell-Nolte-general}  and the fact that the variance is non-negative;  \eqref{eq:Using_cumulant_identity_inverse} follows by using the identity in \eqref{eq:TRE_for_Cumulants}; and \eqref{eq:SEcond_Derivative_log}    follows by using  that $\frac{ {\rm d}^2\log (f_Y(y))}{ {\rm d} y^2}=\frac{f_Y(y) f''_Y(y)-f'_Y(y)^2}{f_Y^2(y)}$. This concludes the proof. 
\end{IEEEproof}

We now consider a few examples.

\begin{example*} Suppose that $X$ is a standard Gaussian random variable.  Then, the conditional expectation is given by 
\begin{equation}
\hat{X}(y)=   \frac{1}{1+\sigma^2} y, \, y \in \mathbb{R}, 
\end{equation}
and the inverse is given by 
\begin{equation}
 \hat{X}^{-1}(x)= (1+\sigma^2) x,  \,    x\in \mathbb{R}. 
 \end{equation} 
 Then, by using \eqref{eq:pdf_general_formula_Hat_X},  for $x \in \mathbb{R}$
 \begin{equation}
 f_{\hat{X}}(x)= f_Y\left( (1+\sigma^2) x \right)   \left (   1+\sigma^2 \right)=\frac{1 }{\sqrt{ 2\pi   \frac{1}{1+\sigma^2}} }  \eu^{-\frac{x^2}{2  \frac{1}{1+\sigma^2}}}.
  \end{equation} 
  In other words, $\hat{X}$ is Gaussian with  variance $\frac{1}{1+\sigma^2}$. 
\end{example*}

\begin{example*} Suppose that $X$ is uniformly distributed on $\{-1,1\}$.   Then,  $\hat{X}(y)=  \tanh \left( \frac{y}{\sigma^2} \right), \, y\in \mathbb{R}$,  
and the inverse and its derivative are given by 
\begin{align}
\hat{X}^{-1}(x)&=  \frac{\sigma^2}{2} \log \left( \frac{1+x}{1-x} \right), x  \in  (-1,1), \\
\frac{{\rm d}}{ {\rm d} x} \hat{X}^{-1}(x)&=  \frac{\sigma^2}{1-x^2}, x  \in  (-1,1). 
\end{align} 
Then, by using \eqref{eq:pdf_general_formula_Hat_X}, for $x \in (-1,1)$
\begin{align}
F_{\hat{X}}(x)&=  F_Y(  \hat{X}^{-1}(x) ) ,\\
 f_{\hat{X}}(x)&= f_Y \left( \hat{X}^{-1}(x) \right)  \frac{\sigma^2}{1-x^2}, 
\end{align}
where $F_{Y}(y)= \frac{1}{2} \Phi \left( \frac{y-1}{\sigma}\right)+\frac{1}{2} \Phi \left( \frac{y+1}{\sigma}\right)$  and $f_Y(y) =\frac{1}{2} \phi_\sigma(y-1)+\frac{1}{2} \phi_\sigma(y+1)$. 
The plot of the cdf of $\hat{X}$ is given in Fig.~\ref{fig:Plot_of_F_hat_vs_F_X_binary}  and is compared to the cdf of $X$.  Since as $\sigma \to 0$ the pdf of $\hat{X}$ starts to concentrate on $1$ and $-1$, it is more convenient to plot $\log f_{\hat{X}}$ instead of $f_{\hat{X}}$.  The plots of the  log of the pdf of  $\hat{X}$ are given in Fig.~\ref{eq:log_f_hat_X_binary}. 

\begin{figure}[h!]
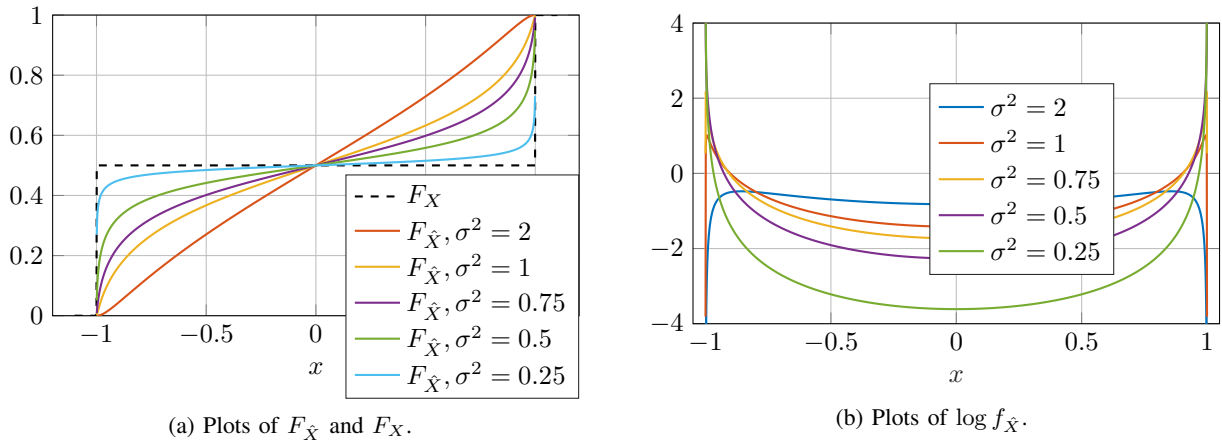
 
	\centering
	  \begin{subfigure}[c]{0.45\textwidth}
	  \center
	\input{CDF_CE_Binary.tex}
	\caption{ Plots of $F_{\hat{X}}$ and $F_{X}$. }
	\label{fig:Plot_of_F_hat_vs_F_X}
	\end{subfigure}
	~
	  \begin{subfigure}[c]{0.45\textwidth}
	  \center
	\input{Log_pdf_CE_Binary.tex}
	\caption{Plots of $\log f_{\hat{X}}$.  }  
	\label{eq:log_f_hat_X_binary}
	  \end{subfigure}
		
		\caption{ The cdf and the pdf of $\hat{X}$.  }
		\label{fig:Plot_of_F_hat_vs_F_X_binary}
		\vspace{-0.5cm} 
\end{figure}%

\end{example*}


The next example computes the cdf numerically for a case in which we do not have a closed-form expression for the conditional expectation. 

\begin{figure}[h!] 
	\centering

	  \center
	\input{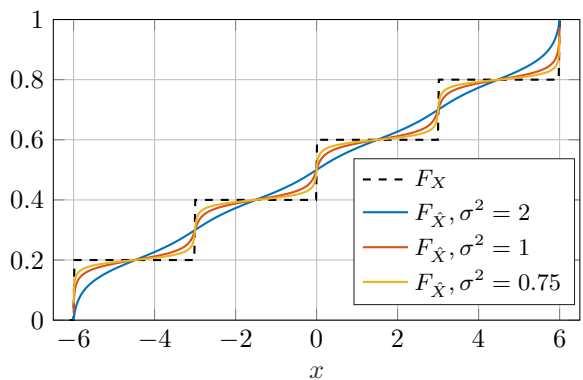}
	\caption{Plots of $F_{\hat{X}}$ and $F_{X}$ when $X$ is uniformly distributed over the set $\{ -6,   -3,     0,     3,     6 \}$. }
	\label{fig:Plot_of_F_hat_vs_F_X_5_point}
%
%
%
	\end{figure}

\begin{example*}
Let $X$  be uniformly distributed on $\{ -6,   -3,     0,     3,     6 \}$.    Fig.~\ref{fig:Plot_of_F_hat_vs_F_X_5_point} shows  the cdf of  $\hat{X}$ for several values of $\sigma^2$. 
To compute the inverse of the conditional expectation, we use the power series  representation of the inverse in Theorem~\ref{thm:Expression_for_inverse_CE}. To implement these examples, we truncate the power series  in Theorem~\ref{thm:Expression_for_inverse_CE} such that the absolute error in the approximation is always below $10^{-4}$. 
\end{example*} 


\subsection{On the Distribution of the Estimation Error $X-\E[X|Y]$ }
\label{sec:Distribution_of_X-EXY}

In this section, we provide a formula for the distribution of $X-\E[X|Y]$. In addition, we also study a distribution  for the case when we have a  \emph{mismatched estimator}.  In particular,  we consider the distribution of $X-\E_{Q}[X|Y]$, where $ \E_{Q}[X|Y]$ is the conditional expectation that assumes that $X$ is distributed according to $Q$ instead of $P_X$. This form of mismatched estimator error has been shown to play an important role in relating estimation and information measures; the interested reader is referred to \cite{MismatchedMSE,chlaily2017information,chen2013mismatched,huleihel2014analysis} and references therein.   

The next theorem provides an expression for the pdf of the estimation error $X-g(Y)$ where $g$ is  some estimator of $X$. 
\begin{theorem}\label{thm:distribution_of_errors} Let $W=X-g(Y)$ where $g$ has a well-defined  inverse and where  $\mathcal{R}_g$ denotes the range of the function $g$. Then,
\begin{equation}
f_W(w) 
= \E \left [ \phi_{\sigma }\left(  g^{-1}(X-w)-X \right)  \hspace{-0.1cm} \left|  \frac{{\rm d} g^{-1}(X-w) }{ {\rm d} w} \right| 1_{\mathcal{R}_g }  \hspace{-0.1cm} \left( X-w \right)   \right], \, w\in \mathbb{R}. \label{eq:ExpressionForTHeError_general_g} 
\end{equation}
Consequently,  for  $W=X-\E[X|Y]=X- \hat{X}(Y)$ 
\begin{equation}
f_W(w) 
= \hspace{-0.01cm}\E  \hspace{-0.05cm} \left[   \hspace{-0.05cm} \phi_{\sigma} \hspace{-0.05cm} \left( \hspace{-0.02cm} \hat{X}^{-1} \hspace{-0.05cm} \left(X-w \right)-X \hspace{-0.02cm}  \right)  \hspace{-0.1cm} \left|  \hspace{-0.01cm}  \frac{{\rm d}  \hat{X}^{-1} \hspace{-0.02cm}(X-w) }{ {\rm d} w}  \hspace{-0.01cm}\right|  1_{ \mathcal{R}_{\hat{X}}  }  \hspace{-0.1cm} \left( X-w \right)    \hspace{-0.05cm} \right]  \hspace{-0.02cm}, \, w\in \mathbb{R}.\label{eq:Distribution_Error_MMSE}
\end{equation}
\end{theorem}
\begin{IEEEproof}
Let $W=X-g(X+N)=X+U$; then  by using the formula for  the sum of correlated variables, we have that 
\begin{equation}
f_W(w)=\E \left[  f_{U|X}(w-X|X)  \right].  \label{eq:GeneralConvolutionExpression}
\end{equation}
To characterize $f_{U|X}(t|x)$, note that given $X=x$ 
\begin{equation}
U=-g(x+N) = -g(N_x) 
\end{equation}
where $N_x=N+x$. 
Therefore,
\begin{align}
f_{U|X}(t|x)&= f_{ - g(N_x) }(t)\\
&= f_{N_x}(g^{-1}(-t)) \left| \frac{ {\rm d} }{ {\rm d} t } g^{-1}(-t) \right|     1_{ \mathcal{R}_{g}  }(t)  \label{eq:Change_Variable_ErrorFormula}\\
&=  \phi_{\sigma} \left(   g^{-1}(-t)-x \right)  \left| \frac{ {\rm d}}{{\rm d} t } g^{-1}(-t) \right|  1_{ \mathcal{R}_{g}  }(t) , \label{eq:PDF_U_given_X_err}
\end{align} 
where in \eqref{eq:Change_Variable_ErrorFormula} we have used a change of variable formula.  Inserting \eqref{eq:PDF_U_given_X_err} into \eqref{eq:GeneralConvolutionExpression} concludes the proof of \eqref{eq:ExpressionForTHeError_general_g}. 
\end{IEEEproof} 

\begin{remark}
Both of the expressions in Theorem~\ref{thm:distribution_of_errors} can be further simplified or rewritten.  
In particular, the expression in \eqref{eq:ExpressionForTHeError_general_g} can be further rewritten by using  $\frac{{\rm d}}{ {\rm d} t} g^{-1}(t)=\frac{1}{g'( g^{-1}(t) ) }$. Furthermore, the expression in \eqref{eq:Distribution_Error_MMSE} can be rewritten by using the expression for $\left|  \hspace{-0.01cm}  \frac{{\rm d}  \hat{X}^{-1} \hspace{-0.02cm}(w) }{ {\rm d} w}  \hspace{-0.01cm}\right|  $ in \eqref{eq:Using_HN_identity_inverse} or \eqref{eq:SEcond_Derivative_log}. For example, by using \eqref{eq:Using_HN_identity_inverse}, we have that 
\begin{equation}
f_W(w) 
=  \sigma^2 \hspace{-0.01cm}\E  \hspace{-0.05cm} \left[    \hspace{-0.01cm}  \frac{ \hspace{-0.05cm} \phi_{\sigma} \hspace{-0.05cm} \left( \hspace{-0.02cm} \hat{X}^{-1} \hspace{-0.05cm} \left(X-w \right)-X \hspace{-0.02cm}  \right) }{ \mathsf{Var}\left( X|Y= \hat{X}^{-1} \hspace{-0.02cm}(X-w) \right)}  \,1_{ \mathcal{D}_{\hat{X}}   } \left(X-w \right)  \hspace{-0.05cm} \right]  \hspace{-0.02cm}, \, w\in \mathbb{R}.
\end{equation}

\end{remark}

\begin{example*}Suppose that $X$  is distributed according to $P_X(1)=1-P_X(-1)=p \in (0,1)$. The conditional expectation of the random variable is given in \eqref{eq:tahn_p_classifier}, and the inverse is given in \eqref{eq:tahn_p_classifier_inverse}.  The derivative of the inverse is given by $\frac{{\rm d}}{ {\rm d} x} \hat{X}^{-1}(x)=  \frac{\sigma^2}{1-x^2}, x  \in  \mathcal{D}_{\hat{X}} $ where $\mathcal{D}_{\hat{X}} =(-1,1)$. Therefore, by using \eqref{eq:Distribution_Error_MMSE},  the distribution of the error is given by 
\begin{align}
 f_W(w) 
&=   \phi_{\sigma} \left(   \frac{\sigma^2}{2} \log \left( \frac{2-w}{w} \frac{1-p}{p} \right) -1 \right)   \frac{\sigma^2 p}{1-(1-w)^2}   1_{ (0,2) }(w)  \notag \\
&\quad +\phi_{\sigma} \left(   \frac{\sigma^2}{2} \log \left( \frac{-w}{2+w} \frac{1-p}{p} \right) +1 \right)   \frac{\sigma^2 (1-p)}{1-(1+w)^2}   1_{  (-2,0) }(w) , \, w\in \mathbb{R},  \label{eq:pdf_erro_binary}
\end{align}
which in the case of $p=\frac{1}{2}$ reduces to 
\begin{equation}
 f_W(w) =  \frac{1}{2}  \phi_{\sigma} \left(  \frac{\sigma^2}{2} \log \left( \frac{2-|w|}{|w|}  \right)-1 \right)    \frac{\sigma^2}{1-(1-|w|)^2}    1_{  (-2,2) }(w), \, w\in \mathbb{R}.
\end{equation}
Plots of the pdf of $W$ are shown in Fig.~\ref{fig:Plot_of_error_pdf}. 
Fig.~\ref{fig:Plot_of_pdf_error_vs_p} shows plots of the pdf in \eqref{eq:pdf_erro_binary}  for several examples of $p$ and for a fixed $\sigma$, and Fig.~\ref{fig:Plot_of_pdf_error_vs_s} shows the pdf   in \eqref{eq:pdf_erro_binary}   for several examples of $\sigma$ and for a fixed $p$. 

\begin{figure}[h!] 
	\centering
	  \begin{subfigure}[c]{0.45\textwidth}
	  \center
	\input{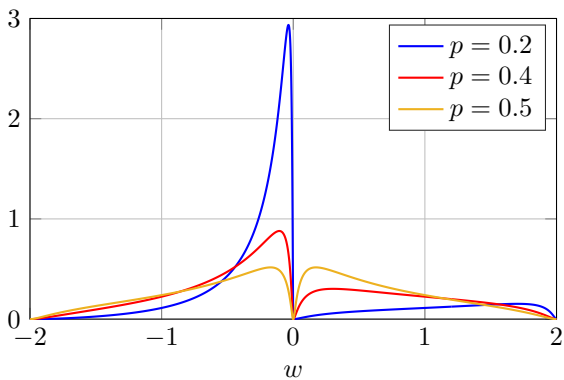}
	\caption{ Plots of $f_W$ for $\sigma=1.5$ and $p=0.2,0.4$ and $0.5$. }
	\label{fig:Plot_of_pdf_error_vs_p}
	\end{subfigure}
	~
	  \begin{subfigure}[c]{0.45\textwidth}
	  \center
	\input{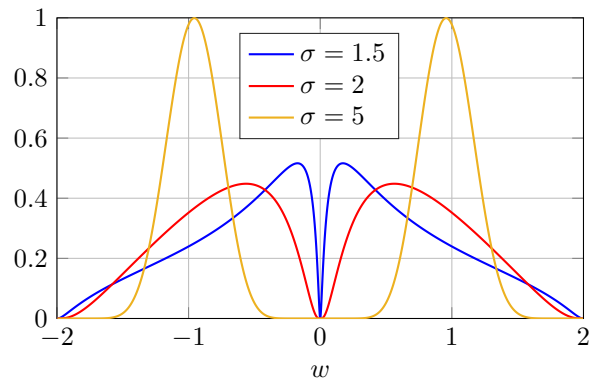}
	\caption{Plots of $f_W$ for $p=\frac{1}{2}$ and $\sigma=1.5,2$ and $5$.  }  
	\label{fig:Plot_of_pdf_error_vs_s}
	  \end{subfigure}
		
		\caption{ Examples of the pdf of  $W$.  }
		\label{fig:Plot_of_error_pdf}
\end{figure}%
\end{example*}

We next consider a mismatched example. 

\begin{example*}
Suppose that $X$  is distributed according to $P_X(1)=1-P_X(-1)=p\in (0,1)$.   We consider two examples of mismatched distributions.  In the  first example, we consider the case in which it is assumed that $X$ is distributed according to $Q_1(1)=1-Q_1(-1)=q$ and the estimator $\E_{Q_1}[X|Y]$ is used. In this case, the distribution of the error is given
 \begin{align}
 f_{W_1}(w) 
&=   \phi_{\sigma} \left(   \frac{\sigma^2}{2} \log \left( \frac{2-w}{w} \frac{1-q}{q} \right) -1 \right)   \frac{\sigma^2 p}{1-(1-w)^2}    1_{  (0,2) }(w)  \notag \\
&\quad+\phi_{\sigma} \left(   \frac{\sigma^2}{2} \log \left( \frac{-w}{2+w} \frac{1-q}{q} \right) +1 \right)   \frac{\sigma^2 (1-p)}{1-(1+w)^2}     1_{  (0,2) }(w) , \, w\in \mathbb{R}.  
\end{align}
In the second example, we assume that $Q_2=\mathcal{N}(0,1)$ and use the estimator $\E_{Q_2}[X|Y]= \frac{1}{1+\sigma^2} Y$ (i.e., a linear estimator).   In this case, the distribution of the error is given by 
 \begin{align}
 f_{W_2}(w) 
&=   \phi_{\sigma} \left(   (1+\sigma^2)(1-w) -1 \right)  (1+\sigma^2) p  +\phi_{\sigma} \left(  (1+\sigma^2)(-1-w) +1 \right)   (1+\sigma^2) (1-p), \, w \in \mathbb{R}.
\end{align}
Fig.~\ref{fig:Plot_of_error_mismatched_pdf} shows two examples of $f_{W_1}$ and $f_{W_2}$, one in a low noise regime and the other in a high noise regime. 

\begin{figure}[h!]
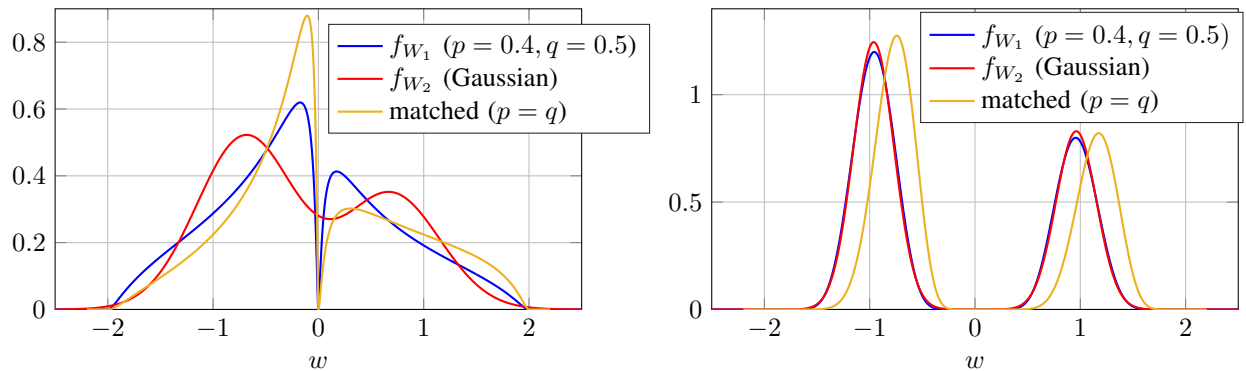
 
	\centering
	  \begin{subfigure}[c]{0.45\textwidth}
	  \center
	\input{pdf_error_mismatch_low_sig.tex}
	\caption{Plots of $f_W$ for $\sigma=1.5$ and $p=0.4$ and $q=0.5$ (low noise). }
	\end{subfigure}
	~
	  \begin{subfigure}[c]{0.45\textwidth}
	  \center
	\input{pdf_error_mismatch_hihg_sig.tex}
	\caption{Plots of $f_W$ for $\sigma=5$ and $p=0.4$ and $q=0.5$ (high noise). }  
	  \end{subfigure}
		
		\caption{ Examples of the pdf of  $W$ (mismatched case).}
		\label{fig:Plot_of_error_mismatched_pdf}
		\vspace{-0.5cm} 
\end{figure}%

\end{example*}

\section{Empirical Bayes for Conditional Moments and Conditional Cumulants} 
\label{sec:Applications}

In this section, we demonstrate how some of the proposed identities can be used to estimate conditional moments and conditional cumulants.  

\subsection{Empirical Bayes for Higher-Order Conditional Moments }
\label{subsec:EB}

 An interesting application of the original TRE identity is the idea of \emph{empirical Bayes} proposed by Robbins in \cite{robbins1956empirical}.  Consider an i.i.d. sequence $Y_1,\ldots, Y_n$ according to $P_Y$, and assume that we have perfect knowledge of $\sigma$.  Because the conditional estimator in the TRE formula depends only on the marginal distribution of the output $Y$, from the sequence observations $Y_1,\ldots, Y_n$, we can build an empirical estimate  of $\E[X|Y=y]$
 by mimicking the TRE formula 
 \begin{equation}
 \widehat{m}(y)=y + \sigma^2 \frac{\widehat{f}'_Y(y)}{ \widehat{f}_Y(y)}, \, y\in \mathbb{R} ,
 \end{equation}   
 where $\widehat{f}_Y(y)$ and $\widehat{f}'_Y(y)$ are estimates of   $f_Y(y)$ and $f_Y'(y)$, respectively.  
 In other words, we are able to estimate  $\E[X|Y=y]$ without the knowledge of the prior distribution on $X$. The interested reader is referred to \cite[Chapter 6.1]{efron2016computer}  for a historical account and the impact of the empirical Bayes formula.
 
Let   $\widehat{f}_Y^{(k)}(y)$ denote the estimate of $ f_Y^{(k)}(y)$ based on a  random sample $Y_1,\ldots, Y_n$. Now, inspired by the new generalization of the TRE in \eqref{eq: GeneralizedHigherOrderTweedy},  define the estimator of  $\E[X^k|Y=y]$   as follows: 
\begin{equation}
\widehat{m}_k(y)= \sigma^{2k}   \frac{ \sum_{m=0}^k  {{k} \choose {m}}   
 \widehat{f}_Y^{(k-m)}(y)   \frac{(-i)^m}{\sigma^m}   H_{e_m} \left( i \frac{y}{\sigma} \right) }{  \widehat{f}_Y(y)  },  \, y\in \mathbb{R}.
 \label{eq:CondExpecEstimator}
\end{equation} 

To estimate the derivatives $f_Y^{(k)}(y)$, we use the following steps:
\begin{enumerate}
\item  Estimate $f_Y$ by using a \emph{kernel density estimator} 
\begin{equation}
    \widehat{f}_Y(y)=\frac{1}{n} \sum_{i=1}^n \frac{1}{a} k \left( \frac{y-Y_i}{a}\right),  \, y\in \mathbb{R} ,
\label{eq:kernelDensityEstimator}
\end{equation}
where $a>0$ is the \emph{bandwidth parameter}. We take the kernel to be $k(y)=\phi(y)=\frac{1}{\sqrt{2 \pi}} \eu^{-\frac{y^2}{2}}$; and 
\item Estimate $f_Y^{(k)}(y)$ by  taking the derivative of  \eqref{eq:kernelDensityEstimator}  $k$ times. 
\end{enumerate}

 The above estimators  are inspired by the estimators of the score function and the Fisher information studied in \cite{bhattacharya1967estimation, dmitriev1974estimation} and \cite{WeiFisher2020}. 
 
The next results show that the estimator in \eqref{eq:CondExpecEstimator} is consistent. 
 
\begin{theorem} \label{thm:ConsistentEB} Let  $a= \frac{1}{n^u}  $ and $t_n = \frac{\sigma^2  \sqrt{ w  \log(n)} }{3}$ for some $ u \in \left(0, \frac{1}{2k +4}  \right)$ and $w \in (0,u)$. Moreover, assume that $\E[X^2]<\infty$.  Then,  for  every $k \in \mathbb{N}$ and  $\sigma^2>0$ 
\begin{equation}
\lim_{n \to \infty}  \mathbb{P} \left[  \sup_{ |y| \le t_n }  \left| \widehat{m}_k(y)-  \E[X^k|Y=y] \right| \ge    \frac{C_{k,\sigma}}{n^{u-w}} \right]= 0, \notag
\end{equation}
where  $C_{k,\sigma}$ is a constant that depends only on $k$ and $\sigma$. 
\end{theorem} 
\begin{IEEEproof}
See Appendix~\ref{app:thm:ConsistentEB}. 
\end{IEEEproof}

\subsection{Empirical Bayes for Conditional Cumulants} 
\label{sec:EB_for_CC}
In the last section, it was shown that the conditional moments can be estimated by only using the copies of $Y$.  This was possible in view of the fact that the conditional moments depend on the joint distribution $P_{XY}$ only through the marginal $P_Y$.   As was shown in Section~\ref{sec:Conditional_Cumulants} the same property holds for the conditional cumulants.   Using this observation, next it is shown that one can estimate conditional cumulants based only on the sequence of  i.i.d. observations $Y_1,\ldots, Y_n$, and no knowledge of the prior distribution $P_X$ is needed. 

The proposed estimators for the conditional cumulant  $\widehat{\kappa}_{X|Y=y}(k+1)$ for order $k+1$ is constructed by mimicking the new identity in Proposition~\ref{prop:Cumulant_derivative_of_CE} and is given by 
\begin{equation}
\widehat{\kappa}_{X|Y=y}(k+1)=  \sigma^{2k} D_h^{(k)}  \widehat{m}(y), \,  y\in \mathbb{R},  \label{eq:Estimation_cumulants}
\end{equation}
where  $\widehat{m}(y)$ is the estimator of the conditional expectation studied in Section~\ref{subsec:EB}, and $D_h^{(k)}$ is the Lanczos' derivative operator in \eqref{eq:Lanczos_derivative}.

The next lemma  provides various approximation results for the Lanczos' derivative operator, which will be used to show that the estimator in \eqref{eq:Estimation_cumulants} is consistent.   

\begin{lem}\label{lem:Lancos_approximations} Fix some $h>0$ and an interval $I=(-b,b)$  for some $b>0$.   Let $M_{k+2} =\sup_{x \in I_h} | f^{(k+2)}(x)|$ where $I_h=[ -b-h,b+h]$. Then, for $k \in \mathbb{N}$  and $x\in I$
\begin{align}
\left| D_h^{(k)} f(x) -  f^{(k)}(x)  \right | \le    \alpha_k  M_{k+2} h^2,  \label{eq:bound_Lancos_derivative_error}
\end{align}
where $\alpha_k=  \frac{c_k}{(k+2)! }   \frac{2}{ \sqrt{  2k+1}}  $. 
Moreover, suppose that   $\sup_{x \in I_h} |f (x)-g(x)| \le \epsilon$.   Then, for $k \in \mathbb{N}$  and $x\in I$
\begin{align}
\left| D_h^{(k)} g(x) -  f^{(k)}(x)  \right | \le   \alpha_k  M_{k+2} h^2    +    \frac{\beta_k   }{ h^k} \epsilon  , \label{eq:bound_Lancos_derivative_error_mismatch}
\end{align} 
where $  \beta_k= (k+2)! \alpha_k$. 
\end{lem}
\begin{IEEEproof}
  See Appendix~\ref{app:lem:Lancos_approximations}. 
\end{IEEEproof}

The next results shows that the estimator in \eqref{eq:Estimation_cumulants} is consistent.

\begin{theorem}\label{thm:EB_for_CC} Let  $a= \frac{1}{n^u}  $ and $t_n = \frac{\sigma^2  \sqrt{ w  \log(n)} }{3}$ for some $ u \in \left(0, \frac{1}{8}  \right)$ and $w \in (0,u)$. Moreover, assume that $\E[X^2]<\infty$.  Then,  for  every $k \in \mathbb{N}$ and  $\sigma^2>0$ 
\begin{equation}
\lim_{n \to \infty} \mathbb{P} \left[  \max_{|y| \le  \frac{t_n}{2}}   \left| \kappa_{X|Y=y}(k+1) - \widehat{\kappa}_{X|Y=y}(k+1) \right|   \ge    \frac{  \tilde{C}_{k,\sigma}  \,  t_n}{n^{ \frac{2 (u-w)}{2+k}} }    \right ]=0,
\end{equation}
where $ \tilde{C}_{k,\sigma}$ is a constant that depends only on $k$ and $\sigma$. 
\end{theorem}
\begin{IEEEproof}
The parameters of the estimator in \eqref{eq:Estimation_cumulants} are set as follows. Let the parameters needed to estimate $ \widehat{m}(y) $ be the same as in Theorem~\ref{thm:ConsistentEB}.  Moreover, let $\epsilon_n=\frac{C_{1,\sigma}}{n^{u-w}} $  where $C_{1,\sigma}$ is defined in Theorem~\ref{thm:ConsistentEB} and choose $h= \epsilon_n^{ \frac{1}{k+2}}$.  Furthermore, assume that $n$ is large enough such that $h \le \frac{t_n}{2}$.    Then, by using Lemma~\ref{lem:Lancos_approximations},  we arrive at 
\begin{align}
 \left(  \alpha_k  M_{k+2}     +   \beta_k \right) \epsilon^{ \frac{2}{2+k}}   
    & \ge \max_{|y| \le t_n -h}  | \kappa_{X|Y=y}(k+1) - \widehat{\kappa}_{X|Y=y}(k+1)|  \\
    &\ge \max_{|y| \le \frac{t_n}{2} }  | \kappa_{X|Y=y}(k+1) - \widehat{\kappa}_{X|Y=y}(k+1)|  ,
     \label{eq:Error_bound_1_kumulants}
\end{align} 
where 
\begin{equation}
\epsilon_n= \max_{|y| \le t_n}  |   \E[X|Y=y]  -   \widehat{m}(y)  |. \label{eq:epsilon_definition_kumulant_estimation}
\end{equation} 
Moreover, the parameter $M_{k+2}$ can be bounded as follows: 
\begin{align}
M_{k+2}&= \max_{|y| \le t_n}  \left| \frac{ {\rm d}^{k+2}}{ {\rm d} y^{k+2}}  \E[X|Y=y]  \right| \\
&= \max_{|y| \le t_n}  \left|  \frac{ \kappa_{X|Y=y}(k+2)}{ \sigma^{ 2k +4}}   \right| \label{eq:Using_Derivative_Identity_For_estimation_bound} \\
&\le   \max_{|y| \le t_n}  \frac{ a_{k+2} |y|^{k+2}+b_{k+2}}{ \sigma^{ 2k +4}}  \label{eq:Bound_On_kappa_application} \\
&=    C_1 |t_n|^{k+2}+ C_2,  \label{eq:Just_Bounding_with Constants}
\end{align} 
where in \eqref{eq:Using_Derivative_Identity_For_estimation_bound} we have used the identity in Proposition~\ref{prop:Cumulant_derivative_of_CE}; and  \eqref{eq:Bound_On_kappa_application} follows by using the bound in Proposition~\ref{prop:propositon_bounds_cumulants}.

Combining \eqref{eq:Error_bound_1_kumulants} and \eqref{eq:Just_Bounding_with Constants},  we arrive at
\begin{equation}
\max_{|y| \le \frac{t_n}{2} }  | \kappa_{X|Y=y}(k+1) - \widehat{\kappa}_{X|Y=y}(k+1)|   \le  C_3  t_n    \epsilon_n^{ \frac{2}{2+k}}. \label{eq:Final_bound_cumulant_estiamtion}
\end{equation} 

Since \eqref{eq:epsilon_definition_kumulant_estimation} leads to \eqref{eq:Final_bound_cumulant_estiamtion},  one obtains that 
\begin{align}
&\mathbb{P} \left[  \max_{|y| \le \frac{t_n}{2} }  | \kappa_{X|Y=y}(k+1) - \widehat{\kappa}_{X|Y=y}(k+1)|   \ge  C_3  t_n    \epsilon_n^{ \frac{2}{2+k}}   \right ]  \notag\\
&\le \mathbb{P} \left[  \max_{|y| \le t_n}  |   \E[X|Y=y]  -   \widehat{m}(y)  | \ge \epsilon_n  \right ].  \label{eq:Confidance_bound_on_Cumulatns}
\end{align} 
The proof is now concluded by using Theorem~\ref{thm:ConsistentEB} where it was shown that \eqref{eq:Confidance_bound_on_Cumulatns} goes to zero as $n \to \infty$. 
\end{IEEEproof}

\section{Conclusion and Outlook} 
\label{sec:ConclusionANDoutlook}
This work has derived a general derivative identity for a conditional mean estimator. This identity has been used to recover several known derivative identities, such as Hatsel and Nolte identity for the conditional variance and recursive Jaffer's identity. Moreover, several new identities have been derived, the most notable of which include: a new integral version of Jaffer's identity, the identity that connects higher-order conditional moments and the derivatives of the conditional expectation via Bell-polynomials, and the identity that shows that the derivatives of the conditional expectation are proportional the conditional cumulants. All of the derived identities are summarized in Table~\ref{table:Identities} (top of the next page). On the application side, the identities have been used to study the distribution of the conditional expectation and the distribution of the estimation error. Moreover, the identities have been used to extend the notion of  empirical Bayes to the higher-order conditional moments and conditional cumulants. 

An interesting future direction would be to see if the main identity in Theorem~\ref{thm:MainIdentity} can shed light on the vector generalization of the single crossing property  in \cite{BustinMMSEparallelVectorChannel}.

\begin{table*}[h]
\caption{List of Identities.} \label{table:Identities}
\center
\begin{tabular}{|>{\centering\arraybackslash}m{0.75in}| p{3.5cm}| l |l|}
\cline{2-3}
\multicolumn{1}{c|}{}                       & \textbf{Name} & \textbf{Identity}            \\ \hline
\multirow{2}{0.8in}{\centering\textbf{Two Main Identities}} & TRE Identity \eqref{eq:TweediesFormulaGaussian} &  $ \E \left[  \X | \Y= {\bf y} \right ]={\bf y}+ \boldsymbol{\mathsf{K}}_{\N}  \frac{   \nabla_{ {\bf y}}  f_\Y({\bf y})  }{ f_\Y({\bf y})} $ \\ \cline{2-3}
                                            &  Identity \eqref{eq:MainIdentity} &  $\boldsymbol{\mathsf{J}}_\mathbf{y}  \E \left[  \U | \Y= {\bf y} \right ]=   \boldsymbol{\mathsf{K}}_{\N}^{-1}   \boldsymbol{\mathsf{Cov}} ( \X, \U | \Y={\bf y})$, \, $\U \leftrightarrow \X  \leftrightarrow \Y$  \\                 \hline
                       \multirow{1}{0.6in}{\centering\textbf{Variance }}                              &  Hatsell and Nolte \eqref{eq:Hatsell-Nolte-general} &  $ \boldsymbol{\mathsf{J}}_\mathbf{y}  \E \left[  \X | \Y= {\bf y} \right ]  =  \boldsymbol{\mathsf{K}}_{\N}^{-1}  \boldsymbol{\mathsf{Var}}(\X | \Y={\bf y})   $  \\    \hline
                  \multirow{4}{0.8in}{\centering\textbf{Recursive Identities}} & Jaffer's Identity  \eqref{eq:RecursiveScalarForm} &  $ \E  \hspace{-0.05cm} \left[ X^{k+1} |Y =y  \right] = \sigma^2 \frac{{\rm d}}{ {\rm d} y }  \E  \hspace{-0.05cm}\left[ X^{k} |Y =y \right]  \hspace{-0.05cm}+   \E  \hspace{-0.05cm}\left[ X^{k} |Y =y \right]  \E  \hspace{-0.05cm}\left [ X |Y =y \right ]
     ) 
  $ \\ \cline{2-3}
                                            & Integral version of Jaffer's Identity  \eqref{eq:ScalarVersionOfTheIdenity} &  $  \E  \hspace{-0.05cm} \left[  X^{k}     |Y= y\right ] =   \hspace{-0.05cm} \sigma^{2k} \eu^{-   \frac{1}{\sigma^2}  \hspace{-0.05cm} \int_0^{y}   \hspace{-0.05cm} \E \left[      X    |Y= t \right ] {\rm d} t}   \hspace{-0.05cm} \frac{  {\rm d}^k} { {\rm d} y^k }   \eu^{ \frac{1}{\sigma^2}  \hspace{-0.05cm} \int_0^{y}   \hspace{-0.05cm} \E \left[     X      |Y= t\right ] {\rm d} t}  $  \\             \cline{2-3}
    & Vector Jaffer's Identity (Version 1)  \eqref{eq:Vector_Jaffer_Ver1} &  $  \E   \hspace{-0.05cm} \left[ (\X \X^\mathsf{T} )^{k}  |\Y ={\bf y} \right]  \hspace{-0.05cm}= \boldsymbol{\mathsf{K}}_{\N} \boldsymbol{\mathsf{J}}_\mathbf{y}   \E   \hspace{-0.05cm} \left[ (\X \X^\mathsf{T} )^{k-1} \X |\Y ={\bf y} \right]    \hspace{-0.05cm}$  \\ 
     & $ { } $ &   \quad \quad $ +  \E  \hspace{-0.05cm} \left[  \X  |\Y ={\bf y} \right]  \hspace{-0.05cm} \E  \hspace{-0.05cm} \left[   \X^\mathsf{T}  (\X \X^\mathsf{T} )^{k-1}  |\Y ={\bf y} \right] $  \\                 \cline{2-3}
                                             & Conditional Expectation via Higher-Order Derivatives \eqref{eq:Higher_Moments_Bell_Polynomial} &  $   \E  \hspace{-0.05cm} \left[  X^{k}     |Y= y\right ]  = \sigma^{2k} \mathsf{B}_{k}\left (   \E^{(0)}  \hspace{-0.05cm} \left[  \frac{X}{\sigma^2}    |Y= y\right ],\ldots,   \E^{(k-1)}  \hspace{-0.05cm} \left[  \frac{X}{\sigma^2}    |Y= y\right ]    \right)$  \\                 \cline{2-3}
                                             & Derivatives of the Conditional Expectation \eqref{eq:Higher_Order_Derivatives_of_CE} &  $ \frac{ {\rm d}^k}{ {\rm d} y^k}\E \hspace{-0.05cm} \left[  X     |Y= y\right ] = \hspace{-0.05cm}  \sigma^2 \sum_{m=1}^{k+1}  c_m  \mathsf{B}_{k+1,m} \left( \E  \hspace{-0.05cm} \left[  \frac{X}{\sigma^2}    |Y= y\right ] ,\ldots,  \E  \hspace{-0.05cm} \left[ \left (\frac{X}{\sigma^2} \right)^{k-m+2}     |Y= y\right ]  \right)  $     \\                 \cline{2-3}
                                             & Generalized TRE Identity  \eqref{eq: GeneralizedHigherOrderTweedy} &  $ \E[   X^k |Y=y ] =\sigma^{2k}     \frac{  \frac{{\rm d}^k}{ {\rm d} y^k }  \left(  f_Y(y)  \eu^{  \frac{y^2}{2 \sigma^2}  } \right) }{ f_Y(y)  \eu^{  \frac{y^2}{2 \sigma^2}  }  } = \sigma^{2k}   \frac{ \sum_{m=0}^k  {{k} \choose {m}}   
 f_Y^{(k-m)}(y)   \frac{(-i)^m}{\sigma^m}   H_{e_m} \left( i \frac{y}{\sigma} \right) }{  f_Y(y)  }$  \\ \hline
\multirow{4}{0.8in}{\centering\textbf{Identities for Conditional Cumulants}} & Conditional Cumulants and Conditional Expectation \eqref{eq:Cumulants_and_Conditional} &  $ \kappa_{X|Y=y}(k+1)= \sigma^{2k} \frac{{\rm d}^{k}}{ {\rm d} y^{k} } \E[ X | Y=y] $  \\ \cline{2-3}
                                & Recursion for Conditional Cumulants  \eqref{eq:Cumulants_Recursive_Identity}         & $ \sigma^2 \frac{{\rm d}}{ {\rm d} y } \kappa_{X|Y=y}(k+1)=  \kappa_{X|Y=y}(k+2)$ \\                 \cline{2-3}
    & Conditional Cumulants Generating Function and Conditional Expectation \eqref{eq:Cum_gen_function_and_Cond_Expect} &  $  \frac{{\rm d}^{k+1}}{ {\rm d} t^{k+1} }  K_X(t|Y=y)  =\frac{{\rm d}^{k+1}}{ {\rm d} y^{k+1} }  K_X(t|Y=y)+  \frac{{\rm d}^{k}}{ {\rm d} y^{k} } \E[ X | Y=y]  $   \\                  \cline{2-3}      
    & Multivariate Identity for Conditional Cumulants and Conditional Expectation  \eqref{eq:Multivaraite_Cumulant_Expecations_identity}         & $   {\bf k}_{s_j}^\mathsf{T}   \frac{\partial^{j-1}   \E[ \X |\Y={\bf y}]  }{ \partial y_{s_1} \dots  \partial y_{s_{j-1}}   }   =\sum_{p_1=1}^n \dots  \sum_{p_j=1}^n  \prod_{i=1}^j   k_{s_i,p_i}   \kappa_{\X|\Y={\bf y}}(p_1,\ldots, p_j) $ \\                 \hline       
\multirow{4}{0.8in}{\centering\textbf{Identities for Distributions and the Information Density}} & Inverse TRE Identity \eqref{eq:Invers_TRE} &  $ f_\Y({\bf y})= c  \exp \left(   \oint_{\bf 0}^{\bf y}  \boldsymbol{\mathsf{K}}_{\N}^{-1} (\mathbf{t}-   \E[\X | \Y={\bf t}])      \boldsymbol{\cdot} {\rm \mathbf{d} } \mathbf{t} \right)$ \\ \cline{2-3}
                                            & Gradient of the Information Density \eqref{eq:Gradient_Info_Density} &  $  \nabla_{\bf y} \iota({\bf x};{\bf y})= \boldsymbol{\mathsf{K}}_{\N}^{-1}  ( {\bf x} -  \E[  \X   | \Y={\bf y}]     )     )$   \\                 \cline{2-3}
                                             & Gradient of the Conditional Distribution \eqref{eq:Indentity_Conditiona_Probaiblity} &  $ \nabla_{\bf y} \log \left(   \mathbb{P}[\X \in \mathcal{A}| \Y={\bf y}] \right)   = \boldsymbol{\mathsf{K}}_{\N}^{-1}    (  \E[  \X  | \Y={\bf y}, \X \in \mathcal{A}] -  \E[  \X   | \Y={\bf y}]     ) 
  $   \\       \cline{2-3}
   & Hessian of Information Density \eqref{eq:Hessian_of_Conditional_distribuiton_continious} &  $\boldsymbol{\mathsf{D}}^2_{\bf y} \iota({\bf x};{\bf y}) =  -   \boldsymbol{\mathsf{K}}_{\N}^{-1}  \boldsymbol{\mathsf{Var}}(\X | \Y={\bf y})  \boldsymbol{\mathsf{K}}_{\N}^{-1}  $  \\         \cline{2-3}    
    & Hessian  of the Conditional Distribution  \eqref{eq:Hessian_of_Conditional_distribuiton_continious} &  $\boldsymbol{\mathsf{D}}^2_{\bf y}  \log \left(   \mathbb{P}[ \X \in \mathcal{A} | \Y={\bf y}] \right) =\boldsymbol{\mathsf{K}}_{\N}^{-1} \left( \boldsymbol{\mathsf{Var}}(\X | \Y={\bf y}, \X \in \mathcal{A}) - \boldsymbol{\mathsf{Var}}(\X | \Y={\bf y})     \right) \boldsymbol{\mathsf{K}}_{\N}^{- 1 }  $  \\       \cline{2-3} 
    & Higher-Order Derivatives of Information Density \eqref{eq:higher_order_derivatives_of_Information_density} &  $\frac{{\rm d}^k}{ {\rm d} y^k} \iota_{P_{X Y}}(x;y)= - \frac{1}{\sigma^{2k}} \kappa_{X|Y=y}(k),  k\ge 3 $  \\              
            \hline
\end{tabular}
\end{table*}



\begin{appendices} 

\section{Proof of Theorem~\ref{thm:MainIdentity}} 
\label{app:sec:MainThm}
First,   observe that 
\begin{align}
 \frac{{\rm d}}{ {\rm d} y_m } \phi_{ \boldsymbol{\mathsf{K}}_{\N}} ({\bf y}-\X)
 &= - \frac{1}{2}   \phi_{ \boldsymbol{\mathsf{K}}_{\N}} ({\bf y}-\X)  \frac{{\rm d}}{ {\rm d} y_m }  ({\bf y}-\X)^\mathsf{T} \boldsymbol{\mathsf{K}}_{\N}^{-1} ({\bf y}-\X)\\
 &=    \phi_{ \boldsymbol{\mathsf{K}}_{\N}} ({\bf y}-\X)   {\bf e}_m^\mathsf{T}   \boldsymbol{\mathsf{K}}_{\N}^{-1} (\X-{\bf y}).  \label{eq:derivativeOfNormalDistribution}
\end{align} 
Second,   for the moment assume that the the interchange of expectation and differentiation is allowed, and  observe the following sequence of steps: 
\begin{align}
\frac{{\rm d}}{ {\rm d} y_m } \E \left [ \U    | \Y={\bf y} \right]  
&=  \frac{{\rm d}}{ {\rm d} y_m } \E \left[ \U  \frac{   \phi_{ \boldsymbol{\mathsf{K}}_{\N}} ({\bf y}-\X) }{ f_\Y({\bf y})}  \right ]  \label{eq:ApplyingBayesTheorem}\\
&=   \E \left[  \U   \frac{{\rm d}}{ {\rm d} y_m }  \frac{   \phi_{ \boldsymbol{\mathsf{K}}_{\N}} ({\bf y}-\X) }{ f_\Y({\bf y})}  \right ]  \label{eq:Interchange_Differentiation_and_Expectation}\\
&=   \E \left[ \U     \frac{     \frac{{\rm d}}{ {\rm d} y_m} \phi_{ \boldsymbol{\mathsf{K}}_{\N}} ({\bf y}-\X) }{ f_\Y({\bf y})}  \right ]  -   \E \left[ \U   | \Y={\bf y}   \right ]   \frac{     \frac{{\rm d}}{ {\rm d} y_m } f_\Y({\bf y})  }{ f_\Y({\bf y})} \\
&= \E \left[  \U  {\bf e}_m ^\mathsf{T}   \boldsymbol{\mathsf{K}}_{\N}^{-1} \X  |\Y= {\bf y} \right ]   -  \E \left[ \U      |\Y= {\bf y} \right ]  {\bf e}_m^\mathsf{T}   \boldsymbol{\mathsf{K}}_{\N}^{-1} {\bf y}  -   \E \left[  \U   | \Y={\bf y}   \right ]   \frac{     \frac{{\rm d}}{ {\rm d} y_m } f_\Y({\bf y})  }{ f_\Y({\bf y})} \label{eq:applyingDerivativeOfNOrmalDistribution} \\
&= \E \left[ \U  {\bf e}_m ^\mathsf{T}   \boldsymbol{\mathsf{K}}_{\N}^{-1} \X  |\Y= {\bf y} \right ] -  \E \left[  \U     |\Y= {\bf y}   \right ]  {\bf e}_m^\mathsf{T}   \left(   \boldsymbol{\mathsf{K}}_{\N}^{-1} {\bf y} 
  +      \frac{  \nabla_{ {\bf y}}    f_\Y({\bf y})  }{ f_\Y({\bf y})}  \right)\\
&= \E \left[  \U  {\bf e}_m ^\mathsf{T}   \boldsymbol{\mathsf{K}}_{\N}^{-1} \X |\Y= {\bf y} \right ] -  \E \left[  \U   |\Y= {\bf y} \right ]  \E \left[ {\bf e}_m^\mathsf{T} \boldsymbol{\mathsf{K}}_{\N}^{-1}  \X | \Y={\bf y} \right] \label{eq:applicationOfTRE}\\ 
&= \E \left[  \U   \X^\mathsf{T}   |\Y= {\bf y} \right ]    \boldsymbol{\mathsf{K}}_{\N}^{-1}  {\bf e}_m  -  \E \left[  \U   |\Y= {\bf y} \right ]  \E \left[  \X^\mathsf{T} | \Y={\bf y} \right] \boldsymbol{\mathsf{K}}_{\N}^{-1}  {\bf e}_m\\
&=  \boldsymbol{\mathsf{Cov}}(\U, \X | \Y={\bf y})   \boldsymbol{\mathsf{K}}_{\N}^{-1} {\bf e}_m, 
\end{align} 
where  the equalities follow from:  \eqref{eq:ApplyingBayesTheorem}   using Bayes' formula; \eqref{eq:applyingDerivativeOfNOrmalDistribution}  using the expression in \eqref{eq:derivativeOfNormalDistribution}; and  \eqref{eq:applicationOfTRE} using the TRE identity   in \eqref{eq:TweediesFormulaGaussian}.   

Now using the definition of Jacobian in \eqref{eq:Def_Jacobian}, we have that
\begin{align}
 \boldsymbol{\mathsf{J}}_\mathbf{y}  \E \left [ \U    | \Y={\bf y} \right]=  \left(   \boldsymbol{\mathsf{Cov}}(\U, \X | \Y={\bf y})   \boldsymbol{\mathsf{K}}_{\N}^{-1} \right)^{\mathsf{T}} \\
 =   \boldsymbol{\mathsf{K}}_{\N}^{-1}   \boldsymbol{\mathsf{Cov}}^{\mathsf{T}} (\U, \X | \Y={\bf y})  \\
  =   \boldsymbol{\mathsf{K}}_{\N}^{-1}   \boldsymbol{\mathsf{Cov}} ( \X, \U | \Y={\bf y}) , 
\end{align} 
where we have used the symmetry of $ \boldsymbol{\mathsf{K}}_{\N}$ and the property that $ \boldsymbol{\mathsf{Cov}}^{\mathsf{T}} (\U, \X | \Y={\bf y})= \boldsymbol{\mathsf{Cov}} ( \X,\U | \Y={\bf y})$. 

Therefore, to conclude the proof, we require to show that the interchange of differentiation and expectation in \eqref{eq:Interchange_Differentiation_and_Expectation} is permitted.  A sufficient condition for the interchange of differentiation and expectation  is given by the Leibniz integral rule, which requires verifying that
\begin{equation}
 \E \left[ \left \|  \U   \frac{{\rm d}}{ {\rm d} y_m }  \frac{   \phi_{ \boldsymbol{\mathsf{K}}_{\N}} ({\bf y}-\X) }{ f_\Y({\bf y})} \right \|  \right ] <\infty.  \label{Eq:Leibniz integral rule}
\end{equation} 
To this end, note that
\begin{align}
 \E \left[ \left \|  \U   \frac{{\rm d}}{ {\rm d} y_m }  \frac{   \phi_{ \boldsymbol{\mathsf{K}}_{\N}} ({\bf y}-\X) }{ f_\Y({\bf y})} \right \|  \right ] 
 &=  \E \left[ \left \|  \U   \left( \frac{     \frac{{\rm d}}{ {\rm d} y_m} \phi_{ \boldsymbol{\mathsf{K}}_{\N}} ({\bf y}-\X) }{ f_\Y({\bf y})}  - \frac{  \phi_{ \boldsymbol{\mathsf{K}}_{\N}} ({\bf y}-\X)     \frac{{\rm d}}{ {\rm d} y_m } f_\Y({\bf y})  }{ f_\Y^2({\bf y})} \right)    \right \|  \right ] \\
 &\le   \E \left[ \left \|  \U   \frac{     \frac{{\rm d}}{ {\rm d} y_m} \phi_{ \boldsymbol{\mathsf{K}}_{\N}} ({\bf y}-\X) }{ f_\Y({\bf y})}  \right \|  \right]  + \E \left[ \left \|  \U \frac{  \phi_{ \boldsymbol{\mathsf{K}}_{\N}} ({\bf y}-\X)     \frac{{\rm d}}{ {\rm d} y_m } f_\Y({\bf y})  }{ f_\Y^2({\bf y})}     \right \|  \right ],  \label{eq:Liebnitz_rule_+triangle}
\end{align} 
where the last step follows by using the triangle inequality. 
We now bound each term of \eqref{eq:Liebnitz_rule_+triangle} separately.  The first term in \eqref{eq:Liebnitz_rule_+triangle} is bounded by 
\begin{align}
\E \left[ \left \|  \U   \frac{     \frac{{\rm d}}{ {\rm d} y_m} \phi_{ \boldsymbol{\mathsf{K}}_{\N}} ({\bf y}-\X) }{ f_\Y({\bf y})}  \right \|  \right] &= \E \left[ \left \|  \U   \frac{    \phi_{ \boldsymbol{\mathsf{K}}_{\N}} ({\bf y}-\X)   {\bf e}_m^\mathsf{T}   \boldsymbol{\mathsf{K}}_{\N}^{-1} (\X-{\bf y}) }{ f_\Y({\bf y})}  \right \|  \right] \\
&= \E \left[ \left \|  \U       {\bf e}_m^\mathsf{T}   \boldsymbol{\mathsf{K}}_{\N}^{-1} (\X-{\bf y})   \right \|  | \Y= {\bf y}\right] \label{eq:Applying_Bayes_for_111} \\
&\le  \E \left[ \left \|  \U       {\bf e}_m^\mathsf{T}   \boldsymbol{\mathsf{K}}_{\N}^{-1} \X    \right \|  | \Y= {\bf y}\right] +   \E \left[ \left \|  \U       {\bf e}_m^\mathsf{T}   \boldsymbol{\mathsf{K}}_{\N}^{-1} {\bf y}   \right \|  | \Y= {\bf y}\right] \\
&\le  \|   {\bf e}_m^\mathsf{T}   \boldsymbol{\mathsf{K}}_{\N}^{-1}\| \E \left[ \left \|  \U  \|  \|    \X  \right \|  | \Y= {\bf y}\right]   + \|     {\bf e}_m^\mathsf{T}   \boldsymbol{\mathsf{K}}_{\N}^{-1} {\bf y} \|   \E \left[ \left \|  \U    \right \|  | \Y= {\bf y}\right] . \label{eq:First_Term_Libnitz}
\end{align} 
where   \eqref{eq:Applying_Bayes_for_111} follows by using sing Bayes' formula; and \eqref{eq:First_Term_Libnitz} follows by  using the Cauchy-Schwarz inequality. 

The second term in \eqref{eq:Liebnitz_rule_+triangle} can be rewritten as 
\begin{align}
 \E \left[ \left \|  \U \frac{  \phi_{ \boldsymbol{\mathsf{K}}_{\N}} ({\bf y}-\X)     \frac{{\rm d}}{ {\rm d} y_m } f_\Y({\bf y})  }{ f_\Y^2({\bf y})}     \right \|  \right ]
 &= \E \left[ \left \|  \U \frac{     \frac{{\rm d}}{ {\rm d} y_m } f_\Y({\bf y})  }{ f_\Y({\bf y})}     \right \| | \Y={\bf y}  \right ]\\
  &=  \frac{    |  \frac{{\rm d}}{ {\rm d} y_m } f_\Y({\bf y}) |  }{ f_\Y({\bf y})}    \E \left[ \left \|  \U   \right \| | \Y={\bf y}  \right ]. \label{eq:Secon_Term_Libnitz}
\end{align} 
Therefore, by combining \eqref{eq:First_Term_Libnitz} and \eqref{eq:Secon_Term_Libnitz},  the condition in  \eqref{Eq:Leibniz integral rule} holds if
\begin{align}
 \E \left[ \left \|  \U  \|  \|    \X  \right \|  | \Y= {\bf y}\right]&<\infty, {\bf y}\in \mathbb{R}^n, \\
\E \left[ \left \|  \U   \right \| | \Y={\bf y}  \right ]&<\infty,  {\bf y}\in \mathbb{R}^n. 
\end{align} 
This concludes the proof.

\section{Proof of Lemma~\ref{lem:recurenceRelationship}}
\label{app:lem:recurenceRelationship}

First, we multiply both side  of \eqref{eq:RecurcenceRelationship} by $\eu^{ \int_0^x f_1(t) {\rm d} t}$, which leads to 
\begin{align}
f_k(  x ) \eu^{ \int_0^x f_1(t) {\rm d} t}  \
&= \hspace{-0.1cm}  \frac{{\rm d}}{ {\rm d} x} \hspace{-0.05cm}   f_{k-1}(x) \eu^{ \int_0^x \hspace{-0.1cm}  f_1(t) {\rm d} t} \hspace{-0.1cm} + \hspace{-0.1cm}  f_{k-1}(x) \hspace{-0.01cm} f_{1}(x) \eu^{ \int_0^x \hspace{-0.1cm}  f_1(t) {\rm d} t} \notag\\
&=  \hspace{-0.1cm}  \frac{{\rm d}}{ {\rm d} x} \hspace{-0.05cm}  \left(   f_{k-1}(x) \eu^{ \int_0^x \hspace{-0.1cm}  f_1(t) {\rm d} t}   \right) , \label{eq:DerivativeOfTheRecurance}
\end{align} 
where the last expression is due to the product rule.  Now, let
\begin{equation}
h_k(x)=f_k(x) \eu^{ \int_0^x f_1(t) {\rm d} t}. 
\end{equation} 
Then, the  expression in \eqref{eq:DerivativeOfTheRecurance} implies that 
\begin{equation}
h_k(x)=  \frac{{\rm d}}{ {\rm d} x}   h_{k-1}(x), \label{eq:DefinitionOfhk}
\end{equation}
which by using the recursion $k-1$ times implies that
\begin{equation}
h_k(x)=     \frac{{\rm d}^{k-1}}{ {\rm d} x^{k-1}} h_1(x)=   \frac{{\rm d}^{k-1}}{ {\rm d} x^{k-1}}   \left(  f_1(x) \eu^{ \int_0^x f_1(t) {\rm d} t } \right).  \label{eq:Definitionofhkfunction}
\end{equation} 
Furthermore, by applying the chain rule in \eqref{eq:Definitionofhkfunction}, we arrive at  
\begin{equation}
h_k(x)= \frac{{\rm d}^{k}}{ {\rm d} x^{k}}   \eu^{ \int_0^x f_1(t) {\rm d} t } .  
\end{equation}
The proof of \eqref{eq:SolutionToRecursion} is concluded by using  the definition of  $h_k(x)$ in \eqref{eq:DefinitionOfhk}.   

To show the expression in \eqref{eq:Bell-polynomial_representation} recall the Fa\`a di Bruno formula for the higher-order chain rule \cite[Thm.~11.4]{charalambides2018enumerative}: given two $k \in \mathbb{N}$ differentiable functions $g(t)$ and $\xi(t)$
\begin{equation}
\frac{{\rm d}^{k}  \xi(  g(t)  ) }{ {\rm d} t^{k}} = \sum_{m=1}^k  \xi^{(m)}( g(t))  \mathsf{B}_{k,m} \left(g^{(1)}(x), \ldots, g^{(k-m+1)}(x) \right). \label{eq: Fa\`a di Bruno formula} 
\end{equation} 
Next, by letting    $g(x) =\int_0^x f_1(t) {\rm d} t$, the expression in \eqref{eq:SolutionToRecursion} can be re-written as
\begin{align}
f_k(x)&=  \eu^{- g(x) }  \frac{  {\rm d}^k} { {\rm d} x^k } \eu^{ g(x)} \\
&=   \eu^{- g(x) } \sum_{m=1}^k  \eu^{g(x) } \mathsf{B}_{k,m} \left(g^{(1)}(x), \ldots, g^{(k-m+1)}(x) \right)  \label{eq:applying_Faa_di_Bruno_formula}\\
&=  \mathsf{B}_{k} \left (g^{(1)}(x), \ldots, g^{(k)}(x) \right) \label{eq:applying_complete_Ball_polynomial_def} \\
&= \mathsf{B}_{k} \left(f^{(0)}_1(x), \ldots, f^{(k-1)}_1(x) \right),  \label{eq:Applying derivative of g(t)}
\end{align} 
where  \eqref{eq:applying_Faa_di_Bruno_formula} follows by using the Fa\`a di Bruno formula in \eqref{eq: Fa\`a di Bruno formula}; \eqref{eq:applying_complete_Ball_polynomial_def} follows from \eqref{eq:Def_Bell_poly}; and  \eqref{eq:Applying derivative of g(t)} follows by noting that $g^{(m)}(x)=f^{(m-1)}(x), m\in \mathbb{N}$. 

To show \eqref{eq:Derivative_Solution_Bell}, we use the inversion formula for the Bell polynomial \cite[Rem.~11.3]{charalambides2018enumerative} which asserts the following:  if 
\begin{equation}
y_k=   \mathsf{B}_{k}(t_1,\ldots, t_k),
\end{equation} 
then
\begin{equation}
t_k= \sum_{m=1}^k (-1)^{m-1}(m-1)!  \mathsf{B}_{k,m}(y_1,\ldots, y_{k-m+1}). 
\end{equation} 
Setting $y_k=f_k(x)$ and $t_k=f_1^{(k-1)}(x)$ leads to 
\begin{align}
f_1^{(k-1)}(x) 
&=\sum_{m=1}^k (-1)^{m-1}(m-1)!  \mathsf{B}_{k,m}(f_1(x),\ldots, f_{k-m+1}(x)).
\end{align} 
The proof is concluded by using a change of variable from $k-1$ to $k$.

\section{Proof of Proposition~\ref{prop:propositon_bounds_cumulants}} 
\label{app:proof_propositon_bounds_cumulants}

First, observe that
\begin{align}
|\kappa_{X|Y=y}(k)| &\le k^k \E[ |X -\E[X|Y] |^k |Y =y] \label{eq:Bounds_on_cumulants}\\
& \le 2^{k-1} k^k  \E[ |X |^k |Y =y], \label{eq:Lp_esimates}
\end{align} 
where in \eqref{eq:Bounds_on_cumulants} we have used the bound in \cite[eq.~(4)]{dubkov1976properties}; and in \eqref{eq:Lp_esimates} we have used $| a+b|^k \le 2^{k-1} (| a|^k+|b|^k), k \ge 1$. 

We now show a bound on $\E[ |X |^k |Y =y]$, which is a generalization of the bound shown in \cite[Proposition~1.2]{fozunbal2010regret}. First, 
\begin{align}
\E[|X|^k|Y=y]
& =\int_{  f_{Y|X}(y|x) \le f_Y(y) } |x|^k \frac{ f_{Y|X}(y|x)}{ f_Y(y) } {\rm d} P_X(x) + \int_{  f_{Y|X}(y|x) > f_Y(y) } |x|^k \frac{ f_{Y|X}(y|x)}{ f_Y(y) } {\rm d} P_X(x) \\
&\le \E[|X|^k]  + \int_{  f_{Y|X}(y|x) > f_Y(y) } |x|^k \frac{ f_{Y|X}(y|x)}{ f_Y(y) } {\rm d} P_X(x) . \label{eq:Prep_bound}
\end{align} 
Next, we bounds the second term in \eqref{eq:Prep_bound} 
\begin{align}
&f_{Y|X}(y|x) > f_Y(y) \notag\\
 &\Rightarrow  (y-x)^2 \le 2 \sigma^2 \log \left( \frac{1}{\sqrt{2 \pi \sigma^2} f_Y(y) }  \right) \notag\\
&\Rightarrow  |y-x|  \le \sqrt{ 2 \sigma^2 \log \left( \frac{1}{\sqrt{2 \pi \sigma^2} f_Y(y) }  \right)}  \notag\\
&\Rightarrow  |x|  \le |y|+ \sqrt{ 2 \sigma^2 \log \left( \frac{1}{\sqrt{2 \pi \sigma^2} f_Y(y) }  \right)}  \label{eq:applying_Triangle_Inequaltiy}\\
&\Rightarrow  |x|^k  \le 2^{k-1} \left( |y|^k+ \left( 2 \sigma^2 \log \left( \frac{1}{\sqrt{2 \pi \sigma^2} f_Y(y) }  \right) \right)^{ \frac{k}{2}}  \right)  \label{eq:Lp-type_Bounds}\\
&\Rightarrow  |x|^k  \le 2^{k-1} \left( |y|^k+ 2\left(  y^2 +\E[X^2] \right)^{ \frac{k}{2}}  \right) \label{eq:Applying_Jensen's_Inequality_to_cumulants}\\
&\Rightarrow  |x|^k  \le  2^{k-1}(2^{ \max (\frac{k}{2}-1,1)}+2) |y|^k + 2^{ \max (\frac{k}{2}-1,1)+k} \E^{ \frac{k}{2}}[X^2] ,   \label{eq:Lp_last_bound}
\end{align} 
where \eqref{eq:applying_Triangle_Inequaltiy} follows from the reverse triangle inequality; \eqref{eq:Lp-type_Bounds} follows by using the bound  $|a +b|^k \le2^{k-1} (|a|^k+|b|^k), k \ge 1$; \eqref{eq:Applying_Jensen's_Inequality_to_cumulants} follows by using Jensen's inequality to 
\begin{align}
f_Y(y)&=\frac{1}{\sqrt{2\pi \sigma^2}} \E \left[ \eu^{-\frac{(y-X)^2}{2 \sigma^2}} \right]\\
&\ge \frac{1}{\sqrt{2\pi \sigma^2}} \eu^{-\frac{ \E \left[  (y-X)^2 \right]}{2 \sigma^2}} \\
&\ge \frac{1}{\sqrt{2\pi \sigma^2}} \eu^{-\frac{  y^2 +\E[X^2]}{\sigma^2}}; \text{ and} 
\end{align} 
\eqref{eq:Lp_last_bound} follows form the bound  $|a +b|^k \le2^{ \max(k-1,1)} (|a|^k+|b|^k), k \ge 0$.

Combining \eqref{eq:Prep_bound} and \eqref{eq:Lp_last_bound} concludes the proof.

\section{Proof of the Expression in \eqref{eq:Kappa_for_X_uniform_sphere}} 
\label{app:Proof_Cumulatns_shell}

The first ingredient of the proof is the following derivative  expression:
\begin{equation}
\frac{{\rm d}}{{\rm d } t} \frac{I_\nu(t)}{I_{\nu-1}(t)}= 1-\frac{2\nu-1}{t}   \frac{I_\nu(t)}{I_{\nu-1}(t)} - \left( \frac{I_\nu(t)}{I_{\nu-1}(t)} \right)^2, \label{eq:Derivative_Ratio_Bessel_Functions}
\end{equation} 
which follows form the identities $\frac{{\rm d}}{{\rm d } t} I_\nu(t)= \frac{\nu}{t} I_{\nu-1}(t) - \frac{\nu}{t}I_{\nu}(t)$ 
  and $\frac{{\rm d}}{{\rm d } t} I_\nu(t)= \frac{\nu}{t} I_\nu(t)+  I_{\nu+1}(t)$  \cite{watson1995treatise}. 

Next, consider the case of  $s_1 \neq s_2$. Then, 
\begin{align}
\frac{\partial \E[ X_{s_2} |\Y={\bf y}]  }{ \partial y_{s_1}    } 
& = R  y_{s_2} \frac{\partial  }{ \partial y_{s_1}    }   \frac{1 }{ \| {\bf y} \| }   \frac{I_{\frac{n}{2}}( R \| {\bf y}\|  ) }{ I_{\frac{n}{2}-1}(R \| {\bf y}\|) } \\
&=    \frac{ R  y_{s_2} y_{s_1} }{ \| {\bf y} \| }   \frac{I_{\frac{n}{2}}( R \| {\bf y}\|  ) }{ I_{\frac{n}{2}-1}(R \| {\bf y}\|) }  +    \frac{ R  y_{s_2}  }{ \| {\bf y} \| }  \frac{\partial  }{ \partial y_{s_1}    }    \frac{I_{\frac{n}{2}}( R \| {\bf y}\|  ) }{ I_{\frac{n}{2}-1}(R \| {\bf y}\|) } \\
&=    \frac{ R  y_{s_2} y_{s_1} }{ \| {\bf y} \| }   \frac{I_{\frac{n}{2}}( R \| {\bf y}\|  ) }{ I_{\frac{n}{2}-1}(R \| {\bf y}\|) }  +    \frac{ R^2  y_{s_2}  y_{s_1} }{ \| {\bf y} \| ^2}   \left(1 -\frac{n-1}{R \| {\bf y}\|}  \frac{I_{\frac{n}{2}}( R \| {\bf y}\|  ) }{ I_{\frac{n}{2}-1}(R \| {\bf y}\|) }  - \left( \frac{I_{\frac{n}{2}}( R \| {\bf y}\|  ) }{ I_{\frac{n}{2}-1}(R \| {\bf y}\|) }  \right)^2  \right),
\end{align} 
where the last equality follows by using the derivative expression in \eqref{eq:Derivative_Ratio_Bessel_Functions}. 

The proof for $s_1 = s_2$ follows along the similar lines. 

\section{Computation of \eqref{eq:Info_density_Gaussian} and \eqref{eq:Conditional_Variance_of_info_gaussian}}
\label{app:sec:Computation_of_conditional_var_of_info-dens}
If $X$ is standard normal, then
\begin{equation}
f_Y(y)=\frac{1}{\sqrt{ 2 \pi (1+\sigma^2)}} \eu^{-\frac{y^2}{ 2(1+\sigma^2)}}, \, y\in \mathbb{R}. 
\end{equation} 
Consequently, the information density is given by 
\begin{align}
 \iota_{P_{X Y}}(x;y)&=\log   \frac{f_{X|Y}(x|y)}{f_X(x)} \\
&= \log \frac{ f_{Y|X}(y|x)}{f_Y(y)} \\
&=  c-  \frac{ (y-x)^2}{2 \sigma^2} +  \frac{y^2}{2( 1+\sigma^2)}, 
\end{align}
where $c = \frac{1}{2}\log \frac{1+\sigma^2}{ \sigma^2}$. Hence,  the conditional variance is now given by
\begin{align}
\mathsf{Var}\left(    \iota_{P_{X Y}}(X; Y )  | X =x\right)
&= \mathsf{Var}\left(   - \frac{ (Y-X)^2}{2 \sigma^2} +  \frac{Y^2}{2( 1+\sigma^2)}  | X =x\right)\\
&=\mathsf{Var}\left(   - \frac{ N^2}{2 \sigma^2} +  \frac{(X+N)^2}{2( 1+\sigma^2)}  | X =x\right)\\
&=\mathsf{Var}\left(   - \frac{ N^2}{2 \sigma^2} +  \frac{ N^2 }{2( 1+\sigma^2)} + \frac{ N X  }{ 1+\sigma^2}  | X =x\right)\\
&=\mathsf{Var}\left(  - \frac{ 1}{2 \sigma^2( 1+\sigma^2)}    N^2+ \frac{ N X  }{ 1+\sigma^2}  | X =x\right)\\
&=  \frac{ \mathsf{Var}\left(N^2 \right)  }{4 \sigma^4( 1+\sigma^2)^2}    + \frac{ x^2 \mathsf{Var}( N )   }{( 1+\sigma^2)^2}  \label{eq:Covariance_N2_and_N}\\
&=  \frac{ 2 \sigma^4 }{4 \sigma^4( 1+\sigma^2)^2}    + \frac{ x^2  \sigma^2   }{( 1+\sigma^2)^2} , \label{eq:Variance_N2_and_N}
\end{align} 
where in \eqref{eq:Covariance_N2_and_N} we have used that ${\rm Cov}( N^2, N)=0$; and \eqref{eq:Variance_N2_and_N} follows by using that $\mathsf{Var}\left(N^2 \right)=2  \sigma^4 $ and $ \mathsf{Var}( N ) =\sigma^2$. 
This concludes the proof.

\section{Proof of Theorem~\ref{thm:multivariate_CGF_CE}}
\label{app:proof:thm:multivariate_CGF_CE}

It is sufficient to characterize only the second order partial derivatives $\frac{\partial }{ \partial y_j   \partial y_i} K_{\X}( {\bf t} |\Y={\bf y})$ and then apply a simple induction.  

 Let $U= \eu^{ {\bf t}^\mathsf{T} \X}$. Then, by using Theorem~\ref{thm:MainIdentity}, we have that the gradient of the cumulant generating function can be expressed as 
\begin{align}
\nabla_{{\bf y}}  K_{\X}( {\bf t} |\Y={\bf y}) 
&=\frac{ \nabla_{{\bf y}} \E[\eu^{ {\bf t}^\mathsf{T} \X}| \Y={\bf y} ]   }{ \E[\eu^{ {\bf t}^\mathsf{T} \X}| \Y={\bf y} ] } \\
&=\frac{ \boldsymbol{\mathsf{K}}_{\N}^{-1}   \boldsymbol{\mathsf{Cov}} ( \X, U | \Y={\bf y}) }{ \E[\eu^{ {\bf t}^\mathsf{T} \X}| \Y={\bf y} ] } \\
&=\frac{ \boldsymbol{\mathsf{K}}_{\N}^{-1} \left(  \E[  \X \eu^{ {\bf t}^\mathsf{T} \X} | \Y={\bf y}  ]  -\E[ \X |\Y={\bf y}] \E[  \eu^{ {\bf t}^\mathsf{T} \X}  | \Y={\bf y}]  \right) }{ \E[\eu^{ {\bf t}^\mathsf{T} \X}| \Y={\bf y} ] }  \\
&= \boldsymbol{\mathsf{K}}_{\N}^{-1} \left( \nabla_{\bf t}   \log \left(  \E[  \eu^{ {\bf t}^\mathsf{T} \X}  | \Y={\bf y}] \right)   -\E[ \X |\Y={\bf y}]   \right) \label{eq:Gradeint_Log_Property} \\
&= \boldsymbol{\mathsf{K}}_{\N}^{-1}  \nabla_{\bf t} K_{\X}( {\bf t} |\Y={\bf y})   -  \boldsymbol{\mathsf{K}}_{\N}^{-1}  \E[ \X |\Y={\bf y}],
\end{align} 
where in \eqref{eq:Gradeint_Log_Property} we have used that $\frac{\E[  \X \eu^{ {\bf t}^\mathsf{T} \X} | \Y={\bf y}  ]}{\E[\eu^{ {\bf t}^\mathsf{T} \X}| \Y={\bf y} ]} = \frac{ \nabla_{\bf t} \E[  \eu^{ {\bf t}^\mathsf{T} \X} | \Y={\bf y}  ]}{\E[\eu^{ {\bf t}^\mathsf{T} \X}| \Y={\bf y} ]}=\nabla_{\bf t}   \log \left(  \E[  \eu^{ {\bf t}^\mathsf{T} \X}  | \Y={\bf y}] \right)$. 

Consequently, the partial derivative with respect to $i$ is given by 
\begin{equation}
\frac{\partial }{ \partial y_i } K_{\X}( {\bf t} |\Y={\bf y}) 
 = {\bf k}_i^\mathsf{T}  \nabla_{\bf t} K_{\X}( {\bf t} |\Y={\bf y}) -  {\bf k}_i^\mathsf{T}   \E[ \X |\Y={\bf y}],  \label{eq:PartialDerivative_of_cumulant_gen_fun}
\end{equation}
where we have used that  ${\bf k}_i^\mathsf{T} =\mathbf{e}_i^\mathsf{T} \boldsymbol{\mathsf{K}}_{\N}^{-1} $    (i.e., the $i$-th row of $\boldsymbol{\mathsf{K}}_{\N}^{-1}$).

Next, by differentiating with respect to $j$, we arrive at 
\begin{align}
\frac{\partial^2 }{ \partial y_j   \partial y_i} K_{\X}( {\bf t} |\Y={\bf y})   &= \frac{\partial }{ \partial y_j   }  \left( {\bf k}_i^\mathsf{T}   \nabla_{\bf t} K_{\X}( {\bf t} |\Y={\bf y}) -  {\bf k}_i^\mathsf{T}      \E[ \X |\Y={\bf y}]  \right)\\
&=  {\bf k}_i^\mathsf{T}   \nabla_{\bf t}   {\bf k}_j^\mathsf{T}   \nabla_{\bf t} K_{\X}( {\bf t} |\Y={\bf y}) -  {\bf k}_i^\mathsf{T}   \frac{\partial }{ \partial y_j   }   \E[ \X |\Y={\bf y}]  ,\label{eq:using_partial_derivative_Expression}
\end{align} 
where   \eqref{eq:using_partial_derivative_Expression} comes from using the partial derivative formula in  \eqref{eq:PartialDerivative_of_cumulant_gen_fun}.

\section{Proof of Proposition~\ref{prop:PartialDerivattives_CE}}
\label{app:prop:PartialDerivattives_CE}

To see this note that $\frac{\partial^{j}  K_{\X}( {\bf t} |\Y={\bf y})  }{ \partial y_{s_1} \dots  \partial y_{s_{j}}   } |_{{\bf t}={\bf 0}} =0$ for every $j$
\begin{align}
\frac{\partial^{j}  K_{\X}( {\bf t} |\Y={\bf y})  }{ \partial y_{s_1} \dots  \partial y_{s_{j}}   } &=\frac{\partial^{j}  \log\left( \E[\eu^{ {\bf t}^\mathsf{T} \X}| \Y={\bf y} ] \right)  }{ \partial y_{s_1} \dots  \partial y_{s_{j}}   }\\
&=\frac{\partial^{j-1}   }{ \partial y_{s_1} \dots  \partial y_{s_{j-1}}}  \frac{\frac{\partial   }{ \partial y_{s_j}}    \E[\eu^{ {\bf t}^\mathsf{T} \X}| \Y={\bf y} ] }{ \E[\eu^{ {\bf t}^\mathsf{T} \X}| \Y={\bf y} ] }\\
&=  \frac{ g({\bf y})}{ ( \E[\eu^{ {\bf t}^\mathsf{T} \X}| \Y={\bf y} ] )^{ \max(1,2(j-1))}}. \label{eq:Gen_Quotion_Proof_Cumulant}
\end{align}  
We do not attempt to find the exact expression for  $g({\bf y})$ and only observe that it contains terms of the form  $\frac{\partial^{k}   }{ \partial y_{s_1} \dots  \partial y_{s_{k}}}   \E[\eu^{ {\bf t}^\mathsf{T} \X}| \Y={\bf y} ] $ for some $k \ge 1$. We next show that $\frac{\partial^{k}   }{ \partial y_{s_1} \dots  \partial y_{s_{k}}}   \E[\eu^{ {\bf t}^\mathsf{T} \X}| \Y={\bf y} ] $ evaluated at ${\bf t}= {\bf 0}$ is equal to zero, which through  \eqref{eq:Gen_Quotion_Proof_Cumulant} will lead to the desired conclusion.      To do this we use  the  Taylor expression for the multivariate moment generating function  given by \cite[Ch.~6]{kolassa2006series}  
\begin{align}
\E[\eu^{ {\bf t}^\mathsf{T} \X}| \Y={\bf y} ] =1+  \sum_{j=1}^\infty \sum_{ {\bf s} \in S(j)} \frac{1}{j!} \mu({\bf y}; s_1,...,s_j) t_{s_1} \dots t_{s_j} , 
\end{align} 
where  $\mu({\bf y}; s_1,...,s_j)=\E[ X_{s_1} \dots X_{s_j} | \Y={\bf y} ] $ and $S(j)=\{1,\ldots, n\}^j$ is set of all vectors of integers with $j$ components and all entries between $1$ and $n$.  Now we have that
\begin{align}
 \frac{\partial^{k}   }{ \partial y_{s_1} \dots  \partial y_{s_{k}}}   \E[\eu^{ {\bf t}^\mathsf{T} \X}| \Y={\bf y} ] \Big |_{ {\bf t}={\bf 0} } 
 & =  \sum_{j=1}^\infty \sum_{ {\bf s} \in S(j)} \frac{1}{j!}  \frac{\partial^{k}   }{ \partial y_{s_1} \dots  \partial y_{s_{k}}}   \mu({\bf y}; s_1,...,s_j) t_{s_1} \dots t_{s_j}   \Big |_{ {\bf t}={\bf 0} } \\
 &=0.
\end{align} 
This concludes the proof.   

  \section{Proof of Proposition~\ref{prop:Hessian_of_log_dist}}
\label{app:prop:Hessian_of_log_dist}
The proof of \eqref{eq:Hessian_of_Conditional_distribuiton_continious} follows by applying  Hatsell and Nolte identity in \eqref{eq:Hatsell-Nolte-general}  to the expressions in  Proposition~\ref{prop:Gradient_log_pmf_pdf} together with the identity  $  \boldsymbol{\mathsf{D}}^2_\mathbf{x} f({\bf x})=   \boldsymbol{\mathsf{J}}_\mathbf{x}  \nabla_{\bf x} f(\mathbf{x}) $. 

To show the general case observe that 
\begin{align}
 \boldsymbol{\mathsf{D}}^2_{\bf y}   \log \left(   \mathbb{P}[\X \in \mathcal{A} | \Y={\bf y}] \right) 
  &= \boldsymbol{\mathsf{J}}_\mathbf{y}  \nabla_{\bf y}    \log \left(   \mathbb{P}[\X \in \mathcal{A} | \Y={\bf y}] \right) \label{eq:Hessian_Jacobian_Gradient_idenity}\\
    &= \boldsymbol{\mathsf{J}}_\mathbf{y}  \boldsymbol{\mathsf{K}}_{\N}^{-1}    (  \E[  \X  | \Y={\bf y}, \X \in \mathcal{A}] -  \E[  \X   | \Y={\bf y}]     ) \label{eq:Using_Gradient_Indeity_cond_dist} \\
     &= \boldsymbol{\mathsf{J}}_\mathbf{y}     \left(  \frac{ \E[  \X \mathsf{1}_{ \mathcal{A} }(\X)  | \Y={\bf y}]}{ \mathbb{P}[\X \in \mathcal{A}| \Y={\bf y}]} -  \E[  \X   | \Y={\bf y}]    \right )   \boldsymbol{\mathsf{K}}_{\N}^{- 1} \label{eq:Jacobian_of_linear_Transf}\\
       &=  \Big(   \frac{   \boldsymbol{\mathsf{K}}_{\N}^{-1}   \boldsymbol{\mathsf{Cov}} ( \X, \X \mathsf{1}_{ \mathcal{A} }(\X)  | \Y={\bf y})  } { \mathbb{P}[\X \in \mathcal{A}| \Y={\bf y}]}     \notag\\
     &\quad -   \frac{ \boldsymbol{\mathsf{K}}_{\N}^{-1}   \boldsymbol{\mathsf{Cov}} ( \X,  \mathsf{1}_{ \mathcal{A} }(\X)  | \Y={\bf y})  \E^\mathsf{T}[  \X \mathsf{1}_{ \mathcal{A} }(\X)  | \Y={\bf y}] }{ \mathbb{P}^2[\X \in \mathcal{A}| \Y={\bf y}] }  \notag  \\
     &\quad  - \boldsymbol{\mathsf{K}}_{\N}^{-1}  \boldsymbol{\mathsf{Var}}(\X | \Y={\bf y})  \Big)   \boldsymbol{\mathsf{K}}_{\N}^{- 1 } \label{eq:Using_Main_Identity}\\
     &=  \boldsymbol{\mathsf{K}}_{\N}^{-1} \Big(   \E[ \X \X^{\mathsf{T}} | \Y={\bf y}, \X \in \mathcal{A}]   \notag\\
 &\quad -  \E[ \X   | \Y={\bf y}, \X \in \mathcal{A}]  \E[  \X^{\mathsf{T}} | \Y={\bf y}, \X \in \mathcal{A}] \notag\\
 & \quad - \boldsymbol{\mathsf{K}}_{\N}^{-1}  \boldsymbol{\mathsf{Var}}(\X | \Y={\bf y})  \Big) \label{eq:Rewritting_ConditionalCovariances}  \\
     &=  \boldsymbol{\mathsf{K}}_{\N}^{-1} \left( \boldsymbol{\mathsf{Var}}(\X | \Y={\bf y}, \X \in \mathcal{A}) - \boldsymbol{\mathsf{Var}}(\X | \Y={\bf y})     \right) \boldsymbol{\mathsf{K}}_{\N}^{- 1 }, \label{eq:Using_Definition_Conditiona_Variance}
\end{align} 
 where \eqref{eq:Hessian_Jacobian_Gradient_idenity} follows by using the identity  $  \boldsymbol{\mathsf{D}}^2_\mathbf{x} f({\bf x})=   \boldsymbol{\mathsf{J}}_\mathbf{x}  \nabla_{\bf x} f(\mathbf{x}) $; \eqref{eq:Using_Gradient_Indeity_cond_dist} follows by using  \eqref{eq:Indentity_Conditiona_Probaiblity};  \eqref{eq:Jacobian_of_linear_Transf} follows by using the property that $  \boldsymbol{\mathsf{J}}_{\bf x}   \boldsymbol{\mathsf{K}}_{\N}^{-1} f(\mathbf{x}) =  \boldsymbol{\mathsf{J}}_{\bf x}  f(\mathbf{x})   \boldsymbol{\mathsf{K}}_{\N}^{- \mathsf{T} } =\boldsymbol{\mathsf{J}}_{\bf x}  f(\mathbf{x})   \boldsymbol{\mathsf{K}}_{\N}^{- 1 } $; \eqref{eq:Using_Main_Identity} follows by using identity in \eqref{eq:MainIdentity} with $\U= \X \mathsf{1}_{ \mathcal{A} }(\X)$ and the quotient rule for differentiation;
\eqref{eq:Rewritting_ConditionalCovariances} follows by   rewriting the first covariance term as 
\begin{align}
 \frac{     \boldsymbol{\mathsf{Cov}} ( \X, \X \mathsf{1}_{ \mathcal{A} }(\X)  | \Y={\bf y})  } { \mathbb{P}[\X \in \mathcal{A}| \Y={\bf y}]}  
 = \E[ \X \X^{\mathsf{T}} | \Y={\bf y}, \X \in \mathcal{A}]   - \E[ \X   | \Y={\bf y}]  \E[  \X^{\mathsf{T}} | \Y={\bf y}, \X \in \mathcal{A}] , 
\end{align}
and the second covariance term as
\begin{align}
   &\frac{   \boldsymbol{\mathsf{Cov}} ( \X,  \mathsf{1}_{ \mathcal{A} }(\X)  | \Y={\bf y})  \E^\mathsf{T}[  \X \mathsf{1}_{ \mathcal{A} }(\X)  | \Y={\bf y}] }{ \mathbb{P}^2[\X \in \mathcal{A}| \Y={\bf y}] }\\
   &\quad=  \E[ \X   | \Y={\bf y}, \X \in \mathcal{A}]  \E[  \X^{\mathsf{T}} | \Y={\bf y}, \X \in \mathcal{A}] -\E[ \X   | \Y={\bf y}]  \E[  \X^{\mathsf{T}} | \Y={\bf y}, \X \in \mathcal{A}];  \text{ and} 
\end{align} 
and \eqref{eq:Using_Definition_Conditiona_Variance} follows from the definition of conditional variance. 
This concludes the proof.

  \section{Proof of Theorem~\ref{thm:ConsistentEB}}
  \label{app:thm:ConsistentEB}
  
  First, assume that 
\begin{equation}
 \sup_{|y| \le t_n} |  \widehat{f}_Y^{(m)}(y)-f^{(m)}_Y(y) | \le \epsilon_n,   m \in [0, \ldots, k] . \label{eq:EpsilonAssumption}
\end{equation} 
Second, using the generalized TRE identity in  \eqref{eq:generalTREhermite}  and the expression for $\widehat{m}_k(y)$ in \eqref{eq:CondExpecEstimator}, we have that 
\begin{equation}
\frac{1}{\sigma^{2k} }\left|  \E[   X^k |Y=y ]    -\widehat{m}_k(y)  \right| 
 \le     \sum_{m=0}^k 
  \frac{  {{k} \choose {m}}  \left|  (-i)^m H_{e_m} \left( i \frac{y}{\sigma} \right) \right|}{\sigma^m}    
      \left|    \frac{  \widehat{f}_Y^{(k-m)}(y)   }{ \widehat{f}_Y(y)}  -  \frac{ f_Y^{(k-m)}(y)   }{f_Y(y)}    \right| , \label{eq:afterTriangleInequality}
\end{equation} 
where the second absolute value term can be bounded as 
\begin{align}
  \left|    \frac{  \widehat{f}_Y^{(k-m)}(y)   }{ \widehat{f}_Y(y)}  -  \frac{ f_Y^{(k-m)}(y)   }{f_Y(y)}    \right|
  &  =    \left|    \frac{  \widehat{f}_Y^{(k-m)}(y) f_Y(y)- f_Y^{(k-m)}(y)  \widehat{f}_Y(y) }{ \widehat{f}_Y(y) f_Y(y)}     \right| \\
    & \le  \left|    \frac{  \widehat{f}_Y^{(k-m)}(y) -  f_Y^{(k-m)}(y)  }{ \widehat{f}_Y(y) }     \right|  +   \left|    \frac{ (   f_Y(y)- \widehat{f}_Y(y) )  f_Y^{(k-m)}(y) }{ \widehat{f}_Y(y)  f_Y(y)}     \right| \\
    & \le  \frac{\epsilon_n}{ \widehat{f}_Y(y) } +  \epsilon_n  \frac{  \left|    f_Y^{(k-m)}(y)  \right|  }{ \widehat{f}_Y(y)  f_Y(y)}    ,  \label{eq:BoundOnABSvalue}
\end{align} 
where in the last step we have used the assumption in \eqref{eq:EpsilonAssumption}.

The following lemma, whose proof can be found in Appendix~\ref{app:auxiliary_Results_approximation}, will be useful to further bound the terms in \eqref{eq:BoundOnABSvalue}. 
\begin{lem} \label{lem:auxiliary_Results_approximation} Let $\E[X^2]<\infty$ and assume \eqref{eq:EpsilonAssumption} holds.  Let   $ \psi_\sigma(y)= \sqrt{2 \pi \sigma^2}  \eu^{\frac{ y^2 +\E[X^2]}{  \sigma^2}} $ and assume that   
\begin{equation}
\epsilon_n   \psi_\sigma(t_n) \le 1. 
\end{equation}
 Then,   for $|y| \le t_n$
\begin{align}
 \frac{1}{f_Y(y)} &\le \psi_\sigma(y),  \label{eq:B1}  \\
\frac{1}{ \widehat{f}_Y(y) } & \le  \frac{1}{1-\epsilon_n  \psi_\sigma(y) }   \psi_\sigma(y), \label{eq:B2}  \\
 \left|    f_Y^{(k)}(y)  \right|  & \le    \frac{1}{ \sqrt{2 \pi \sigma^2}}  \frac{\sqrt{k!}}{\sigma^k}  , \label{eq:B3} \\
  \left|  H_{e_m} \left( i \frac{y}{\sigma} \right) \right| &\le  \sqrt{ m!} \eu^{ \frac{y^2}{4\sigma^2}} \le  \frac{\sqrt{ m!}}{\sqrt{ 2 \pi \sigma^2}}  \psi_\sigma(y) , \label{eq:B4} \\
\delta_{r,a}&=   \max_{t \in \mathbb{R}}\left| \mathbb{E}[ \widehat{f}_Y^{(r)}(t)]-f_Y^{(r)}(t) \right | \le   a   \sqrt{   \frac{ 4  (r+1)! }{ 3 \pi\sigma^{2(r+1)}}  },  \label{eq:B5}  \\
 v_r&= \int  \left | k^{(r+1)}(x) \right|  {\rm d} x  \le  \sqrt{(r+1)!} \sqrt{ \frac{2}{3}} \label{eq:B6}  . 
\end{align} 
\end{lem}

Combining \eqref{eq:afterTriangleInequality}, \eqref{eq:BoundOnABSvalue} and {\eqref{eq:B1}-\eqref{eq:B4}},  we arrive at
\begin{align}
 \sup_{ |y| \le  t_n} \left|  \E[   X^k |Y=y ]    -\widehat{m}_k(y)  \right| 
 & \le  C_{1,\sigma, k}     \frac{\epsilon_n  \psi_\sigma^2(t_n)   }{1-\epsilon_n  \psi_\sigma(t_n) }  \left(1+ C_{2,\sigma, k}   \psi_\sigma(t_n)    \right) \\
 &=\varepsilon_n,  \label{Eq:error_between_condtionals}
\end{align} 
where
\begin{equation}
 C_{1,\sigma, k} =   \sum_{m=0}^k 
 \frac{  {{k} \choose {m}}\sqrt{m!}}{\sigma^m}   ,  \,   C_{2,\sigma, k}=  \sqrt{\frac{2}{3}}  \max_{ m \in 0,\ldots, k}   \frac{ \sqrt{m!} }{\sigma^m}  . 
\end{equation} 

Now, choose $\epsilon_n >   \max_{m \in [0,\ldots, n]} \delta_{m,a} $ and observe that 
\begin{align}
\mathbb{P} \left [  \sup_{ |y| \le  t_n}   \left|  \E[   X^k |Y=y ]    -\widehat{m}_k(y)  \right|   \ge \varepsilon_n   \right]  
& \le  \sum_{m=0}^k  \mathbb{P} \left[   \sup_{ |y| \le  t_n}  |  \widehat{f}_Y^{(m)}(y)-f^{(m)}_Y(y) | >\epsilon_n  \right] \label{eq:Implication_for_error}\\
&  \le  \sum_{m=0}^k 2 \eu^{-2n \frac{a^{2m+2} (\epsilon_n - \delta_{m,a})^2}{ v_m^2} } ,\label{Eq:using_schuster_results}
\end{align} 
where in \eqref{eq:Implication_for_error} we have used the fact that \eqref{eq:EpsilonAssumption} leads to \eqref{Eq:error_between_condtionals};  and in \eqref{Eq:using_schuster_results} is due to the bound in  \cite[p.1188]{schuster1969estimation}.  Therefore, to complete the proof,  we need choose $\epsilon_n, a$ and $t_n$ such that 
\begin{align}
&\lim_{n \to \infty} \epsilon_n  \psi_\sigma^3(t_n)=0,  \\
& \lim_{n \to \infty}  t_n \to \infty,\\ 
& \epsilon_n  >   \delta_{m,a}, \forall \, m \in [0,\ldots, n], \\
&  \lim_{n \to \infty} n a^{2m+2} (\epsilon_n - \delta_{m,a})^2= \infty  , \forall \,   m \in [0,\ldots, k],
\end{align} 
The following choice satisfies the above equations:
\begin{align}
a&=  \frac{1}{n^u},  u \in \left(0, \frac{1}{2k +4}  \right), \\
\epsilon_n&= 2\delta_{m,a}=2  a  \max_{m \in [0,\ldots, k] }   \sqrt{   \frac{ 4  (k+1)! }{ 3 \pi\sigma^{2(k+1)}}  }, \\
t_n&= \frac{\sigma^2  \sqrt{ w  \log(n)} }{3} ,  w \in (0,u),
\end{align} 
and, for large enough $n$, result in 
\begin{equation}
 \varepsilon_n \le  \frac{C_{k,\sigma} }{n^{u-w}}, 
\end{equation}
where the constant $C_{k,\sigma}$  is independent of $n$.

  \section{Proof of Lemma~\ref{lem:auxiliary_Results_approximation}} 
  \label{app:auxiliary_Results_approximation} 
  
  To show \eqref{eq:B1}, we use Jensen's inequality and the bound $(a+b)^2 \le 2 (a^2+b^2)$
  \begin{align}
  f_Y(y) &=  \E \left[  \frac{1}{ \sqrt{2 \pi \sigma^2}}  \eu^{- \frac{(y-X)^2}{2 \sigma^2}} \right]\\
  & \ge    \frac{1}{ \sqrt{2 \pi \sigma^2}}  \eu^{- \frac{  \E \left[(y-X)^2 \right]}{2 \sigma^2}}\\
   & \ge    \frac{1}{ \sqrt{2 \pi \sigma^2}}  \eu^{- \frac{ y^2 +\E[X^2]}{  \sigma^2}} = \frac{1}{ \psi_\sigma(y) }. 
  \end{align} 
  
  To show \eqref{eq:B2} we use the assumption in \eqref{eq:EpsilonAssumption} and the bound in  \eqref{eq:B1}
  \begin{align}
      \frac{ 1 }{ \widehat{f}_Y(y)}& =   \frac{ 1 }{ \widehat{f}_Y(y)-  f_Y(y)  +   f_Y(y) }\\
     & \le  \frac{ 1 }{    f_Y(y)  -\epsilon_n} \label{eq:TheUglyBound2}\\
     & =   \frac{ 1 }{   1- \frac{\epsilon_n}{  f_Y(y) } } \frac{1}{  f_Y(y) }\\
     &\le \frac{ 1 }{   1- \epsilon_n \psi(y) } \psi(y). 
     \end{align} 
     
Next,  recall the following inequality \cite[eq.~22.14.17]{abramowitz1965handbook}: 
\begin{equation}
\left| H_{\eu_{r+1}}(x)   \right| \le  \sqrt{(r+1)!} \eu^{ \frac{x^2}{4} }. \label{eq:BoundOnHermitePolynomial}
\end{equation} 
First, the  above  inequality  immediately leads to the proof of  \eqref{eq:B4}.   Second,  to show \eqref{eq:B3},  observe the following sequence of steps:
\begin{align}
\left|  f_Y^{(k)}(t)  \right|&=  \left| \frac{ {\rm d}^{k}}{  {\rm d} t^{k}} \E \left[  \frac{1}{ \sqrt{2 \pi \sigma^2}}  \eu^{- \frac{(t-X)^2}{2 \sigma^2}} \right]  \right| \\
 &= \left| \E \left[  \frac{1}{ \sqrt{2 \pi \sigma^2}}  \frac{ {\rm d}^{k}}{  {\rm d} t^{k}}  \eu^{- \frac{(t-X)^2}{2 \sigma^2}} \right] \right| \\
  &= \left|  \E \left[  \frac{1}{ \sqrt{2 \pi \sigma^2}}  \frac{(-1)^k}{\sigma^k} H_{e_k} \left( \frac{t-X}{ \sigma}  \right)  \eu^{- \frac{(t-X)^2}{2 \sigma^2}} \right] \right| \\
    &\le  \E \left[  \frac{1}{ \sqrt{2 \pi \sigma^2}}  \frac{1}{\sigma^k} \left|  H_{e_k}  \left( \frac{t-X}{ \sigma}  \right) \right|   \eu^{- \frac{(t-X)^2}{2 \sigma^2}} \right] \\
  & \le  \E \left[  \frac{1}{ \sqrt{2 \pi \sigma^2}}  \frac{1}{\sigma^k}  \sqrt{k!} \eu^{ \frac{(t-X)^2}{4 \sigma^2} } \eu^{- \frac{(t-X)^2}{2 \sigma^2}} \right]  \label{eq:UsingHermitePolynomial}\\
  & \le   \frac{1}{ \sqrt{2 \pi \sigma^2}}  \frac{\sqrt{k!}}{\sigma^k}  ,  \label{eq:bound_exp(-x) <=1}
 \end{align}  
 where the inequality in \eqref{eq:UsingHermitePolynomial} follows from the inequality in \eqref{eq:BoundOnHermitePolynomial}; and the  inequality in \eqref{eq:bound_exp(-x) <=1} follows by using $ \eu^{-3 \frac{(t-X)^2}{4 \sigma^2}} \le 1$. 
 
 To show \eqref{eq:B5}, first  note that  
\begin{equation}
 \mathbb{E}[\widehat{f}_Y^{(r)}(t)]=  \int  \frac{1}{a} \frac{ {\rm d}^r }{ {\rm d} t^r}  k \left( \frac{t-y}{a} \right)   f_Y(y) {\rm d} y  = \int  \frac{1}{a}   k \left( \frac{t-y}{a} \right)   f_Y^{(r)}(y) {\rm d} y  ,
\end{equation}
where we have used integration by parts. Then, the bias can be bounded as follows: 
\begin{align}
 \left| \mathbb{E}[\widehat{f}_Y^{(r)}(t)]-f_Y^{(r)}(t) \right | 
 &=  \left| \int  \frac{1}{a}   k \left( \frac{t-y}{a} \right)  \left(  f_Y^{(r)}(y) -f_Y^{(r)}(t) \right){\rm d} y  \right| \\
  &=   \left| \int     k \left( y \right)  \left(  f_Y^{(r)}(t+a y) -f_Y^{(r)}(t) \right){\rm d} y   \right| \\
    &\le  \sup_{t \in \mathbb{R}} \left| f_Y^{(r+1)}(t)  \right| \int     k \left( y \right)  a  |y|  {\rm d} y      \\
        &= a \sqrt{ \frac{2}{\pi}}     \sup_{t \in \mathbb{R}} \left| f_Y^{(r+1)}(t)  \right| \\
        & \le  \frac{1}{ \sqrt{2 \pi \sigma^2}}  \frac{\sqrt{k!}}{\sigma^k}, 
         \label{eq:BoundOnDifferenceOfExpectedValue}
 \end{align} 
 where the last bound is due to \eqref{eq:B3}. 
 
 Finally, to show \eqref{eq:B6},  we use the inequality in \eqref{eq:BoundOnHermitePolynomial}
\begin{align}
v_r&= \int  \left | k^{(r+1)}(x) \right|  {\rm d} x \\
&=  \int         \left| H_{\eu_{r+1}}(x)  k(x)  \right| {\rm d} x \\
&\le   \int   \sqrt{(r+1)!} \eu^{ \frac{x^2}{4} } k(x) {\rm d} x \\
&=  \sqrt{(r+1)!} \sqrt{ \frac{2}{3}}. 
\end{align}
This concludes the proof.

\section{Proof of Lemma~\ref{lem:Lancos_approximations}}
\label{app:lem:Lancos_approximations}

The proof largely follows the derivations in \cite{groetsch1998lanczos} and \cite{rangarajan2005lanczos} and explicitly states the constants of the approximation.

To show \eqref{eq:bound_Lancos_derivative_error}, we use  Taylor's remainder theorem  
\begin{align}
f(x+ht)&= f(x)+ ht f'(x) + \ldots +  \frac{ (h t)^k}{k!} f^{(k)}(x)  + \frac{(h t)^{k+1}}{ (k+1)!} f^{(k+1)}(x) + \frac{(ht)^{k+2}}{ (k+2)!}  f^{(k+2)}(\tau),
\end{align} 
for some $\tau$ between $x$ and $x+ht$, which leads to 
\begin{align}
D_h^{(k)} f(x)&= \frac{c_k}{h^k }  \int_{-1}^1 f(x+h t)  P_k(t) {\rm d} t \\
&= f^{(k)}(x)   +   \frac{c_k h^2}{(k+2)! }  \int_{-1}^1 f^{(k+2)}(\tau)  t^{k+2} P_k(t) {\rm d} t,
\end{align}
where we have used the following properties of the Legendre polynomials \cite[Ch.~22]{abramowitz1965handbook}: 
\begin{align}
  \int_{-1}^1 t^m P_k(t) {\rm d} t&= 0, \,  m\in [0:k-1],\\
    \int_{-1}^1 t^k P_k(t) {\rm d} t&= \frac{k!}{c_k} ,  \\
        \int_{-1}^1 t^{k+1} P_k(t) {\rm d} t&= 0 . 
\end{align}

Consequently,
\begin{align}
|D_h^{(k)} f(x)-  f^{(k)}(x) | 
 &\le   \frac{c_k h^2}{(k+2)! }  \int_{-1}^1 |  f^{(k+2)}(\tau)  t^{k+2} P_k(t) |  {\rm d} t\\
&\le   \frac{c_k h^2}{(k+2)! }  M_{k+2}  \int_{-1}^1 |   P_k(t) |  {\rm d} t \label{eq:Bound_f_tau}\\
&\le   \frac{c_k h^2}{(k+2)! }  M_{k+2}   \frac{2}{ \sqrt{  2k+1}}  \label{eq:Lagender_Polynomial_bound_jensen}\\
&= \alpha_k  M_{k+2} h^2 ,
\end{align} 
where \eqref{eq:Bound_f_tau} follows by using that   $\tau$  is between $x$ and $x+ht$ and hence $|f^{(k+2)}(\tau) | \le M_{k+2}$;  and \eqref{eq:Lagender_Polynomial_bound_jensen} follows by using Jensen's inequality and $\int_{-1}^1 |   P_k(t) |^2  {\rm d} t = \frac{2}{2k+1}$ to show that   
\begin{equation}
\int_{-1}^1|P_k(t)| {\rm d} t   \le   2 \sqrt{ \int_{-1}^1|P_k(t)|^2  \frac{1}{2} {\rm d} t  } = \frac{2}{ \sqrt{  2k+1}}.  \label{eq:norm_Legender_polynomial}
\end{equation} 
This completes the proof of the bound in \eqref{eq:bound_Lancos_derivative_error}.
 
To show \eqref{eq:bound_Lancos_derivative_error_mismatch} observe the following sequence of steps:
\begin{align}
 | D_h^{(k)} g(x) -  f^{(k)}(x) |  
& \le  | D_h^{(k)} f(x) -  f^{(k)}(x) | +  | D_h^{(k)} g(x) -  D_h^{(k)} f(x) | \label{eq:Apply_Triangular_Inequality_Lancos}\\
& =  \alpha_k  M_{k+2} h^2   +  \left|  \frac{c_k}{h^k }  \int_{-1}^1 \left( g(x+h t) - f(x+h t) \right) P_k(t) {\rm d} t      \right|\\
& \le  \alpha_k  M_{k+2} h^2   +   \frac{c_k}{h^k }  \epsilon  \int_{-1}^1|P_k(t)| {\rm d} t   \label{eq:bound_Lancos_derivative_error_applying}   \\
& \le \alpha_k  M_{k+2} h^2    +   \frac{c_k}{h^k }  \epsilon  \sqrt{ \frac{2}{2k+1}} \label{eq:Lagender_Polynomial_bound_jensen_v2}  \\
& = \alpha_k  M_{k+2} h^2    +    \frac{\beta_k   \epsilon}{ h^k}  ,
\end{align} 
where \eqref{eq:Apply_Triangular_Inequality_Lancos} follows by using the triangular inequality;  \eqref{eq:bound_Lancos_derivative_error_applying} follows by using that $\sup_{x \in I_h} |f (x)-g(x)| \le \epsilon$; and \eqref{eq:Lagender_Polynomial_bound_jensen_v2} follows by using the bound in \eqref{eq:norm_Legender_polynomial}.
This completes the proof.

\end{appendices}

\bibliography{refs}
\bibliographystyle{IEEEtran}

\end{document}